\documentclass[12pt]{iopart}
\ifx\pdfoutput\undefined
   \usepackage[dvips]{graphicx}
   \else
   \usepackage[pdftex]{graphicx}
   \fi
   \usepackage{epstopdf}

\usepackage{hyperref}
\usepackage{amssymb}
\usepackage{graphicx}
\usepackage{color}
\usepackage{sidecap}
\makeatletter

\begin{document}
   \DeclareGraphicsExtensions{.pdf,.gif,.jpg.eps}

\title{A Phenomenological Theory of the Anomalous Pseudogap Phase in Underdoped Cuprates}

\author{T. M. Rice$ ^{1,2,3}$, Kai-Yu Yang$^{1,2,4}$ and F. C. Zhang$^{2}$}

\address{
$^{1}$Institut fur Theoretische Physik, ETH Zurich,CH-8093 Zurich, Switzerland \\
$^{2}$ Center for Theoretical and Computational Physics and Department of Physics, The University of Hong Kong, Hong Kong SAR, China \\
$^{3} $Condensed Matter Physics and Materials Science Department, Brookhaven National Laboratory, Upton, NY 11973, USA \\
$^{4}$ Department of Physics, Boston College, Chestnut Hill, Massachusetts 02467, USA
}
\date{\today}
\begin{abstract} 
The theoretical description of the anomalous properties of the pseudogap phase in the underdoped region of the cuprate phase diagram lags behind the progress in spectroscopic and other experiments. A phenomenological ansatz, based on analogies to the approach to Mott localization at weak coupling in lower dimensional systems, has been proposed by Yang, Rice and Zhang [Phys. Rev. B \textbf{73} (2006),174501]. This ansatz has had success in describing a range of experiments. The motivation underlying this ansatz is described and the comparisons to experiment are reviewed. Implications for a more microscopic theory are discussed together with the relation to theories that start directly from microscopic strongly coupled Hamiltonians. 
\end{abstract}
\maketitle

\tableofcontents

\section{Introduction}
  A quarter century has passed since the unexpected discovery by George Bednorz and Alex Mueller \cite{BM86} of high temperature superconductivity in doped cuprate insulators. Their breakthrough radically altered the field of superconductivity and awoke it from a long period of reduced activity. The great theoretical breakthrough of Bardeen, Cooper and Schrieffer (BCS) occurred some thirty years previously. Their BCS microscopic theory of superconductivity caused by a weak electron-electron attraction arising through the exchange of phonons, in one fell swoop turned superconductivity from the most puzzling and mysterious into the best understood phenomenon in metals. But by the mid-eighties the last details of BCS theory had long been worked out and the highest known transition temperature had languished slightly above $20K$ for many years. Even a value of 30K was looking decidedly optimistic. The initial measurements on cuprates \cite{BM86} reached only $T_{c}$ values in the upper thirties, but this clear jump from the low twenties and the fact, quickly recognized by P. W. Anderson \cite{PWA87}, that the superconductivity appeared in a lightly doped Mott insulator led him to propose that this was a radically different form of superconductivity. In a Mott insulating state the onsite Coulomb repulsion between electrons scales to strong coupling, localizing electrons and dominating the physics on the low energy scale, opposite to the case of conventional superconductors.

P. W. Anderson in his early paper \cite{PWA87} went much further. He realized that the relevant element in the cuprates is the CuO$_2$ planes. In the parent undoped Mott insulator, the Cu$^{2+}$ ions form an antiferromagnetically (AF) coupled square lattice of $S$=1/2 spins, which could possibly realize the resonant valence bond (RVB) liquid of singlet spin pairs. In a RVB state the large energy gain of the singlet pair state, resonating between the many spatial pairing configurations, drives strong quantum fluctuations --- strong enough to suppress long range antiferromagnetic order. Further he showed that the RVB wavefunction could be expressed also as a Gutzwiller projected BCS pair wavefunction. In the undoped state the Gutzwiller projector, which acts to enforce single occupancy on each site, localizes an electron in a half-filled band with a $S$=1/2 moment on every lattice site, leading to a Mott insulator. Hole doping allows electrons to move through empty sites and introduces charge fluctuations of the singlet pairs and superconductivity. This elegant scenario in our view remains to this day to be the most compelling intuitive picture of the spectacular superconductivity of cuprates \cite{PWA04}.  Later refinements change a number of details. The parent insulators are actually AF ordered, but with the sublattice moment reduced by strong quantum fluctuations. The pairs are in a relative \textit{d}-wave not \textit{s}-wave configuration. But these modifications do not change the essence of Anderson's RVB picture.

  To proceed further, we need an explicit model Hamiltonian to describe the low energy physics. The Cu-ions have 4 nearest neighboring (nn) O-neighbors in the plane and usually, though not always, apical O neighbors lying further away above and/or below the CuO$_{2}$ plane. In this configuration the uppermost band is associated with the antibonding  $3d_{x^{2}-y^{2}}$ Cu -- $2p_{x(y)}$ O orbital lying in the plane. In density functional (LDA) band structure calculations \cite{PIRMP89} this band is half filled and lies isolated above the other 3d and 2p bands. The LDA calculations give a metallic ground state so a strong short range Coulomb repulsion needs to be added to obtain the observed Mott insulator in the undoped compounds. The simplest Hamiltonian consists of a single band with both nn and next nearest neighboring (nnn) hopping matrix elements in a tight binding representation and an onsite repulsion. This Hubbard Hamiltonian is a good starting point if we start from the strongly hole doped side of the cuprate phase diagram. In this limit a full Fermi surface in good agreement with the LDA results has been determined in angle resolved photoemission emission spectroscopy (ARPES)  \cite{PlatPRL05} and quantum oscillation experiments \cite{VigNat08}. If we start from the other limit, namely strongly underdoped cuprates, it is convenient to integrate out high-energy configurations. This procedure was applied by Zhang and Rice \cite{ZRPRB87} directly to the overlapping $3d_{x^{2}-y^{2}}$ Cu -- $2p_{x(y)}$ O orbitals in the CuO$_{2}$ plane. They showed that doped holes enter a spin singlet Cu$^{3+}$ oxidation state and move through a background of $S=1/2$ Cu$^{2+}$ ions, thereby justifying a $t-J$ model similar to the strong coupling limit of the Hubbard model. These single band Hubbard and $t-J$ models on a square lattice look at first glance to be deceptively straightforward. In fact we are dealing with a strongly interacting fermion model, which cannot be expanded perturbatively in a small parameter around a well-described limit. The analysis of this strongly correlated model poses a clear challenge to theory.

\begin{figure}[tf]
\centerline
{
\includegraphics[width = 8.0cm, height =4.5cm, angle= 0]
{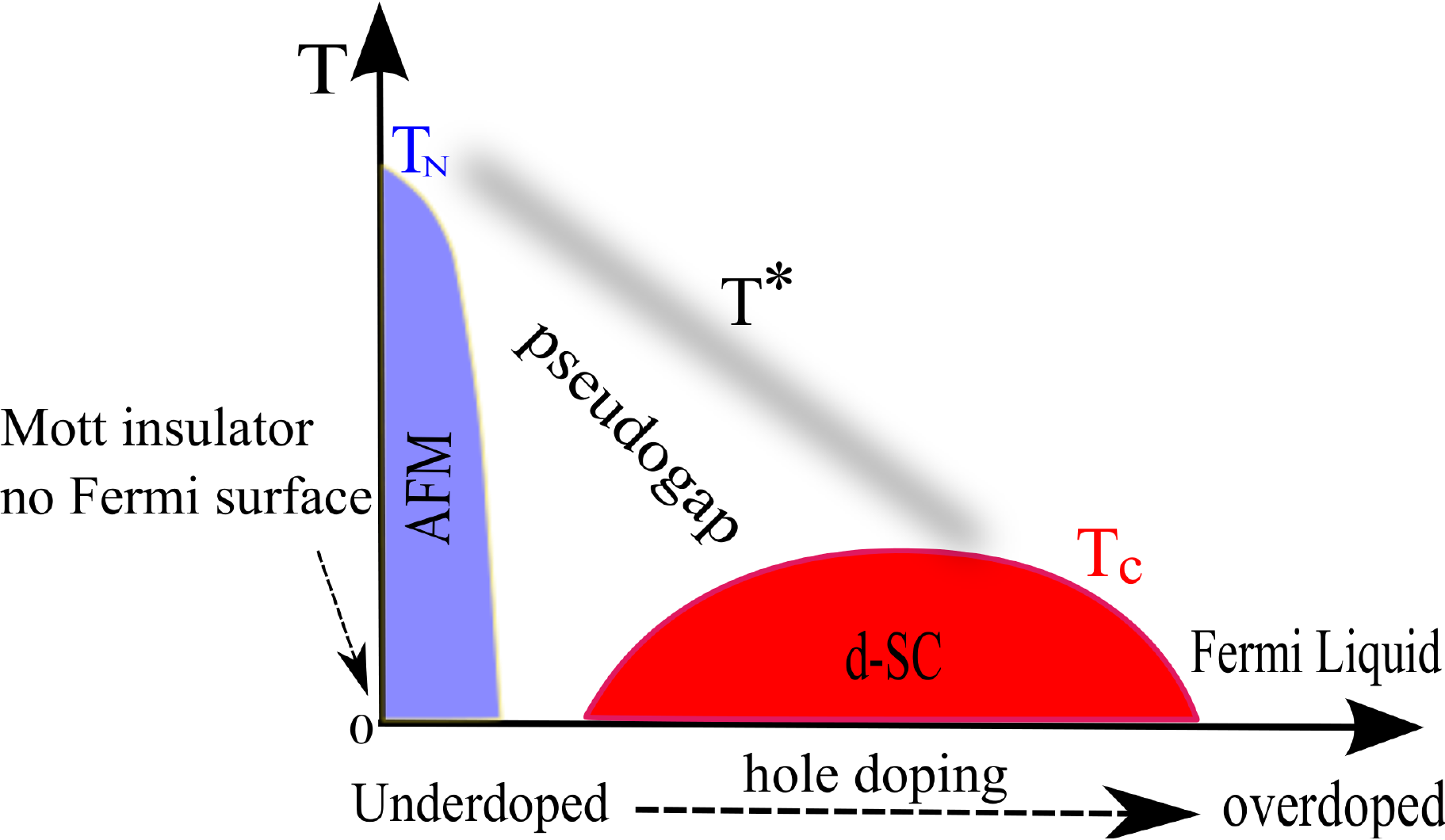}
}
 \caption[]
 {(Color online) The boundary between the antiferromagnetically ordered state (denoted by AFM) and the \textit{d}-wave superconductor (denoted by d-sc.) is uncertain. The overdoped Fermi liquid has a full Fermi surface while the stoichiometric Mott insulator has a charge gap.
 }
\label{fig:phase-diagram}
\end{figure}

  The phase diagram of the cuprates is by now quite well established. The two limits of a standard Landau Fermi liquid metallic state at strong overdoping and an AF ordered Mott insulator in the parent stoichiometric compound, are well understood. In between as schematically illustrated in Fig.\ref{fig:phase-diagram}, we are faced with a complex evolution as the full Fermi surface at overdoping, evolves as the doping is reduced and disappears completely at zero doping.

    Starting from the overdoped side of the phase diagram, two changes appear as the hole density is reduced from the standard metallic state. Superconductivity appears with \textit{d}-wave pairing symmetry below a critical hole density. The normal state also changes. In a recent series of experiments, Nigel Hussey and collaborators studied the angle dependent magnetoresistance oscillations (AMRO) on single layer Tl-cuprates in a high magnetic field \cite{NH07, NH08} --- high enough to suppress superconductivity and follow the normal state down to low temperatures. From these AMRO experiments they determined the temperature ($T$) and angular dependence of the inelastic scattering rate around the Fermi surface. They found that the onset of superconductivity coincided with the onset of a strongly anisotropic linear in T term in the inelastic scattering rate. The anisotropic inelastic scattering rate causing the breakdown of the standard Landau $T^{2}$-behavior, peaks in the antinodal directions and implies the presence of an anomalous and anisotropic  strong scattering vertex for low energy quasiparticles connecting these antinodal regions in $\mathbf{k}$-space. As will be discussed later such anomalous behavior was foreshadowed by functional renormalization group (FRG) calculations on a single band Hubbard model a decade earlier \cite{CH01,CH02}

   Starting from the undoped side a $t-J$ model describes the doped Mott insulator as a dilute density of holes moving in a background of an AF coupled square lattice of  $S$ =1/2 spins. The motion of a hole rearranges the spin configuration leading to strong coupling between the two degrees of freedom. In the past two decades many methods, both numerical and analytical, have been employed to analyze this $t-J$ model, e.g. renormalized mean field theories (RMFT) and numerical Monte Carlo sampling of variational wavefunctions (VMC) for Gutzwiller projected fermionic wavefunctions \cite{TKLee-PRB-06}. Another set of theories implements the Gutzwiller constraint in terms of a gauge theory and slave boson formulation or a slave fermion and Schwinger boson formulation. These methods have been extensively reviewed in a series of recent articles by Lee, Nagaosa and Wen \cite{LNW06},  Edegger, Muthukumar and Gros \cite{EMG08}, Lee \cite{PAL08}, and Ogata and Fukuyama \cite{OgFu08}. An earlier review by Dagotto covered numerical approaches \cite{Da94} and a survey of the current status of various theories has recently been published by Abrahams \cite{Ab10}. In this review we will not cover these approaches again and refer the reader instead to these comprehensive reviews.

  These approaches successfully explain the suppression of long range AF order as holes are introduced. One issue, which remains open, concerns the possible coexistence of \textit{d}-wave superconductivity and long range AF order. Earlier experiments found an intermediate region with a disordered spin glass separating the two ordered phases but recently NMR experiments on multilayer Hg-cuprates have been interpreted as evidence for coexistence of both broken symmetries \cite{Mu06, Mu08}. The multilayered Hg-cuprates are undoubtedly cleaner especially in the inner layers, than the acceptor doped single layer cuprates but suffer from the complications of interlayer coupling between adjacent CuO$_2$ planes with different hole densities.

  Beyond a critical hole density the so called pseudogap phase appears --- the most puzzling and anomalous region of the phase diagram. A comprehensive review of the unusual experimental results that appear in this phase was written by Timusk and Statt a decade ago \cite{TS99}.  The anomalous behavior was first observed in early NMR experiments on underdoped cuprates, which showed a dramatic drop in the uniform spin susceptibility deduced from Knight shift data \cite{Wa90, AL89}. This drop is too large to be explained by the onset of AF spin correlations and instead is evidence for singlet pair formation, in line with the central thesis of the RVB theory. Another spectacular anomaly that characterizes the pseudogap phase appeared in ARPES experiments. They showed that the full Fermi surface in the overdoped region shrinks down to 4 Fermi arcs, centered on each nodal direction \cite{NO98} in the underdoped region. These Fermi arcs mean there is only a partial gapping of the Fermi surface hence the name pseudogap.  We will not review here the full experimental characterization of the pseudogap phase and instead refer the reader to the review by Timusk and Statt \cite{TS99}. This gives a complete account of the experiments up till a decade ago. In recent years further experiments particularly ARPES, scanning tunneling microscopy (STM) and quantum oscillation (QO) experiments have expanded this characterization substantially. These will be discussed later when we come to the comparison between experiment and the phenomenological theory that is the main thrust of this review.

    The theoretical explanation of the whole range of anomalous phenomena that appear in the pseudogap phase is clearly a big challenge. What is required is a consistent theoretical description that covers the full range and variety of the anomalous experimental results. While there has been considerable progress on the analysis based directly on the $t-J$ and Hubbard Hamiltonians, at present, one is far short of this ambitious goal. In this review we will be less ambitious. Our aim is to obtain a comprehensive theory based on a phenomenological approach. A key element is an ansatz for the single electron propagator put forward a few years ago by Yang, Rice and Zhang \cite{YRZ06}. This YRZ ansatz interpolates between the FRG results \cite{CH01,CH02}, which describe the breakdown of standard Landau Fermi liquid behavior as the hole density is reduced in the overdoped metal, and the RMFT for the RVB \cite{RMF} behavior at underdoping. The success of any phenomenological approach is measured by its ability to interrelate the wide variety of anomalous phenomena and to develop intuition about the physics underlying the pseudogap phase. Below we review our recent progress in testing the YRZ ansatz against a series of spectroscopic measurements --- ARPES investigations \cite{YA09}, particularly those showing particle-hole asymmetry along the Fermi arcs \cite{YAPDJ08,YAPDJ10}, AIPES  (Angle Integrated PES) \cite{HA09} results on the hole density of states across the pseudogap region, and STM measurements analyzing quantum interference to obtain the Fermi arcs underlying the coherent Bogoliubov quasiparticle spectra at low temperatures in the superconducting state \cite{KOH08}. Recently Carbotte, Nicol and coworkers \cite{IL09, CAR09,CAR10,LeB10} have used the YRZ ansatz to analyze a series of experiments in the pseudogap phase. They obtained good fits to the evolution of in-plane optical conductivity \cite{IL09}, the specific heat \cite{CAR09}, London penetration depth \cite{CAR10} and Raman scattering \cite{LeB10} with changing doping and temperature in the pseudogap phase. Raman scattering experiments were also analyzed earlier by Bascones and Valenzuela \cite{VB07}. These fits require different dependences for the antinodal RVB gap and the maximum superconducting gap on the Fermi arcs. These differences are central to the two-gap scenario, which has been discussed by many authors.  For a recent review see Hufner et al. \cite{HH08}.

  The analysis by Konik, Rice and Tsvelik (KRT) \cite{KRT06} of Fermi pockets in a 2-dimensional array of 2-leg Hubbard ladders at weak coupling and densities near half-filling, was also an important input in the YRZ ansatz. Recently KRT extended their analysis to arrays of 4-leg Hubbard ladders \cite{KR10}. In this case in a certain parameter range, they derived a low energy model with small Fermi pockets coupled to finite energy \textit{d}-wave Cooperons associated with gapped regions in $\mathbf {k}$-space. They suggested that such models have a wider validity and are relevant to the cuprates. Related suggestions were made earlier on the basis of FRG calculations on the 2-dimensional Hubbard model \cite{CH01,CH02}.
Other interesting systems which show closely related behaviour are the wide Hubbard ladders with many legs, recently reviewed by LeHur and Rice \cite{LeH09}.
  As we will discuss further below, these analyses of related models suggest there is scope to further develop the phenomenological model to better describe the interrelationship between \textit{d}-wave superconductivity and AF fluctuations in the cuprates. The close interplay between these two is at the heart of the RVB proposal. The recent discovery that the static stripe phase, that occurs most prominently in La$_{2-x}$Ba$_{x}$Cu$_{2}$O$_{4}$ at $x=1/8$ \cite{Li07},  consists of spatially oscillating superconductivity coexisting with charge and spin density waves demonstrates just how subtle this interplay is.

An outline of the review follows. In Section 2 we discuss the approach from overdoping concentrating on the manner in which the Landau Fermi liquid description at large doping breaks down as the pseudogap phase is approached. A summary of the general features of the pseudogap phase appears in Section 3. The YRZ ansatz for the single particle propagator is introduced in Section 4 with a comparison to recent spectroscopic measurements. In Section 5 the comparison to experiment is further developed with a review of the recent calculations applying the YRZ ansatz to describe anomalous properties of the pseudogap phase. The ansatz has implications for the underlying microscopic theory of superconductivity and antiferromagnetism in the pseudogap phase, which are summarized in Section 6. There are many theoretical proposals for the pseudogap phase and a comparison to several which are closest to the current theory is the subject of Section 7. The conclusions of the review are in Section 8.

\section{ Breakdown of the Landau Fermi Liquid at Overdoping}
   We start the discussion of the cuprate phase diagram in the overdoped region and examine how the standard Landau Fermi liquid state found in heavily overdoped cuprates breaks down as the hole density, $x$, is reduced. The onset of superconductivity in a Fermi liquid is caused by a divergence of the particle-particle scattering amplitude in the Cooper channel. The cuprates however show more complex behavior as the highly anomalous pseudogap state is approached when $x$ is reduced away from the overdoped region of the phase diagram. A key question then is what is the form of the additional divergent scattering processes that signal the breakdown of Landau theory associated with the crossover into the pseudogap phase. Identifying these additional scattering processes gives valuable insight into the physics of the anomalous pseudogap phase.

\subsection{Theory : Functional Renormalization Group Analysis of the Single Band Hubbard Model in 2-dimension}
    In the overdoped region of the phase diagram, the effects of the onsite repulsion are less dramatic than in the underdoped region, so that a single band Hubbard model description with moderate interactions may be an appropriate starting point. In analyzing the Hubbard model it is important to use a method that treats all the possible instabilities in an unbiased manner. This is necessary if we want to identify the instabilities that drive the anomalous pseudogap phase that appears as  the hole density crosses into the underdoped region. A second desirable feature is good resolution in $\mathbf{k}$-space since we want to examine the fate of the $\mathbf{k}$-space Fermi surface and the \textit{d}-wave pairing instability in the Cooper channel.

   The Functional Renormalization Group (FRG) method satisfies both requirements. It is however limited to moderate values of the interaction by the one loop approximation which is very difficult to improve on. But this limitation as we shall see, still allows a reasonable, if fully quantitative, description of the breakdown of Landau theory. Numerical FRG calculations for the 2-dimensional Hubbard model were started by Zanchi and Schulz \cite{ZS97,ZS00,ZS01} some years ago. They examined the simplest tight binding model with only nn hopping in the kinetic energy. This choice introduces special features such as a perfectly nesting form of the half-filled Fermi surface, which also contains two divergent van Hove singularities. Later Honerkamp, Salmhofer and collaborators \cite{CH01,CH02} extended the calculations to include a nnn hopping term in the kinetic energy. This is necessary to reproduce the observed cuprate Fermi surface. It also removes perfect nesting and moves the van Hove singularities away from half-filling. The kinetic energy includes nn and nnn hopping on a square lattice with matrix elements denoted as $t$ and $t'$ respectively. In the Hubbard model the interaction between electrons is reduced to an onsite Coulomb repulsion U,
\begin{eqnarray}
H = - \sum_{<i,j>,\sigma} t c^{\dag}_{i,\sigma}c_{j,\sigma} -  \sum_{<<i,j>>,\sigma} t' c^{\dag}_{i,\sigma}c_{j,\sigma} + U \sum_{i} n_{i,\uparrow}n_{i,\downarrow}
\label{eq:Hubbard}
\end{eqnarray}

   The FRG method can be formulated in several ways. In the Wilsonian form, successive energy shells are integrated out to arrive at the low energy effective Hamiltonian. In this form the energy cutoff, $\Lambda$ is the flow parameter in the RG equations that determines the effective interaction vertex $V_{\Lambda}(\mathbf{k}_{1},\mathbf{k}_{2},\mathbf{k}_{3})$. This vertex is a function of 3 wave vectors (the fourth is determined by momentum conservation including umklapp processes) leading to a functional RG flow. To handle this flow numerically, Zanchi and Schulz \cite{ZS97} divided the Brillouin zone into discrete patches, assuming a constant value in each patch set. Later Honerkamp and Salmhofer \cite{HS01} developed an alternative formulation using the temperature as the RG flow parameter. This $T$-flow method has the advantage that it includes possible small-Q instabilities, e.g., Pomeranchuk and Stoner instabilities, which are absent in the Wilsonian formulation. However in the parameter range of interest here, such instabilities are never the dominant ones \cite{CH02}. Below we summarize the main results of the Wilsonian FRG calculations \cite{CH01} and refer the reader to the original papers for more details.

\begin{figure}[tf]
\centerline
{
\includegraphics[width = 10.5cm, height =7.5cm, angle= 0]
{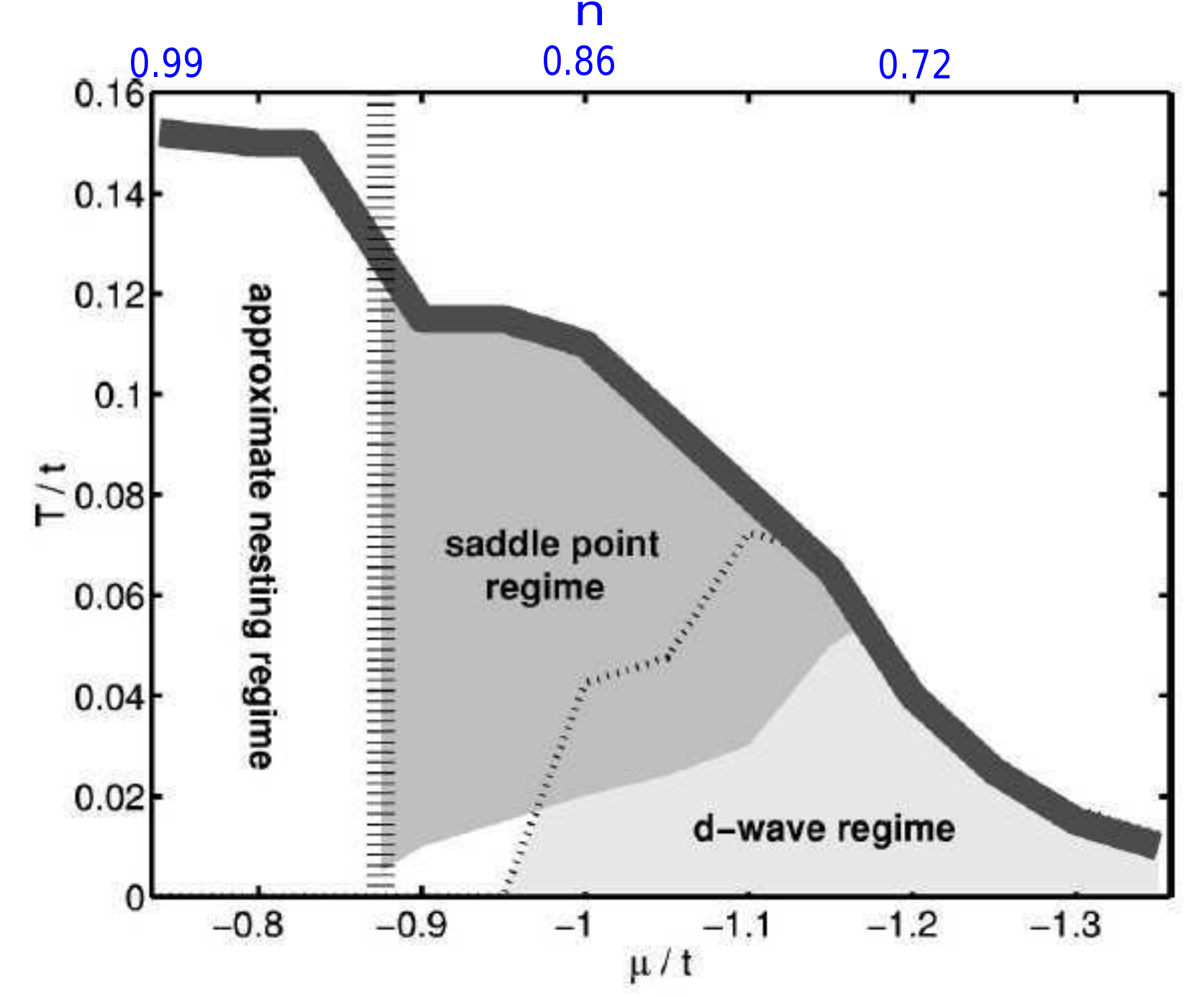}
}
 \caption[]
 {
 (Color online) Phase diagram of the 2-dimensional Hubbard model with varying temperature (T) and chemical potential ($\mu$), hole density ($x$) with $n=1-x$ according to the functional renormalization group flow analysis of Honerkamp and coworkers \cite{CH01}. The model parameters in Eq.\ref{eq:Hubbard} are $t'/t =-0.3$ and $U/t = 3$. The left region is underdoped and the right region overdoped. The solid line represents the temperature scale where an instability in the flow appears. The leading instability is to AF order in the approximate nesting regime at strong underdoping crossing over along the dotted line to \textit{d}-wave pairing at overdoping. Note both appear prominently in the intermediate saddle point regime.
 }
\label{fig:RG-phase}
\end{figure}

\begin{figure}[tf]
\centerline
{
\includegraphics[width = 12.5cm, height =11.5cm, angle= 0]
{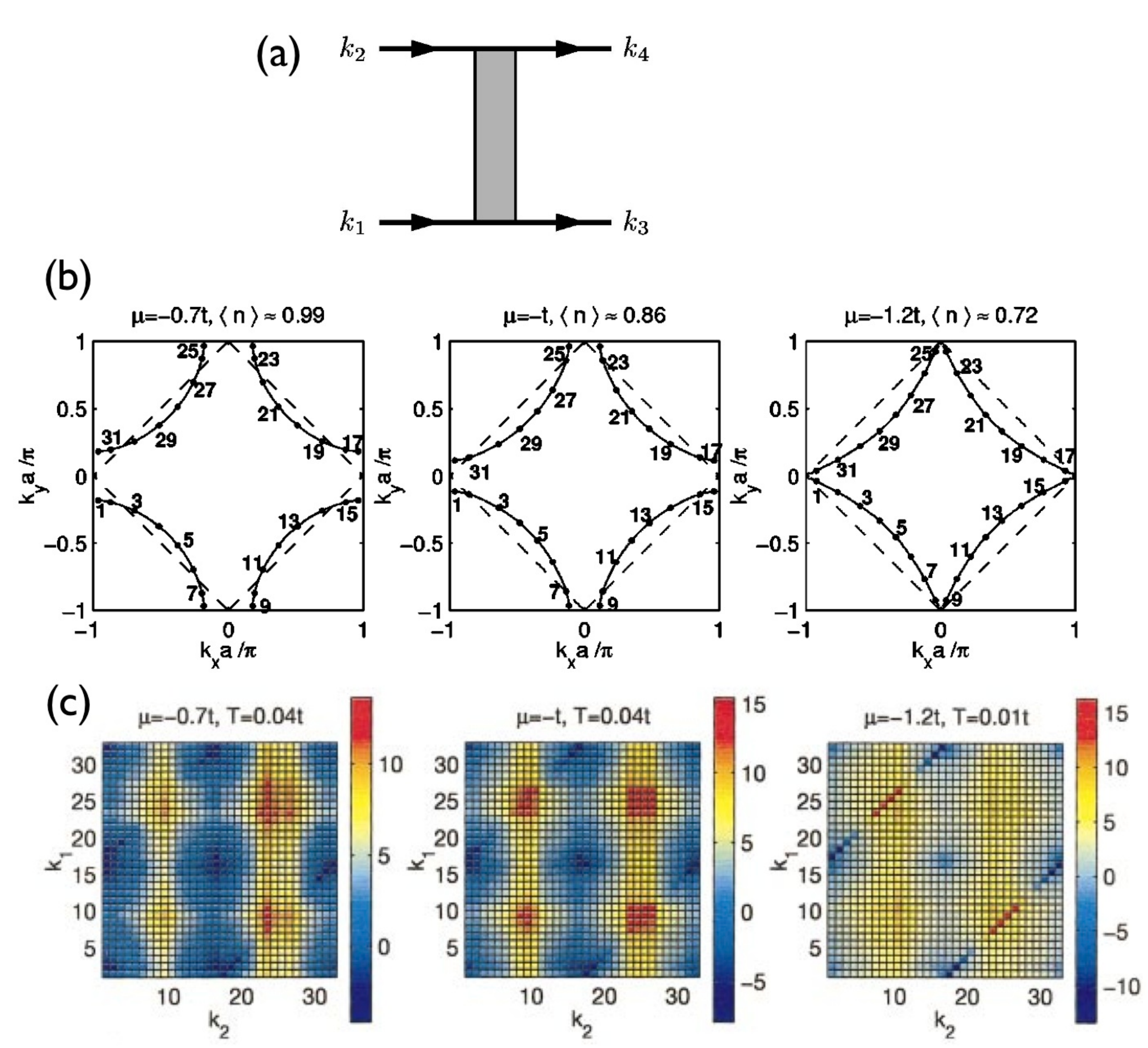}
}
 \caption[]
 {
 (Color online) 
 (a) The vertex $V_{\Lambda}(\mathbf{k}_{1} ,\mathbf{k}_{2}, \mathbf{k}_{3})$ with $\mathbf{k}_{4}$ determined by momentum conservation.
 (b) Fermi surfaces and the 32 patchs for 3 values of the chemical potential. $\langle n \rangle$ denotes the average particle number per site, i.e.,  $\langle n \rangle =1$ corresponds to half- filling. The dots on the FS (solid line) indicate the patch centers, with patch indices given by the numbers. The dashed line denotes the umklapp surface.
 (c) Snapshot of the couplings $V_{\Lambda}(\mathbf{k}_{1} ,\mathbf{k}_{2}, \mathbf{k}_{3})$ with the first outgoing wave vector $\mathbf{k}_{3}$ fixed at point 1 (see panel (b)) when the largest couplings have exceeded the order of the bandwidth for the 3 choices of chemical potential and temperature discussed in the text. The color bars indicate the values of the couplings.
 }
\label{fig:RG-Carsten}
\end{figure}

 In a band structure with $t^{\prime} = -0.3t$, typical for cuprates \cite{OKA95},  the RG flows for the vertex $V_{\Lambda}(\mathbf{k}_{1},\mathbf{k}_{2},\mathbf{k}_{3})$  depend strongly on the hole density. Even with only a moderately strong value of the interaction, $U = 3t$ (approximately half the bandwidth) three regions with qualitatively different behavior are apparent in Fig.\ref{fig:RG-phase}. In the strongly overdoped region with $x = 0.28$ (corresponding to $\mu/t=-1.2$ in Fig.\ref{fig:RG-phase}),  the snapshot of the wave vector dependence of the vertex in Fig.\ref{fig:RG-Carsten}  shows a clean \textit{d}-wave pairing form. The energy scale in the vertex is chosen at the value when the strongest components equal to the bandwidth. Note at this density the Fermi surface touches the antinodal Brillouin zone but this special choice has only a small influence. The largest components of $V_{\Lambda}(\mathbf{k}_{1},\mathbf{k}_{2},\mathbf{k}_{3})$ are those with a zero momentum incoming pair, i.e. Cooper pair scattering with a $d_{x^{2}-y^{2}}$ form factor. In this case the divergence appears in a single channel indicating a transition to \textit{d}-wave superconductivity.

  For a very small doping, $x=0.01$ (corresponding to $\mu/t=-0.7$ in Fig.\ref{fig:RG-phase}), a single channel dominates the pattern at the cutoff scale $\Lambda$ when the strongest components reach a value of order of the bandwidth.  The energy of the cutoff is now much larger and the snapshot of the vertex in Fig.\ref{fig:RG-Carsten} is quite different. The leading interactions are umklapp scatterings involving the patches where Fermi surface intersects what would be the reduced antiferromagnetic Brillouin zone, indicating a transition to AF order.

   In the intermediate density range (intermediate values of $\mu$ in Fig.\ref{fig:RG-phase}) there is a continuous rise in the critical energy scale between over- and underdoping and a more complex pattern appears in the corresponding snapshot (see Fig.\ref{fig:RG-Carsten}). It signals a continuous crossover between the two instabilities, \textit{d}-wave pairing and AF. This comes about because there is mutual reinforcement, not simple competition, between the umklapp and pairing scattering processes. This complicates the interpretation of the strong coupling state that should follow from these leading interactions - a point we will return to later.
    One result is clear, the overdoped region is characterized by the growth of a highly structured pattern in the leading components of the interaction vertex  $V_{\Lambda}(\mathbf{k}_{1},\mathbf{k}_{2},\mathbf{k}_{3})$  at low energy or temperature.  It is not simply a \textit{d}-wave Cooper pair pattern but it contains also umklapp scattering processes which strengthen as the hole density is reduced.

\subsection{Experimental Investigations and Theoretical Models of the Breakdown of Landau Behavior with Decreasing Hole Density in Overdoped Cuprates}
\begin{figure}[tf]
\centerline
{
\includegraphics[width = 10.5cm, height =7.5cm, angle= 0]
{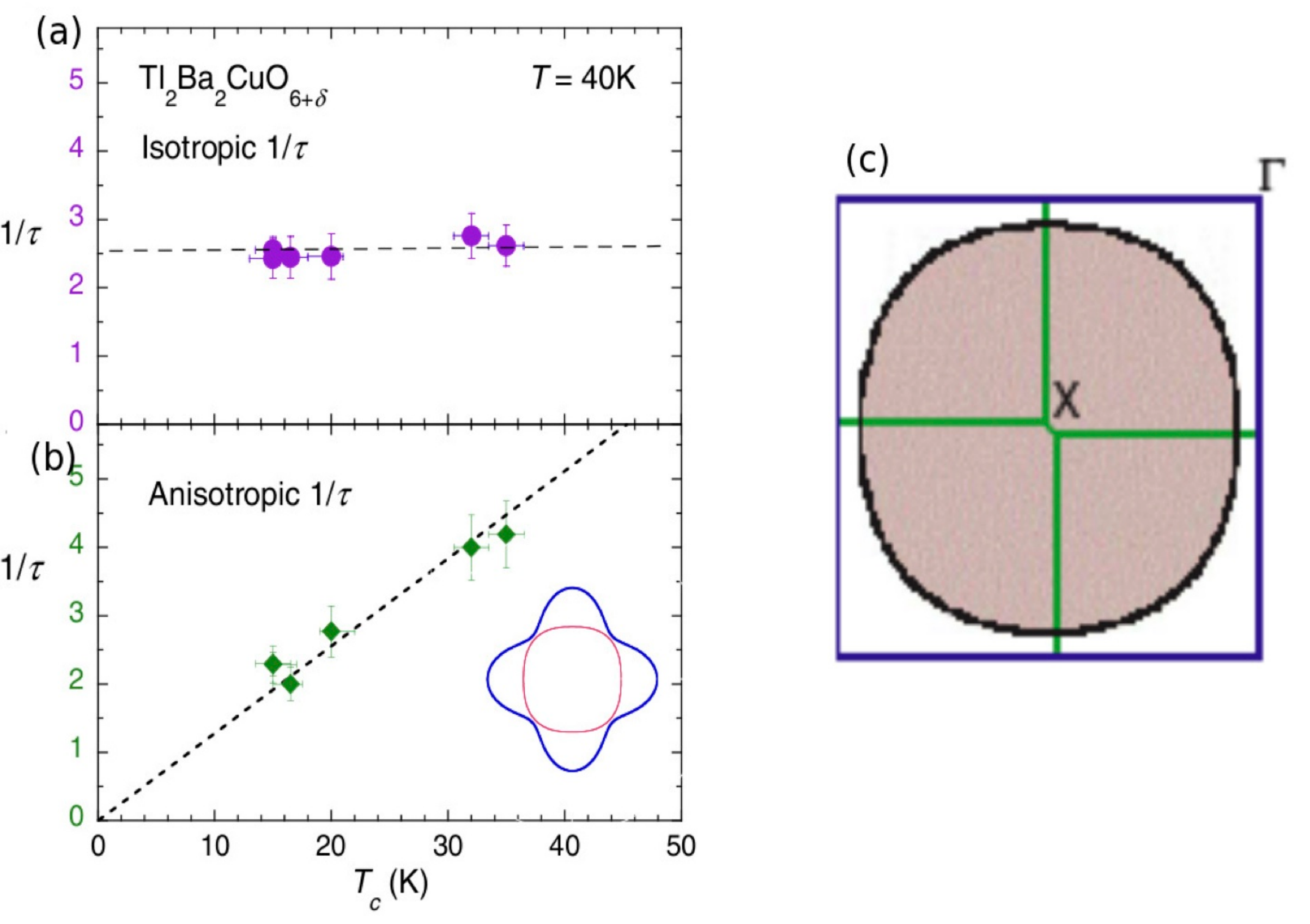}
}
 \caption[]
 {
(Color online) In-plane transport scattering rate, $1/\tau$ as function of hole density (superconducting transition temperature $T_{c}$) in overdoped Tl$_{2}$Ba$_{2}$CuO$_{6+x}$ samples at $T=40K$, from Abdel-Jawad et al. \cite{NH07}. The scattering rate consists of isotropic  and  anisotropic components around the Fermi surface shown in (a) and (b). Inset: A sketch of the in-plane variation of the anisotropic component (thick blue line) around the Fermi surface (thin red line). (c) The Fermi surface in the overdoped region
  \cite{VigNat08}. 
  }
\label{fig:resistivity-OM}
\end{figure}

    Recently Nigel Hussey and collaborators \cite{NH07,NH08} performed an interesting series of transport measurements in the overdoped region of the phase diagram, which have shed new light on how the standard Landau behavior breaks down. They found a striking correlation between charge transport and superconductivity in heavily overdoped high temperature superconducting cuprates. The onset of superconductivity as the doping, $x$, is reduced coincides with the appearance in the normal phase of strong and highly anisotropic quasiparticle scattering processes. The application of a magnetic field, which suppresses superconductivity, revealed that the anisotropic term in the in-plane transport scattering rate was linear in temperature, violating the perturbative quadratic dependence characteristic of a Landau Fermi liquid. The experiments by Hussey and collaborators \cite{NH07,NH08} were carried out on well characterized tetragonal single layer Tl$_{2}$Ba$_{2}$CuO$_{6+x}$ samples. They measured the angle-dependent magnetoresistance oscillations and analyzed their results using a Boltzmann equation to describe transport in a quasi-2D metal with a fourfold anisotropic scattering rate. This analysis also provided detailed Fermi surface information. Subsequent quantum oscillation experiments by Vignolle \textit{et al}. \cite{VigNat08} and ARPES experiments by Plate et al  \cite{PlatPRL05}, confirmed that the form of  Fermi surface in an overdoped  Tl$_{2}$Ba$_{2}$CuO$_{6+x}$ sample with doping $x=0.3$ agrees well with LDA band structure calculations shown in Fig.\ref{fig:resistivity-OM}(a3). The anisotropic term in the quasiparticle scattering rate grew as the doping, $x$ is reduced, proportional to the superconducting transition temperature, $T_{c}$ (in zero field). The scattering rate varies strongly around the Fermi surface and is maximal in the antinodal directions and vanishing along the nodal directions [see Fig.\ref{fig:resistivity-OM}(a2)]. Note this anomalous behavior is not associated with a 2-dimensional van Hove singularity at the Fermi energy, as the measured Fermi surface shows it lies clearly below the Fermi energy in these Tl- cuprates. Instead this anomalous behavior points to a strongly anisotropic scattering vertex at low energies, which grows in strength as the doping $x$ is reduced.

\begin{figure}[tf]
\centerline
{
\includegraphics[width = 10.5cm, height =5.5cm, angle= 0]
{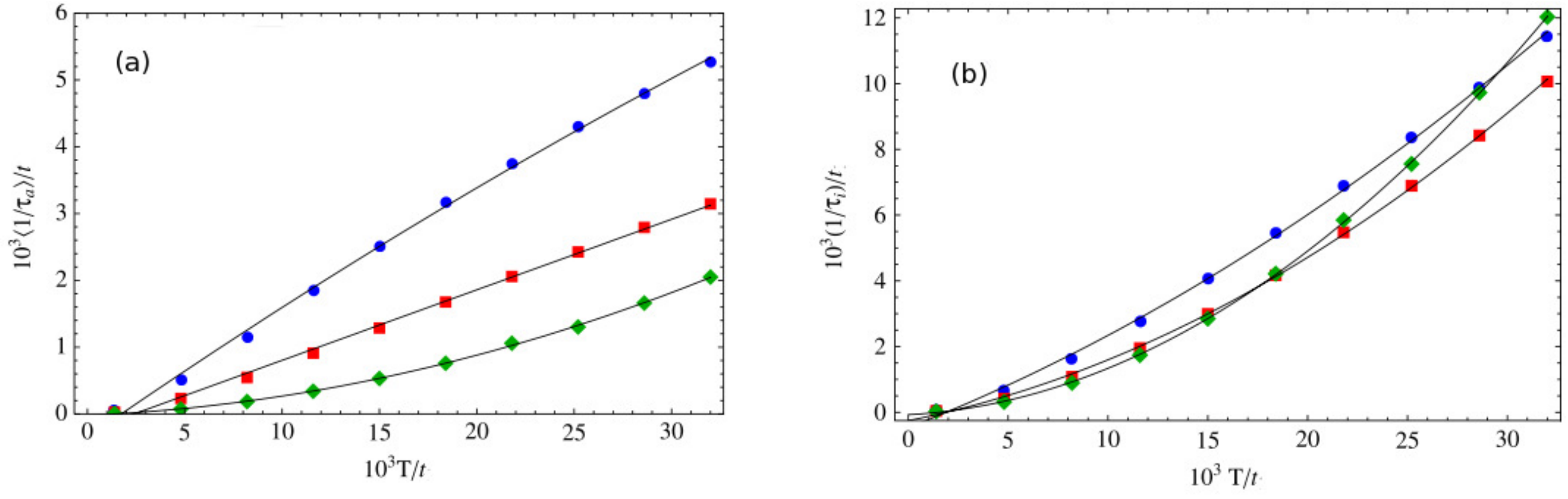}
}
 \caption[]
 {
 (Color online) Quasi-particle scattering rate $1/\tau$ in units of the nn hopping, $t$ calculated by FRG for the 2-dimensional Hubbard model by Ossadnik et al. \cite{OS08}. The $T$-dependence of the anisotropic (a) and isotropic (b) components of the scattering rate at the Fermi surface for hole dopings $x=$0.15 (blue), 0.22 (red), and 0.30 (green) are illustrated in (a) and (b). Only the anisotropic component shows a strong $x$-dependence consistent with experiment, see Fig. \ref{fig:resistivity-OM}
}
\label{fig:RG-OM}
\end{figure}

   As we discussed above, earlier FRG investigations of a 2-dimensional Hubbard model on a square lattice found that in the overdoped region of the phase diagram the structureless high energy scattering vertex evolved into a strongly anisotropic scattering vertex in the particle-particle and particle-hole channels at low energies and temperatures \cite{CH01,CH02,ZS97,ZS00,ZS01,HM00}. Further investigations revealed that the resulting self-energy is also very anisotropic \cite{Z01,CHEur01,KK04,RM05}. This work stimulated Ossadnik and coworkers \cite{OS08} to undertake an extensive FRG study of the doping and temperature dependence of the quasiparticle scattering rate under conditions such that the pairing instability suppressed. They chose a band structure in accord with the experimental Fermi surface [see Fig.\ref{fig:RG-OM}(a1)] and a moderate value of the onsite Coulomb repulsion since the 1-loop approximation in the FRG calculations limits their reliability to at most moderately strong interactions. The FRG calculations for these doping values ($0.15 <x<0.2$) and interaction strength, show a divergent flow at finite scales in the \textit{d}-wave pairing channel, indicating a finite $T_{c}$ for \textit{d}-wave superconductivity. In the experiments \cite{NH08} the pairing instability towards \textit{d}-wave superconductivity is suppressed by a strong external magnetic field. It is not feasible to include a strong magnetic field in the FRG calculations. Instead a suppression of \textit{d}-wave Cooper pairing was achieved by adding in an elastic scattering term in the flow of the 4-point scattering vertices.  Ossadnik \textit{et al} \cite{OS08} found a rising highly structured scattering vertex that gives an anisotropic contribution to the decay rate with an approximately linear, not quadratic, temperature dependence, when the pairing divergence is suppressed. It is worth repeating that this breakdown of standard Landau Fermi liquid behavior is not associated with a divergent density of states from a van Hove singularity at the Fermi energy but instead is due to strong scattering processes at large momentum transfer connecting the antinodal regions in $\mathbf{k}$-space. These processes appear in the FRG flows as a precursor to the Mott insulating behavior at $x=0$, i.e at half-filling.

  The main results of the FRG calculations are illustrated in Fig.\ref{fig:RG-OM}. A comparison to the experimental results in Fig.\ref{fig:resistivity-OM} shows good qualitative agreement. The limitations on the interaction strength in the 1-loop FRG calculations rule out a quantitative comparison but the good qualitative agreement is encouraging and gives us confidence that the FRG calculations capture the main features of the evolution of the 4-point scattering vertex with hole density in the overdoped range.

   Alternative proposals to explain the anomalous quasiparticle decay rate are discussed in the recent review by Hussey \textit{et al} \cite{NH10}. Here we comment on only those put forward by Alexandrov \cite{AL97} and Wilson \cite{JAW08} and by Dell Anna and Metzner \cite{AM07}. The former two postulate the presence of two types of charge carriers, bound electron pairs forming charge 2e bosons and conduction electrons. The charged bosons act as strong scattering sites for the lighter conduction electrons causing the anomalous scattering rate. However it is unclear how to reconcile this proposal with the observed Fermi surface, which agrees with the `a priori' LDA band structure calculations and includes all conduction electrons. The latter proposal postulates the presence of strong nematic fluctuations possibly due to a proximity to a Pomeranchuk instability \cite{AM07}. Scattering off these nematic fluctuations gives a power law exponent 4/3, close to linear, in the scattering rate. Note however that the presence of strong nematic fluctuations is rather questionable since the 2-dimensional van Hove singularities lie clearly below the Fermi energy in the overdoped ($x=0.3$) sample studied by Vignolle \textit{et al} \cite{VigNat08} and it moves lower as $x$ is reduced. This behavior is difficult to reconcile with a Pomeranchuk instability driven by the proximity of 2-dimensional van Hove singularities to the Fermi energy across the extended range of hole densities.  Also the Tl- and the closely related Hg-cuprates do not undergo a crystallographic phase transition but retain their tetragonal crystal structure across the full doping range, from overdoped to underdoped \cite{QH95}. Nonetheless a search for nematic fluctuations by neutron scattering investigations of the spin fluctuation spectrum in these tetragonal cuprates would be worthwhile.

\subsection{Transport Experiments on the Crossover from the Overdoped to  the Underdoped Hole Density  Regions}
\begin{figure}[tf]
\centerline
{
\includegraphics[width = 10.5cm, height =7.5cm, angle= 0]
{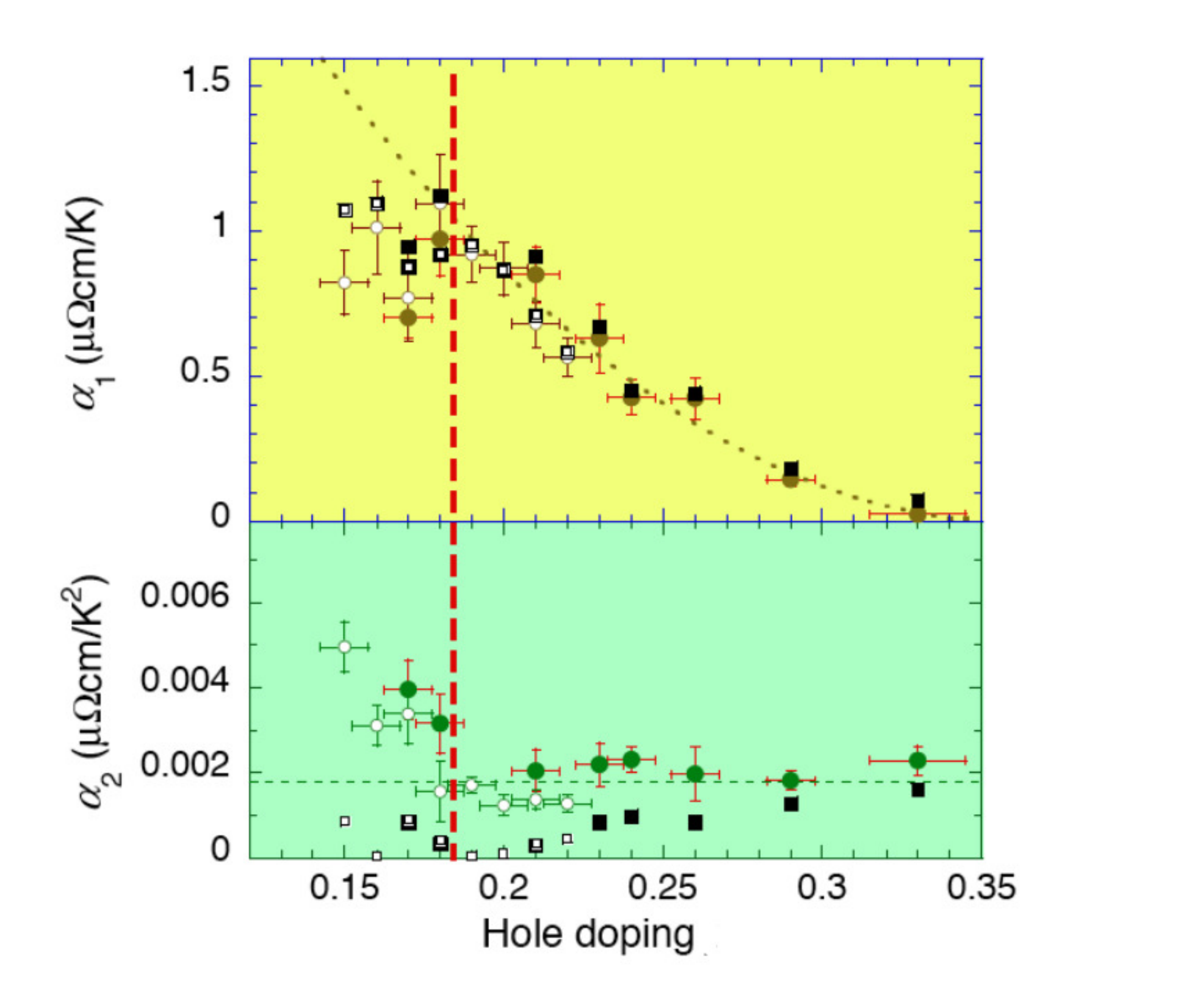}
}
 \caption[]
 {
 (Color online) Doping evolution of the temperature-dependent coefficients of the in-plane resistivity $\rho_{ab}(T)$, in overdoped La$_{2-x}$Sr$_{x}$CuO$_{4}$ quoted by Cooper et al. \cite{CO09}. The resistivity is fit to $\rho_{ab}(T)= \alpha_{0}(x) + \alpha_{1}(x) T + \alpha_{2}(x) T^{2}$ (for symbols see Ref.\cite{CO09})
.  Upper panel: Doping dependence of $\alpha_{1}$. Lower panel: Doping dependence of $\alpha_{2}$. The hole density at the QCP is indicated by the red dashed line.
}
\label{fig:anistropy-scatter}
\end{figure}

   Early NMR and specific heat experiments found that radical changes appear in the electronic properties as the hole density is reduced beyond a critical value. The experimental properties of underdoped cuprates have been comprehensively reviewed by Timusk and Statt \cite{TS99}. We will summarize some key results below. For now we briefly review the very recent experiments by Hussey and collaborators extending their transport measurements to lower densities \cite{CO09}.  The high values of $T_{c}$ in single layer Tl- and Hg-cuprates make them unsuitable for experiments on the normal state at low $T$ near optimal doping. This would require magnetic field strengths beyond those currently available to suppress superconductivity. Hussey and coworkers \cite{CO09} extended their transport measurements to the LSCO (La$_{2-x}$Sr$_{x}$CuO$_{4}$)  cuprates which have lower $T_{c}$ values and can therefore be studied in the normal state in laboratory magnetic fields over a wide range of $x$-values. However detailed angular dependent magneto-resistannce experiments were not possible. Instead the planar resistivity, $\rho_{ab}(T, x)$, was measured and analyzed in the temperature range $T< 200K$ in terms of linear and quadratic $T$-terms, i.e. the resistivity is represented as $\rho_{ab}(T, x) = \alpha_{0}(x) + \alpha_{1}(x) T + \alpha_{2}(x) T^{2}$. Their results for the anomalous linear term, $\alpha_{1}(x)$,  show a similar behaviour to the anisotropic linear-$T$ term reported earlier in the Tl-cuprates, namely a rising value as $x$ is decreased in the overdoped regime in samples with finite $T_{c}$ values, see Fig.\ref{fig:anistropy-scatter}. Interestingly this rise does not continue indefinitely but has a maximum at a critical hole density, $x_{c}  =0.19$. This density marks the transition between the overdoped and underdoped regions of the phase diagram, indicating the presence of a quantum critical point (QCP) at $x= x_{c}$. As Cooper \textit{et al} \cite{CO09} note, this behaviour with $\alpha_{1}(x)$, the linear-$T$ coefficient of the resistivity, finite over a broad density range around $x_{c}$ differs distinctly from the funnel shape region centered on the QCP seen in other metals where the QCP is caused by a symmetry change. In this latter case the linear-$T$ behavior is generally interpreted as a consequence of scale invariant physics near to the QCP. Cooper \textit{et al} \cite{CO09} propose that in the cuprates the QCP at $x= x_{c}$, signals the destruction of the quasiparticles by the increasing interaction strength as $x$ passes through $x_{c}$. The angular dependence of the quasiparticle scattering rate cannot be determined from the LSCO experiments. However we know from the experiments on the Tl-cuprates that the anisotropic interactions couple predominantly the quasiparticles near the antinodal points. This suggests that the change in the quasiparticle spectrum sets in first in the vicinity of the antinodal points.

\section{General Features of the Pseudogap Phase at Underdoping}
   The underdoped cuprates with $x< x_{c}$ show a host of anomalous properties in what is nowadays referred to as the pseudogap phase.  Timusk and Statt \cite{TS99} published a very complete review of experiments in this phase a few years back. Here to motivate the later analysis, we summarize some key results and refer the reader for more details to the Timusk-Statt review \cite{TS99}. A shorter review covering the main issues in the ongoing debate on the physics of the pseudogap appeared later \cite{No05}.

     The anomalous features first appeared in an early comparison of the spin susceptibility, determined from the NMR Knight shift, between underdoped and optimally doped YBa$_{2}$Cu$_{3}$O$_{7-\delta}$ \cite{Wa90,AL89,Wa89,AL91}. Samples with $\delta$ values around 0.3, showed a very substantial drop in the spin susceptibility in the normal state at $T>$T$_{c}$, which contrasted strongly with the constant value in the normal phase in optimally doped samples with  $\delta =0$. This drop is generally interpreted as a consequence of a gap, or partial gap, opening in the magnetic spectra as a result of singlet spin pairing. The electronic specific heat also showed clear changes, with a substantial reduction in the normal state linear-$T$ coefficient and an anomaly at $T_{c}$ \cite{Lo94,Lo01}.

\subsection{ Spectroscopic Studies of the Pseudogap Phase}
  A series of ARPES experiments in the normal phase at $T>T_{c}$ showed that full Fermi surface observed in the overdoped region, was truncated to a set of four Fermi arcs centered on the four nodal points $(\pi/2,\pi/2)$ \cite{NO98}. In the antinodal regions the ARPES spectra in underdoped cuprates show a finite energy gap. The spectral weight at the Fermi energy consists of a set of 4 arcs, disconnected from each other and ending on the square surface in $\mathbf{k}$-space whose sides connect the antinodal points.
  Since the Timusk-Statt review \cite{TS99}, more detailed ARPES experiments which will be described later, have examined these Fermi arcs. A similar set of Fermi arcs was deduced from the interference patterns in recent STM spectra at low voltages and low temperatures in the superconducting state of underdoped samples \cite{KOH08}. These patterns, when interpreted as a consequence of weak disorder scattering, allow the determination of the energy spectra of the coherent Bogoliubov quasiparticles in the superconducting state. These spectra allowed Kohsaka \textit{et al} \cite{KOH08} to observe the Fermi surface underlying these propagating Bogoliubov quasiparticles. It also consists of arcs, similar to those observed directly in the ARPES experiments. Further evidence for a reconstruction of the Fermi surface comes from recent quantum oscillation experiments on underdoped samples at high fields and low temperatures \cite{LeB07}.  The period of the oscillations requires small closed Fermi pockets, quite different to the full LDA Fermi surface observed in overdoped samples. The relationship between these closed electron-like Fermi pockets at high fields and low temperatures and the apparently open Fermi arcs observed by ARPES at zero fields and higher temperatures, remains to be clarified. But the pronounced change in the quantum oscillation periods and the sign change of the Hall effect \cite{Ro10} at high fields and low T, confirm that a drastic change has occurred in the normal phase as the doping $x$ is reduced through the QCP at $x= x_{c}$ .

\subsection{Theoretical Proposals  to Describe the Pseudogap Phase}
\begin{figure}[tf]
\centerline
{
\includegraphics[width = 14.5cm, height =7.5cm, angle= 0]
{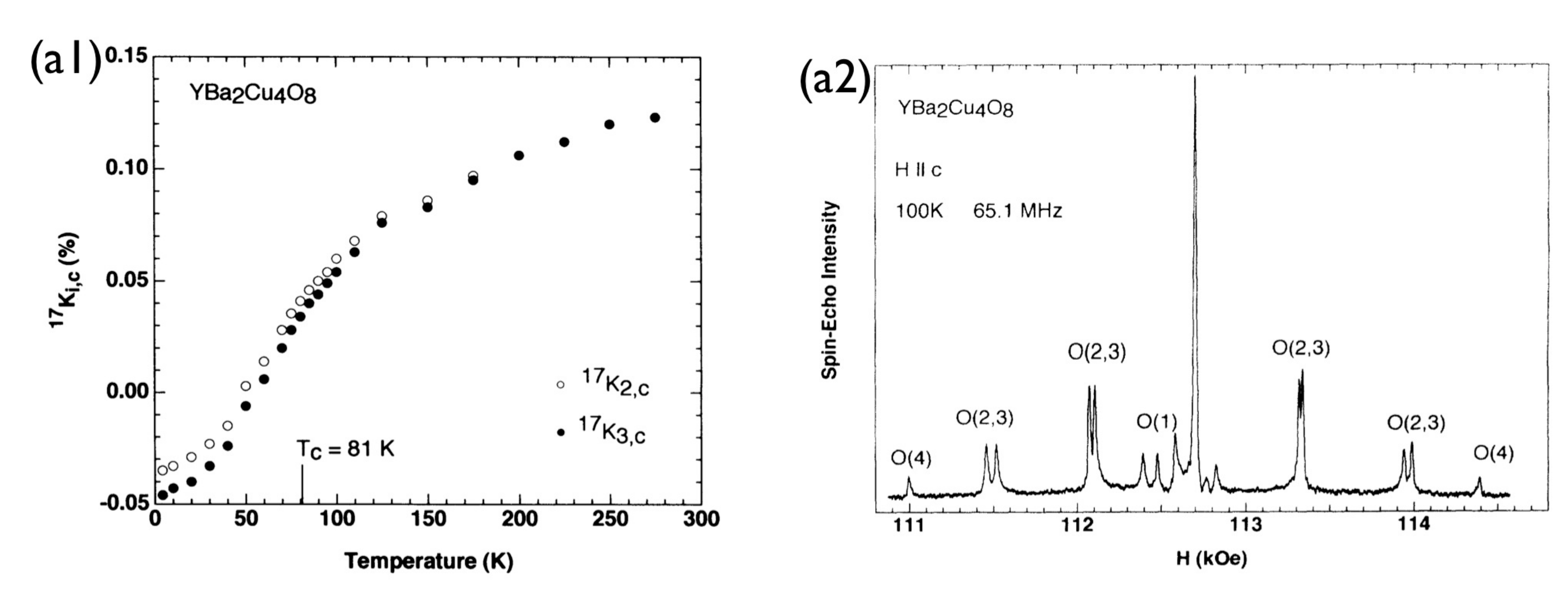}
}
 \caption[]
 {
 (Color online) NMR data obtained in YBa$_{2}$Cu$_{4}$O$_{8}$ with $H || c$.
(a1) Temperature dependence of Knight shift $^{17}K_{i,c}$ for the planar O(2) and O(3) sites \cite{To94}.
 (a2) $^{17}$O NMR spectrum taken at 65.1 MHz and 100K. The small splitting between narrow peaks assigned to the planar O(2) and O(3) sites above and between the chains respectively, is temperature independent. The absence of additional splittings rules out the onset of a superlattice or nematically ordered phase with distinct O sites.
 }
\label{fig:NMR-pseudogap}
\end{figure}

    The observation of these apparently disconnected Fermi arcs rather than connected closed Fermi curves found in standard metals, has stimulated strong theoretical interest and various proposals have been put forward to explain it.  One line of thought focuses on a Fermi surface reconstruction due to the presence of some kind of superlattice order. This requires a breaking of translational symmetry. The simplest such symmetry breaking would be the presence of charge (CDW) or spin (SDW) wave order. Such superlattice order does appear in some underdoped cuprates which show static stripes, e.g. La$_{1.875}$Ba$_{0.125}$CuO$_{4}$ which has been studied by neutron scattering \cite{Li07} and NMR/NQR experiments \cite{Im01}. However it is not a general feature of the underdoped cuprates as can be seen from NMR/NQR experiments. This experimental technique has the advantage that it can be widely applied and as a local measurement it can detect modulations in the local environment of specific nuclei. On the other hand the spin gap phenomenon is a general feature of underdoped cuprates and is clearly apparent in the NMR experiments on Hg-cuprates \cite{Bo97, It98} and on YBa$_{2}$Cu$_{4}$O$_{8}$ \cite{To94}. The latter is especially interesting as well ordered stoichiometric crystals of this compound can be grown. The Hg-cuprates are the least disordered of the randomly acceptor doped cuprates \cite{Es04}. As a result these cuprates are good choices to test for the presence of intrinsic superlattice order. NMR/NQR experiments on all of these cuprates report negative results for CDW and SDW superlattice order \cite{Bo97,It98,To94}, e.g. as shown in Fig.\ref{fig:NMR-pseudogap}. Note, in many strongly underdoped cuprates there is experimental evidence for broken translational and/or rotational symmetry. Vojta has recently published a comprehensive review on this subject \cite{Vo09}. Recent examples are the observation of the onset of a nematic distortion of the magnetic excitation spectrum at $T\sim150K$ in strongly underdoped YBCO by Hinkov and collaborators \cite{Hinkov-science-08} and the short range nematicity observed in STM spatial images reported by Lawler et al \cite{La10}. The key issue in our view is whether these broken symmetries are the cause of the pseudogap phenomena or are instabilities of the pseudogap phase which show up at lower temperatures and doping, perhaps influenced by the strong local electric fields arising from the random acceptor ions. The absence of these broken symmetries in the cleanest and best ordered underdoped cuprates discussed above at small magnetic fields, leads us to espouse the latter point of view. In high magnetic fields anomalies in transport properties (e.g. see Rourke et al. \cite {Ro10}) point towards some form of superlattice ordering when superconductivity is weakened by the magnetic field. We shall return to this issue towards the end in Sect. 6 when discussing the magnetic properties of the pseudogap phase.

  In the absence of direct observations of broken translational symmetry in clean samples, there are suggestions that some form of hidden symmetry breaking could be causing the reconstruction of the Fermi surface. One possibility put forward by Chakravarty and coworkers \cite{Chak01}  is a so called \textit{d}-density wave (DDW) state also known as a staggered flux phase due to the presence of staggered orbital currents around the CuO$_{4}$ plaquetes. All Cu- and O- sites are equivalent in a DDW ordered state so it is not clear if NMR is a definitive test. The doubling of the unit cell in a DDW state reconstructs the Fermi surface into a set of elliptical pockets centered on the nodal points. Chakraverty \textit{et al} \cite{Chak03} showed that the spectral weight of quasiparticles is not uniform around the pockets and is reduced on the side furthest away from the zone centre. As a result the pockets could appear as disconnected Fermi arcs in ARPES experiments. For a recent review see Chakravarty \cite{Ca11}. While the NMR and ARPES experiment do not obviously contradict the presence of DDW order, here are still some open questions.  The staggered orbital currents in this phase give rise to a small AF moment perpendicular to the CuO$_{2}$ planes. Neutron scattering experiments on YBa$_{2}$Cu$_{4}$O$_{8}$ have not been possible due to the small size of the crystals but the YBCO compound OrthoII- YBa$_{2}$Cu$_{3}$O$_{6.5}$, which is underdoped with short range ordered alternating empty and filled CuO chains, can be made in large enough samples to allow neutron scattering experiments. These did not provide evidence for static AF order \cite{CS02}. A further difficulty is the absence of signs of a phase transition in thermodynamic measurements, e.g. the Knight shift measurements on YBa$_{2}$Cu$_{4}$O$_{8}$ show a continuous onset of the spin gap unlike the behavior expected at a symmetry breaking phase transition leading to a strong Fermi surface reconstruction \cite{To94}.

 Lastly, we should comment on recent neutron scattering experiments which find evidence that the onset of the pseudogap is accompanied by the appearance of spontaneous intra-unit-cell orbital currents as the temperature is lowered \cite{FA06,LI10a}. This shows up as an additional contribution, which flips the neutron spin, to the Bragg peaks in neutron scattering. At present there are open questions in the interpretation of the origin of these orbital currents \cite{WE09}. Their relevance to the spin gap and the Fermi surface reconstruction is not immediately obvious. A small anomaly has been found in the uniform magnetic susceptibility at the onset temperature of the these orbital currents \cite{LE09}. Very recently extra modes were reported in neutron scattering experiments on Hg-cuprates and attributed to the orbital currents by Li et al \cite{LI10a}. The importance of these orbital currents remains under debate at present. A recent search by NMR for a weak magnetic field on the Ba sites in YBa$_{2}$Cu$_{4}$O$_{8}$ caused by the orbital currents proved negative \cite{St11}. Note, Varma and collaborators \cite{AJ07} argue that this broken symmetry order plays a key role in the physics of the pseudogap phase.

  We turn now to a brief review of the theoretical proposals to describe the anomalous ARPES spectra of the pseudogap phase without invoking the presence of a broken symmetry order. The simplest possibility is to ascribe the whole phenomenon to a remnant  \textit{d}-wave pairing energy gap in the presence of strong phase fluctuations of the superconducting order parameter. Such fluctuations are controlled by the superfluid stiffness, which in turn is controlled by the superfluid carrier density. Early measurements by Uemura showed this scales with the hole doping, $x$, in the pseudogap region \cite{UE89}. This is consistent with a hole doped Mott insulator, not unexpected given the proximity to the Mott insulator at stoichiometry. Early on Emery and Kivelson \cite{EM95} argued for an extended temperature region of preformed pairs extending above a BKT transition temperature $T_{TBK}\sim x$. In recent years experiments by Ong and collaborators on the Nernst effect \cite{WA06} and diamagnetic contributions to the magnetic susceptibility in the pseudogap region \cite{Lu10}, have confirmed the existence of superconducting fluctuations above $T_{c}$ extending up to a temperature  ~2$T_{c}$ . This temperature however lies substantially below the onset temperature $T^{*}$ for the pseudogap and spin gap particularly at small $x$ where $T_{c}\to 0$ but $T^{*} \to$ const. Recent experiments using a variety of experimental techniques have reached similar conclusions \cite{Al10,Du10,Bi11}. This suggests that there is more to this behavior than simply a region of preformed pairs with strong phase fluctuations. The large value of the antinodal energy gap measured by ARPES and STM at small $x$, relative to $k_{B}T_{c}$, and its scaling with $T^{*}$ rather than $T_{c}$ have led many authors to a two-gap scenario (for a review see \cite{HH08}). In this approach the mechanisms underlying the pseudogap and the superconducting order differ and are to be distinguished. An interesting Ginzburg-Landau formulation of the phase fluctuation scenario has recently been published by Ramakrishnan and collaborators \cite{BanerjeeGL10}.  We will return to this debate in more detail in the next chapter.

    The possibility of singlet spin pairing in the underdoped region was raised very early by Anderson \cite{PWA87}. He introduced the RVB concept which emphasizes the energetic stability of the spin singlet state of a pair of AF coupled $S=1/2$ spins. He argued that a spin liquid state of singlet pairs stabilized by quantum mechanical fluctuations could become superconducting upon doping. While it is now clear that the planar AF $S=1/2$ Heisenberg model has a long range AF ordered ground state, it is also clear that this state contains strong singlet pairing fluctuations and that the AF order is rapidly destabilized by introducing hole doping. The RVB concept is very appealing physically but has proved very difficult to construct a full microscopic theoretical formulation. The difficulty lies in the inherent strong coupling fermionic nature of a hole doped AF. Much progress has been made over the years using gauge theory formulations coupled with slave boson mean field approximations. In addition a variety of numerical techniques, e.g. variational Monte Carlo (VMC), exact diagonalization (ED) of finite clusters, dynamic mean field theory (DMFT), have been employed to gain insight into this system of strongly interacting fermions. A series of recent reviews cover these topics \cite{PWA04,LNW06, EMG08,  PAL08, OgFu08}. 

 In the next section we will look further into the possibility of Fermi surface reconstruction without translational symmetry breaking. This seems counterintuitive at first but there is a well known counterexample, namely 2-leg Hubbard ladders \cite{DR96}, which we will argue in the next section, has strong similarities to the underdoped cuprates.

\section{The YRZ Ansatz for the Single Particle Propagator in the  Pseudogap Phase}

\subsection{Approach to the Mott Insulator in Certain Systems Already at Weak Onsite Repulsion}
 The basic premise behind the YRZ ansatz \cite{YRZ06} is that the crossover from the overdoped metallic state to the pseudogap phase signals the transition from a full Fermi surface metal to a phase which behaves as a doped Mott insulator. An important characteristic of the Mott phase is that its insulating property is not the result of AF long range order or other broken translational symmetry, but only of a strong onsite Coulomb repulsion at a filling of exactly 1 el./site. It follows that precursors to the Mott state need not be accompanied by broken translational symmetry either. In a Mott insulator the Fermi surface is completely suppressed and it is clearly interesting to examine the way in which the Fermi surface evolves in the precursor state. In an early paper Brinkman and Rice \cite{BR70} analyzed the approach to the Mott transition from the metallic side at a density of 1 el./site using the Gutzwiller approximation to treat a projected Fermi sea wavefunction. The volume enclosed by the Fermi surface remains constant in their approach but the coherent quasiparticle weight is progressively reduced and the quasiparticle effective mass diverges as the onsite repulsion approaches the critical value.  In this analysis the exchange spin-spin coupling is ignored. The divergence of the effective mass can be associated to a reduction of the effective Fermi temperature --- the temperature scale at which the entropy saturates at the spin only value, $k_{B}\ln2$/site. Actually the Brinkman-Rice approach works well in the Landau-Fermi liquid state of  $^{3}$He and as Vollhardt showed captures the essential characteristics of the approach to the solid crystalline state when an external pressure is applied \cite{VO84}. This is a special case since the exchange coupling between the spins vanishes due to the very strong short range repulsion between $^{3}$He atoms. The cuprates are very different with a strong superexchange coupling between spins and also an underlying lattice which allows umklapp scattering events, unlike the  $^{3}$He liquid.  The Brinkman-Rice approach does not work for them. This can be seen directly in ARPES experiments which show a partially truncated Fermi surface and an almost constant Fermi velocity in the nodal direction in the underdoped  pseudogap phase.

       A second scenario is an evolution of the Fermi surface analogous to the behavior of Cr alloys \cite{FA94}. In these alloys the Fermi surface is progressively truncated by a series of incommensurate SDW phases as the el./atom ratio is changed and transition to a  the commensurate AF ordered phase with a much reduced Fermi surface and a larger energy gap is stabilized.  Similar behavior can also be ruled out for the cuprates by the absence of long range SDW order in neutron scattering and NMR experiments associated with the pseudogap phase.  The only exception is the so called stripe phase observed in cuprates with the low temperature tetragonal crystal structure around a hole doping on $x = 1/8$. But the recent experiments of Li et al \cite{Li07} as interpreted by Berg et al \cite{Be07} show more complex behavior with an array of interpenetrating SDW and anti-phase superconducting domain walls.

  For an alternative approach, we look to systems where a repulsive interaction causes insulating behavior even at weak values and without breaking translation symmetry. Thus the two requirements for a Mott insulator are satisfied. The simplest such case is the 2-leg Hubbard ladder at half-filling. This system has been extensively discussed in the literature (for a review see Ref\cite{DR96}). The ground state is nondegenerate and insulating for all $U > 0$ with finite energy gaps in the spin, single particle and pair channels. Further translational symmetry along the ladder is preserved.  The pair gap, necessary for a charge gap and insulating behavior, is present only at half-filling. It is interesting to trace its origin in the FRG analysis at weak coupling. It arises because of the presence of additional elastic umklapp scattering processes connecting the 4 Fermi points exactly at half-filling e.g. the two right moving Fermi vectors of the bonding and antibonding bands add up to $\pi$ and so such a pair can backscatter to the two left moving Fermi vectors with an allowed crystal momentum transfer of $2\pi$.

   While one may argue that this is a special feature of one dimension without an analog in higher dimensions, the evolution of wider ladders as the width increases, raises the possibility of analogous behavior in two dimensions. We shall summarize the key results here and refer to a recent review for more details \cite{LeH09}. For N legs with nn hopping and open boundary conditions, there are N bands corresponding to standing waves with transverse wavevectors  $\pi j / (N+1)$ with $j = 1, . . .N$.  At half-filling the bands can be grouped in pairs with equal and decreasing Fermi velocities as follows
\begin{eqnarray}
v_{1} = v_{N} < v_{2} = v_{N-1} < . . . .
\label{eq:ladder}
\end{eqnarray}

\begin{figure}[tf]
\centerline
{
\includegraphics[width = 12.5cm, height =5.5cm, angle= 0]
{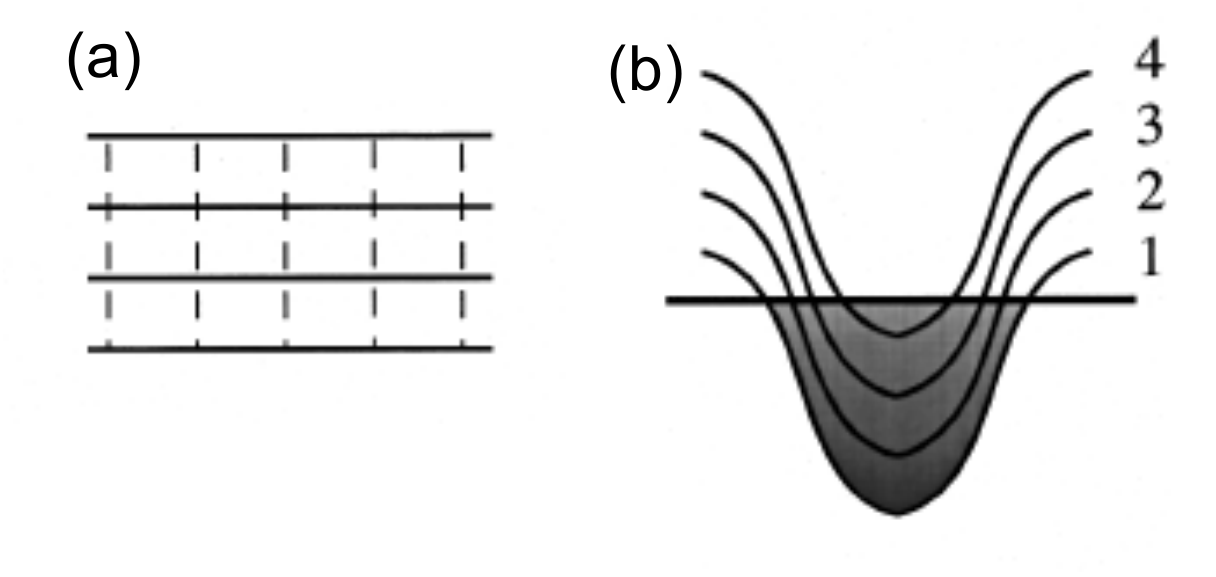}
}
 \caption[]
 {(Color online) Left panel: schematic representation of a 4-leg ladder. Right panel: Dispersion of the 4 bands. Note the Fermi velocities in the outer bands 1 $\&$ 4 are smaller than the central bands 2 $\&$ 3.
 }
\label{fig:4legL}
\end{figure}

The case of a 4-leg ladder is illustrated in Fig.\ref{fig:4legL} which shows that the Fermi velocities in band pair 1 and 4 are smaller than band pair 2 and 3.
In a weak coupling Hubbard model the FRG flow of the interactions inside the band pairs is similar to the flow of a 2-leg ladder. The scale of the divergences depends exponentially on the inverse Fermi velocity of the band pair.  As a result the band pairs have a range of charge gaps with maximum values for the pair with $j = (1,N)$ and minimum values at $j=(N/2, N/2 +1)$ for even values of N. It follows that upon hole doping away from half-filling, the hole density will not be uniformly spread over all the bands. Instead the holes will enter first the central band pair with the largest Fermi velocity and smallest charge gap and then into the next band pair etc. The result will be a successive opening of the bands in pairs with increasing hole density, starting from the band pair, which is nearest to the nodal direction in a 2D representation of the ladder and ending with the band pair nearest the antinodal directions. For the case of odd N, the centre band, $j= (N+1)/2$ remains unpaired and its FRG flow is that of a single chain Luttinger liquid.

  This partial truncation of the Fermi surface in a wider Hubbard ladder lightly doped away from half-filling, contradicts Oshikawa's non-perturbative derivation \cite{OS00} of a relation between the volume enclosed by a Fermi surface in an interacting Fermi liquid and the total electron density in any dimension. This derivation has been quoted \cite{PAL08} in support of the proposal that the observation of only finite Fermi arcs in the pseudogap normal state, is due to strong phase fluctuations of the Cooper pairing amplitude which destroy long range coherence but not the larger energy gaps opening on the full Fermi surface near antinodal. The key step in the derivation is to equate the total wavevector associated with a displacement of the Fermi surface through $2\pi/L$ and the wavevector obtained by applying a corresponding Lieb-Schultz-Mattis transformation in a lattice containing $L^{D}$ sites. However, closer examination reveals that the presence of an underlying lattice allows for more flexibility. This can be seen directly by looking at the simple example of a lightly hole doped 3-leg Hubbard ladder. In this case the bands 1 and 3 are even parity and $j=2$ is odd parity under reflection about the central leg. Both RG calculations at weak coupling \cite{LE00} and exact diagonalizations at strong coupling \cite{RI97},  find the charge gap in the odd parity band is smaller than in the pair of even parity bands at half-filling. As a result all the holes enter the odd band initially up to a finite value of the hole density $x_{c}$, which depends on the interaction strength, U. In this doping range the odd band behaves as a Luttinger liquid with a spanning vector $2 k_{F} = \pi ( 1-3x)$ and only this odd band contributes to the total wavevector resulting from the displacement by $2\pi/L$.  The even bands remain half-filled with a charge gap. The total electron density per unit ladder length corresponds to a total wavevector under a Lieb-Schultz-Mattis transformation with a value $3 \pi(1-x)$. These two wavevectors differ by $2\pi$, i.e. a reciprocal lattice vector. In general, in a crystal two wavevectors can only be equated modulo a reciprocal lattice vector, as is the case here.  This flexibility allows for the possibility of a partial truncation of a Fermi surface in higher dimensions by a commensurate gap even in the absence of broken translational symmetry. In one dimension, i.e. a single chain this flexibility is absent, because the relevant wavevectors are smaller than a reciprocal lattice vector. This flexibility has been noted also by Senthil and coworkers \cite{Se04}  in their analysis of the FL*- model for heavy fermion metals with a small Fermi surface containing only the conduction electrons. As will be discussed further in Section 7 a FL*-model for underdoped cuprates has recently been proposed by Sachdev and coworkers \cite{QS10,MS10} . The FL*-models contain gapless gauge modes which Senthil et al \cite{Se04} argue can contribute to resolve the total wavevector discrepancy.

\begin{figure}[tf]
\centerline
{
\includegraphics[width = 10.0cm, height =6.0cm, angle= 0]
{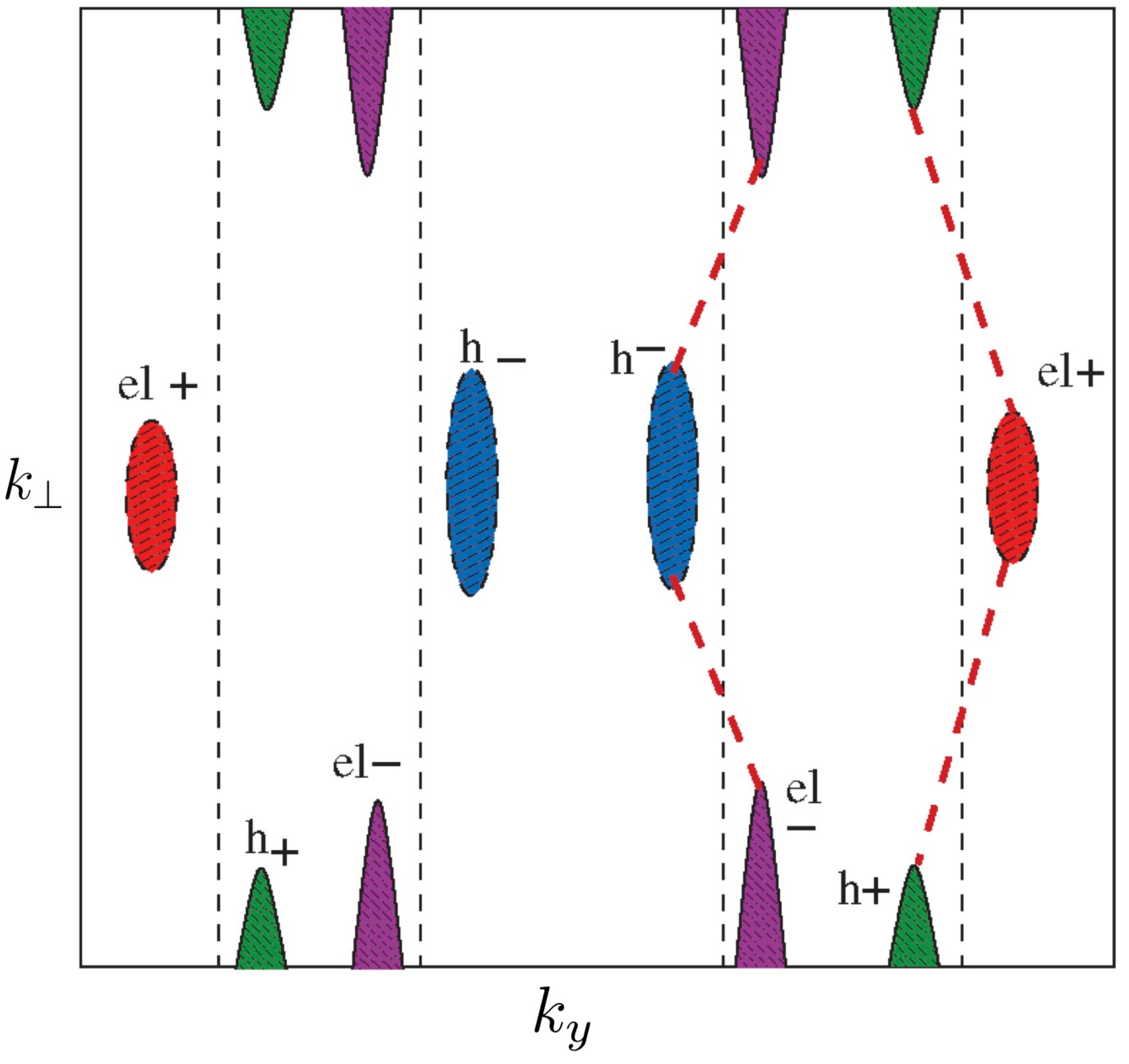}
}
 \caption[]
 {(Color online) The Luttinger surface consists of zeroes (dashed lines at $\mathbf{k} = (k_{y} = \pm k_{F}^{a,b}$) and infinities (at Fermi pockets: electron-like (red) and hole-like (magenta)) predicted from the KRT propagator for a 2-Dimensional array of 2-leg Hubbard ladders \cite{KRT06}. The bonding (antibonding) ladder band is denoted by $+$($-$).
 }
\label{fig:ladder}
\end{figure}

    Another well-controlled model with a partially truncated Fermi surface is a two dimensional array of weakly coupled 2-leg Hubbard ladders oriented along the $y$-axis. Konik, Rice and Tsvelik \cite{KRT06} considered the case of long range inter-ladder hopping whose Fourier transform is sharply peaked in $k_{\perp}$. This choice allows them to treat the inter-ladder hopping by an RPA approach. At half-filling, the individual ladders have a charge and spin gap and the single particle propagator $G(j, k_{y}, 0)$ [$j=ab(\mbox{antibonding}),b(\mbox{bonding})$],  changes sign as $k_{y}$ passes through each Fermi wavevector by passing through a zero rather than an infinity. Note that the single ladder Green's function in Eq.\ref{eq:twoladder} derived by Konik and Ludwig in the weak coupling limit \cite{KL01} has the form
\begin{eqnarray}
G^{0}(j, k_{y},\omega) &=& z_{j}\frac{1}{\omega - \epsilon_{j}(k_{y}) - \frac{\Delta^{2}}{\omega + \epsilon_{j}(k_{y})}} + G_{j, reg}
\label{eq:twoladder}
\end{eqnarray}
where $G_{j, reg} $ is the incoherent part, $z_{j} \sim 1$ is the coherent weight, $\epsilon_{j}(k_{y})$ is the bare band dispersion, $\Delta$ is the quasiparticle gap. The single ladder self energy at zero frequency diverges at the Fermi wavevectors, $k_{y} = \pm k_{F}^{a,b}$, causing zeroes to appear in $G^{0}(j,k_{y}, 0)$ at these wavevectors. Note that the self-energy in 
Eq.\ref{eq:twoladder} is similar to the diagonal self-energy in a superconducting state and there is no off-diagonal component. The KRT propagator for the coupled 2-leg ladders takes the form
\begin{eqnarray}
G(j,k_{y}, k_{\perp},\omega) &=& \frac{1}{ [G^{0}(j,k_{y},\omega)]^{-1} - t_{\perp}(k_{\perp})}
\label{eq:ladder-array}
\end{eqnarray}
For small values of the transverse inter-ladder hopping $t_{\perp}$ the only sign changes in $G (j,\mathbf{k}, 0)$ occur along the lines of zeroes at $\mathbf{k} = (k_{y} = \pm k_{F}^{a,b}, k_{\perp})$ so that the Luttinger sum rule on the area within which the total Green's function
$G (\mathbf{k}, 0) > 0$ is satisfied. Increasing $t_{\perp}$ beyond a finite value causes poles to appear in the propagator leading to the appearance of a set of electron and hole pockets as illustrated in Fig.\ref{fig:ladder} . At half-filling these electron and hole pockets have equal areas so that the Luttinger sum rule is again obeyed. Beyond a certain hole doping the electron pockets get emptied and we are left with a Fermi surface consisting of a set of hole pockets. The Luttinger sum rule continues to hold with sign changes at the fixed lines of zeroes as well as the hole pockets. Note these lines of zeros are fixed in $\mathbf{k}$-space independent of the hole doping and enclose an area of exactly 1 el./site in agreement with the arguments made above. $G^0(j,k_y,\omega)$ remains unchanged upon hole doping close to half filling.
The role of zeroes in Green's function has been first discussed by Dzyaloshinski \cite{Dz03} some years ago.

\subsection{The Yang-Rice-Zhang Ansatz}
Motivated by the FRG results which are valid up to a moderate strength of the onsite coupling U, and by the single particle propagator obtained for a two-dimensional array of weakly linked 2-leg ladders, we proceed to the Yang-Rice-Zhang ansatz for the coherent part of the single particle Green's function for the normal state in a doped Mott insulator. The key postulate is that there is a continuous crossover from the weak to moderate interaction to the strong interaction limit. Such a continuous crossover is common in one-dimensional systems, e.g. a two leg Hubbard ladder. In this case at half-filling the system is always insulating and with strictly short ranged AF order and the main difference between weak and strong coupling is in the separation of the energy of charge and spin excitations as the onsite repulsion is increased. This is the motivation behind the choice of the YRZ ansatz.

 We consider a 2D $t-J$ model, which is in the strong coupling limit of the repulsive Hubbard model, with $t$ the hopping integral, and $J$ the nearest neighbor spin-spin coupling.
The $t-J$ model is in the Gutzwiller projected Hilbert space with no doubly occupied electron states on the same site. We start from the RMFT theory for a RVB state. This approximation treats the effect of strong correlations through renormalization factors, which relate the expectation value of the kinetic and spin exchange energies between the Gutzwiller projected states with those in the corresponding unprojected states \cite{Gutzwiller}.  For a hole doping of $x$, the renormalized $t-J$ Hamiltonian, which operates in the unprojected Hilbert space, takes the form of an effective Hamiltonian \cite{RMF}%
\begin{eqnarray}
H_{eff}=g_{t}(x)H_{K}+g_{s}(x)J\sum_{\left\langle i,j\right\rangle }\mathbf{S}%
_{i}\cdot \mathbf{S}_{j}
 \label{RMT}
\end{eqnarray}%
with the kinetic $H_{K}$, and spin exchange energy terms modified by factors $g_{t}(x)$ and
$g_{s}(x)$, respectively
\begin{eqnarray}
\begin{tabular}{cc}
$g_{t}(x) ={2x}/{(1+x)} \mbox{;}$ &  $ g_{s}(x) ={4}/{(1+x)^{2}}$
\end{tabular}
\label{eq:factor}
\end{eqnarray}%
At half-filling $g_{t}=0$ and $H_{eff}$ has only the spin exchange energy. The RVB mean field approximation factorizes the spin-spin interaction term introducing both Fock exchange, $\chi _{i,j}=\langle c_{i,\sigma }^{\dagger
}c_{j,\sigma }\rangle $ and pairing, $\Delta _{i,j}=\left\langle
c_{i,\uparrow }c_{j,\downarrow }\right\rangle $ expectation values. Note that the factorization procedure is not unique but the resulting spin quasiparticle dispersion
is unique, $E_{\mathbf{k}}=(3g_{s}J/8)(\cos ^{2}k_{x}+\cos
^{2}k_{y})^{1/2}$. Upon hole doping $g_{t}>0$, coherent quasiparticle
poles with a small weight $g_{t}$ appear in the single particle Green's function. By analogy with the KRT form for the doped spin liquid discussed in the previous section, we make the following ansatz for the coherent part of single particle Green's function $G^{RVB}(\mathbf{k%
},\omega )$ in a doped RVB spin liquid, with $\mathbf{k}=(k_{x},k_{y}),$%
\begin{eqnarray}
G^{RVB}(\mathbf{k},\omega ) &=&\frac{g_{t}(x)}{\omega -\xi (\mathbf{k}%
)-\Sigma^{RVB}(\mathbf{k}, \omega) } \mbox{;           } \nonumber \\
\Sigma^{RVB}(\mathbf{k}, \omega) &=& \Delta _{R}^{2}(\mathbf{k})/(\omega +\xi _{0}(\mathbf{k}))  \label{eq:PG}
\end{eqnarray}%
where
\begin{eqnarray}
\xi (\mathbf{k}) &=&\xi _{0}(\mathbf{k})-4\bar{t}^{\prime }(x)\cos
k_{x}\cos k_{y} -2\bar{t}^{\prime \prime }(x)(\cos 2k_{x}+\cos 2k_{y})-\mu _{p} \nonumber \\
\xi _{0}(\mathbf{k}) &=&-2\bar{t}(x)(\cos k_{x}+\cos k_{y})  \nonumber \\
\Delta _{R}(\mathbf{k}) &=&\Delta _{0}(x)(\cos k_{x}-\cos
k_{y})  
\label{eq:parameter2}
\end{eqnarray}%
Eq.\ref{eq:PG} is analogous to Eq.\ref{eq:twoladder} for the coupled 2-leg ladder array, and $%
\xi (\mathbf{k})-\xi _{0}(\mathbf{k})$ is analogous to $t_{\bot
}(k_{\bot })$ in Eq.\ref{eq:ladder-array}. In the renormalized dispersion we include
hopping terms out to 3$^{\mbox{rd}}$ nearest neighbor with coefficients
\begin{eqnarray}
\bar{t}(x) &=&g_{t}(x)t+3g_{s}(x)J  \bar{\chi}/8,  \nonumber \\
\bar{t}^{\prime}(x) &=&g_{t}(x) t^{\prime } \mbox{;               }
\bar{t}^{\prime \prime }(x) = g_{t}(x)t^{\prime \prime }  \label{parameter}
\end{eqnarray}%
where $t, t', t''$ are the hopping integrals with values taken from experiments, and all in units of $t_{0}$ with $t=t_{0}$. The constant $\bar{\chi}$ is taken to have a doping independent value of 0.338 due to the weak doping dependence of $\chi_{i,j} \sim \bar{\chi}$ near half filling.
At or very close to half filling, the Green's function in Eq.\ref{eq:PG} gives a quasi-particle spectrum identical to that in the RVB theory for the $t-J$ model \cite{RMF}. Similar to the self-energy in a 2-leg ladder array in Eq.\ref{eq:twoladder}, the self-energy here resembles the diagonal self-energy in a superconducting state. Its form and the position of the Green's function's zeroes remain unchanged for $x<x_c$. The magnitude of $\Delta_0(x)$ is taken to be the RVB gap in the RMFT \cite{RMF}, with a form
$\Delta_0(x)= \Delta_0 (1- x/0.2)$.  This sets up a quantum phase transition at the density
$x=x_c$ (=0.2). The RVB gap and hence the self energy in Eq.\ref{eq:PG}
vanishes at $x>x_c$. The parameter $\mu _{p}$ represents a shift of the chemical potential chosen so that the Luttinger sum rule (LSR) on the total electron density is satisfied and acts to concentrate the holes in the nodal region at underdoping.

This Green's function can also be rewritten as
\begin{eqnarray}
G^{RVB}(\mathbf{k}, \omega) &=& \sum_{\alpha=\pm}Z^{\alpha}({\mathbf{k}}) /[\omega- E^{\alpha}({\mathbf{k}})] \nonumber \\
 E^{\pm}({\mathbf{k}}) &=& \bar{\xi}({\mathbf{k}}) \pm \sqrt{\bar{\xi}({\mathbf{k}})^{2} + \epsilon({\mathbf{k}})^{2}} \nonumber \\
 Z^{\pm}({\mathbf{k}}) &=& 1/\Big\{1 + [\Delta_{R}({\mathbf{k}})]^{2} / [E({\mathbf{k}})^{\pm} + \xi_{0}({\mathbf{k}})]^{2}  \Big\}
 \label{eq:PG-band}
\end{eqnarray}
where $\bar{\xi}({\mathbf{k}}) = [\xi({\mathbf{k}}) - \xi_{0}({\mathbf{k}})]/2, \epsilon({\mathbf{k}}) ^{2}= \xi({\mathbf{k}}) \xi_{0}({\mathbf{k}}) + \mid \Delta_{R}({\mathbf{k}})\mid ^{2}$. The subindex ($\alpha = \pm$) refers to
the lower or upper energy band.

 \begin{figure}[tbp]
\includegraphics[width=10.0cm,height=7.0cm]{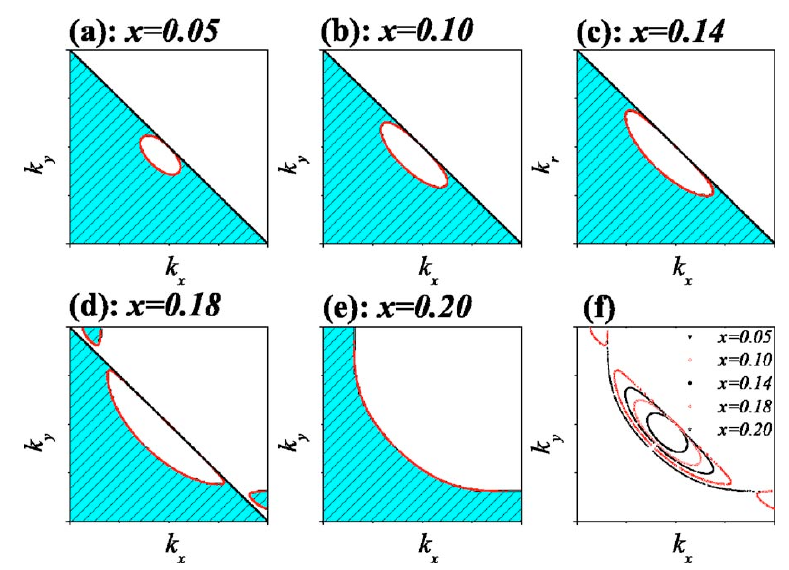}
\caption{
(Color online) The Fermi surface evolution as hole concentration $x$ increases in the YRZ Green's function. Only a quarter of the Brillouin zone (BZ) is shown.  The total area of the 4 Fermi pockets
(red lines in panels (a) - (c)) in the whole BZ is $x/2$ of the area of the BZ \cite{YRZ06}. The black lines in panels (a) - (d) are the lines of Luttinger zeroes. Note that there is an additional electron pocket near the critical doping $x_c$ [see (d)].
} \label{LSR_fig}
\end{figure}

 The YRZ Green's function has the desired properties of line zeroes and hole pockets at small doping and a large Fermi surface at large doping. These are illustrated in Fig.\ref{LSR_fig}.  In the limit $x\rightarrow 0$, $g_{t}(x)\rightarrow 0$ and
$G^{RVB}(\mathbf{k},\omega )$ has vanishing weight. The
quasiparticle dispersion reduces to the spinon dispersion.
At small but finite $x$ the zero frequency Green's function $G^{RVB}(%
\mathbf{k},0)$ that enters the LSR has lines of zeroes when $\xi _{0}(%
\mathbf{k})[=-2\bar t(x)(\cos k_{x}+\cos k_{y})]=0$. The Luttinger contour of zeroes in $G^{RVB}(\mathbf{k},0)$ consists of straight lines connecting the points $(\pm \pi, 0)$ and $(0,\pm
\pi )$, which coincides with the antiferromagnetic Brillouin zone.  More importantly it coincides with the umklapp surface appearing in FRG calculations on the weak coupling 2D $t-t^{\prime }-U$ Hubbard model. As discussed by
Honerkamp \cite{CH01, CH02} and by Laeuchli \textit{et al}. \cite{Laeuchli}, at weak to moderate coupling the umklapp scattering processes in both particle-hole and
particle-particle channels grow strongly in the RG sense at low energies and temperatures so that an energy gap would open up on this umklapp surface below a critical scale with short range rather than long range ordering.

\begin{figure}[tbp]
\includegraphics[width=10.0cm,height=12.0cm]{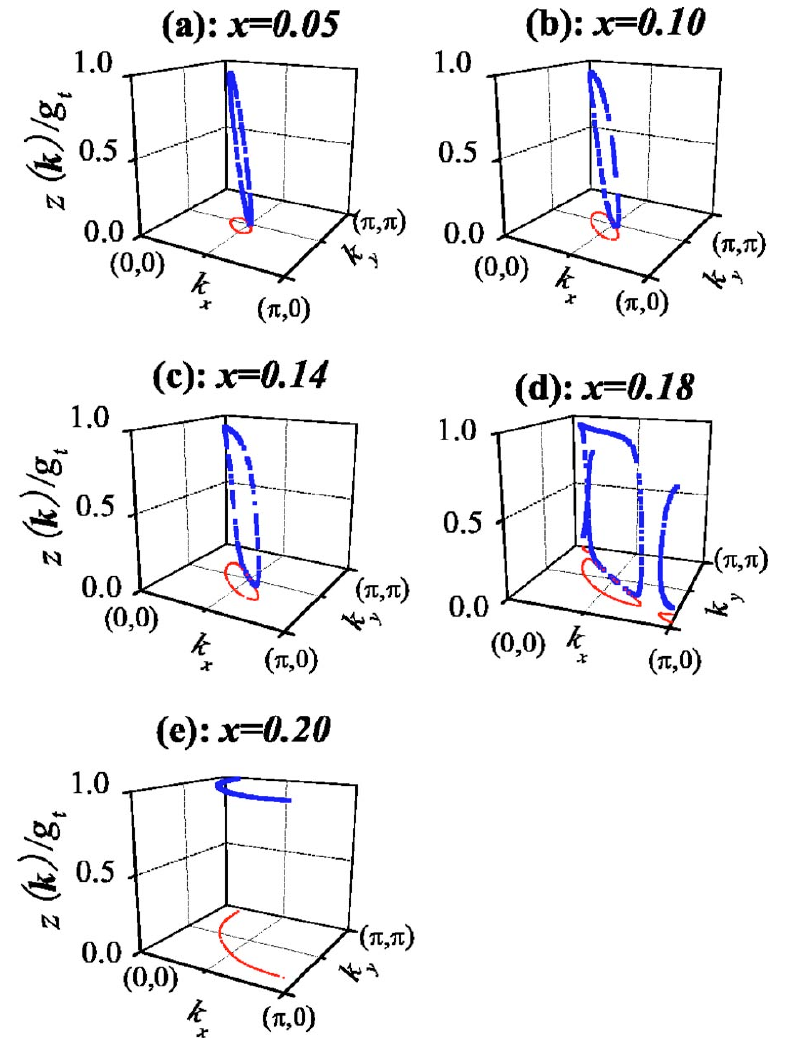}
\caption{ (Color online) The spectral weight distribution
$Z(\mathbf{k})/g_{t}$ around the hole/electron quasiparticle pockets in
the normal pseudogap phase \cite{YRZ06} is shown in blue. The back sides of the hole pockets have a small weight and are difficult to observe in ARPES experiments. The red curves denote the Fermi surface.
}
\label{pocket_weight_fig}
\end{figure}

A second feature of $G^{RVB}(\mathbf{k},\omega )$  is the appearance of hole pockets at finite hole doping as shown in Fig.\ref{LSR_fig}. The hole pockets define Fermi surfaces where
$G^{RVB}(\mathbf{k},0)$ changes sign through infinities. The
total area of the four pockets in units of the Brillouin zone is
equal to the hole density $x$, and  the LSR is
satisfied.

In Fig.\ref{LSR_fig}(a-e) we show that the contours on which
$G^{RVB}(\mathbf{k},\omega =0)$ changes sign at various hole
dopings, with the hopping parameters in Eq.[\ref{parameter}] 
chosen appropriate to Ca$_{2-x}$Na$_{x}$CuO$_{2}$Cl$_{2}$, (see reference
\cite{YRZ06} for details). As $x$ increases, the hole Fermi pocket evolves
gradually into a more normal surface.  There is a quantum
critical point at $x=x_c$ associated with the collapse of the pseudogap and the recovery of the full Fermi surface at $x >x_{c}$. For our choice of parameters, at $x=0.18$ which is smaller than, but close to $ x_c$, additional electron Fermi pocket appear near the antinodes.
 The spectral weight of the quasiparticle pole varies strongly around
the hole pocket as shown in Fig.\ref{pocket_weight_fig}.  It is very small on the outer edge of the pocket closest to the Luttinger surface of zeroes and vanishes completely
at $(\pi/2, \pi/2)$. In the next subsections, we will discuss a number of spectroscopic measurements  (ARPES, AIPES and STM), and show that the YRZ Green's function agrees well with these direct spectroscopic probes.

 The YRZ form for $G^{RVB}$ in the normal phase is close but not identical to the form Tai-Kai Ng \cite{TK-Ng} derived based on the slave-boson mean field theory of the $t-J$ model. In the U(1) gauge theory, an electron is factorized into a spinon and a holon at the mean field level.  Spinons and holons interact with each other via a gauge field.  To describe the gauge fluctuations, Ng introduced a phenomenological attraction between a spinon and holon which is a constant in a spatial distance range. This attraction leads to binding of a spinon and a holon, and therefore to quasiparticle poles and a self-energy form similar to that in the YRZ ansatz. His work is closely related to earlier work of Wen and Lee \cite{Wen-Lee}, who used a phenomenological attraction to bind a spinon and a holon within a SU(2) gauge theory for the $t-J$ model.  In Ng's work, longer distance hopping processes were not included, however. It would be interesting to further develop the microscopic formalism of this type of Green function.

\begin{figure}[tbp]
\includegraphics[width=10.0cm,height=8.0cm]{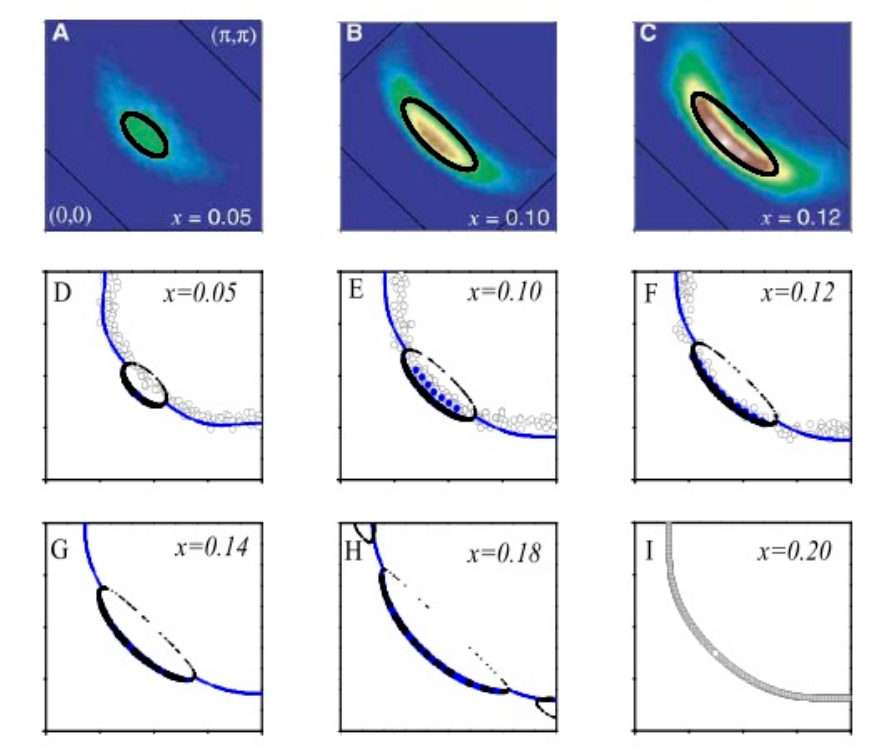}
\caption{
(Color online) (A-C) Comparison between the pockets predicted by the YRZ Green's function \cite{YRZ06} and arcs in ARPES experiments \cite{KMShen-05}.
(D-I) The underlying ``Fermi surface'' defined by the locus of minimum energy below the chemical potential, is shown in blue  \cite{YRZ06}. The open circles are experimental data.
}
\label{comparison_fig}
\end{figure}

\subsection{ARPES Experiments}
In this subsection, we compare the YRZ ansatz with a number of ARPES experiments and find good agreement between theory and experiments.
We start with the ARPES spectra of K. M. Shen et al. on
Ca$_{2-x}$Na$_{x}$CuO$_{2}$Cl$_{2}$ \cite{KMShen-05}.
In Fig.\ref{comparison_fig} we plot contours of the hole
Fermi pockets and also the minimum energy gap lines extending out from the ends of the pockets from the YRZ ansatz and the ARPES results \cite{KMShen-05}. We see good agreement between the two sets of curves. Note that at the outer edges of the hole pockets, the predicted spectral weight of the quasiparticle band  from the YRZ theory is very small (see Fig.\ref{pocket_weight_fig}), and these were not observed in the ARPES experiments of Shen \textit{et al.} \cite{KMShen-05}.

\begin{figure}[tbp]
\includegraphics[width=11.0cm,height=8.0cm]{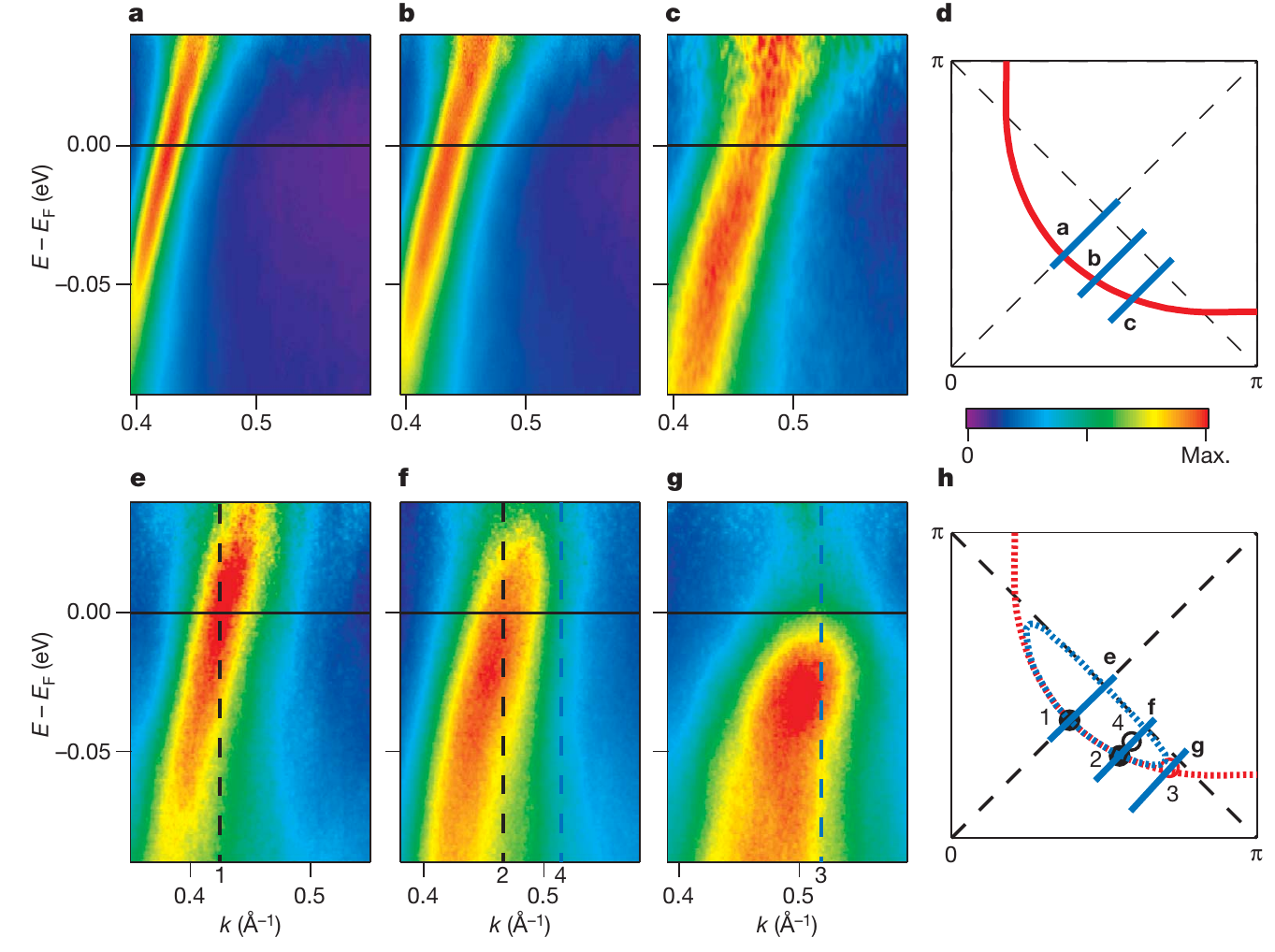}
\caption{(Color online) ARPES spectral plots for optimal doped [panel (a-d) : $T_{c} = 91K$] and underdoped [panel (e-h) : $T_{c} = 65$K] BSCCO samples in the normal state. \cite{YAPDJ08}. A full Fermi surface is observed in the overdoped sample. Particle-hole asymmetry with a gap opening above the chemical potential, appears in the underdoped sample away from the nodal direction in panels f and g.
 }
\label{fig:hong-bo-nature-ph}
\end{figure}
Next, we discuss the particle-hole asymmetry reported in ARPES experiments by H. B. Yang et al. \cite{YAPDJ08, YAPDJ10}. These authors analyzed high resolution ARPES spectra taken on underdoped BSCCO at $T=140K$, carefully dividing out the Fermi function from the spectra both below and above the chemical potential, $\mu$. They found that the quasiparticle energy and spectral weight displayed particle-hole asymmetry as one moves away from the nodal direction in the pseudogap state but not in optimal doped samples.

\begin{figure}[tbp]
\includegraphics[width=10.0cm,height=10.0cm]{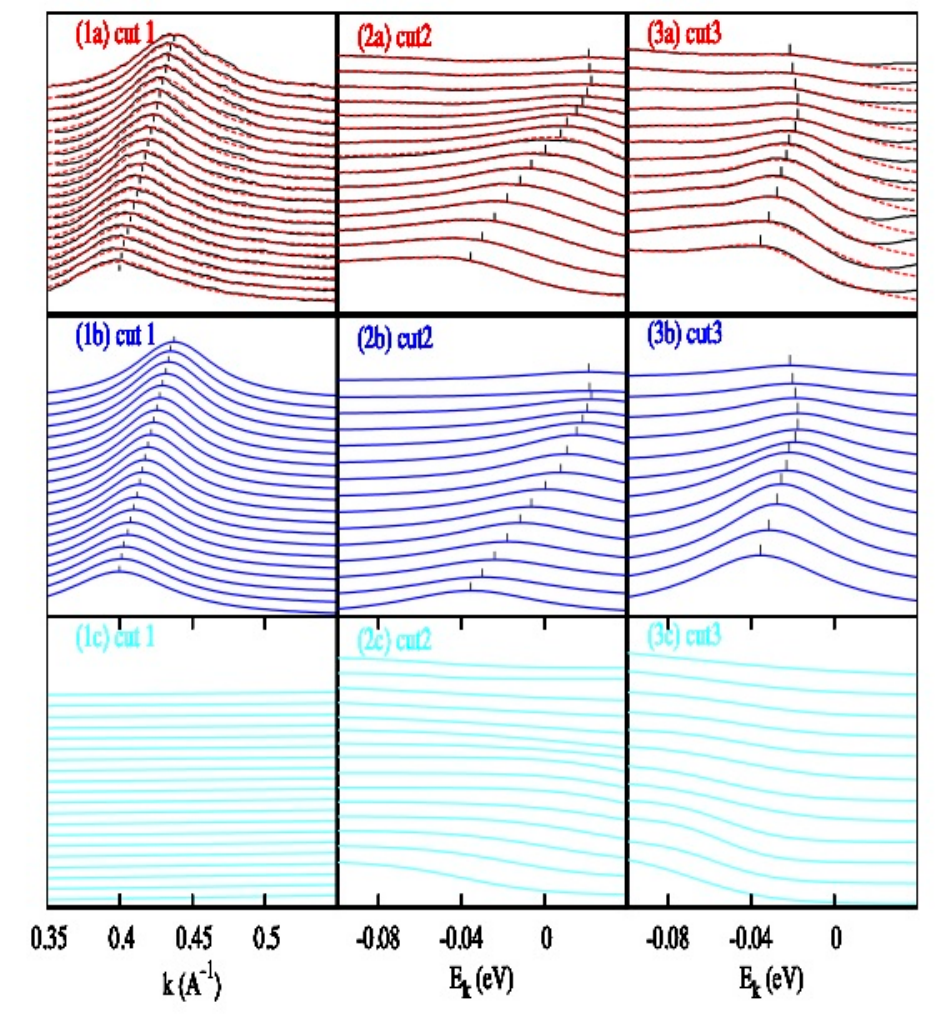}
\caption{(Color online) Fits to the ARPES spectra along the cuts (1-3) [see Fig.\ref{fig:hong-bo-nature-ph}(h)] combining a smooth background (c) and a Lorenzian peak (b) \cite{YA09}. The comparison of the results of the fit to the experimental data is shown in panels (1a-3a). The maximum in the quasiparticle dispersion in cuts 2 and 3 is evidence for the opening of a gap above the chemical potential.
 }
 \label{fig:arpes-fit}
\end{figure}

\begin{figure}[tbp]
\includegraphics[width=14.0cm,height=6.0cm]{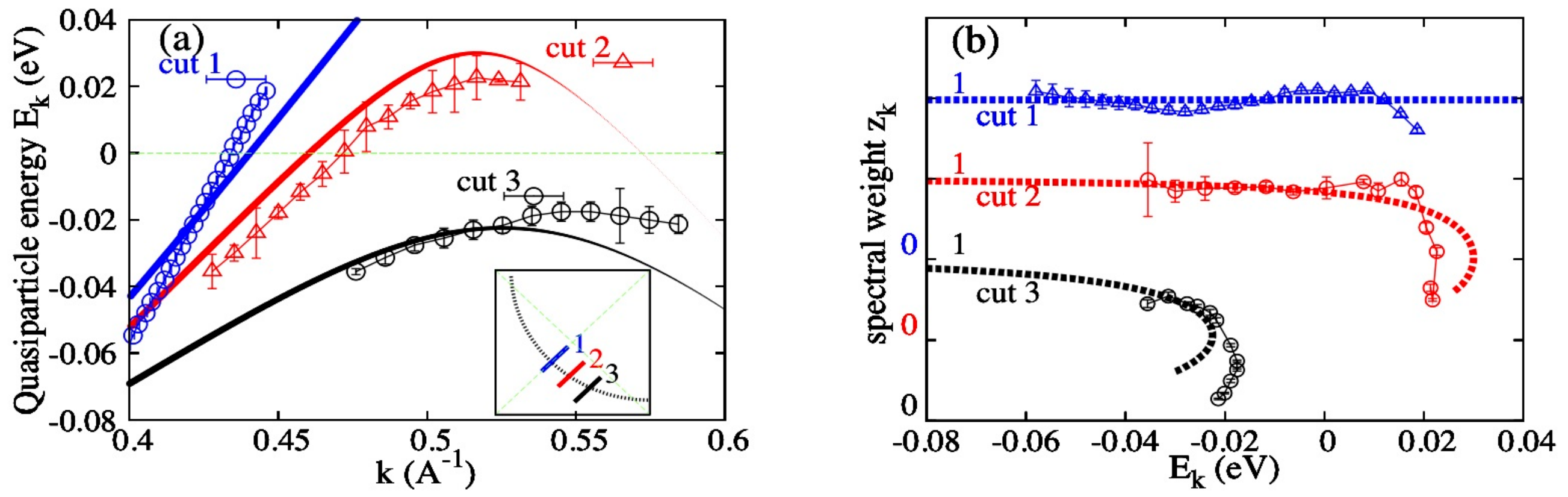}
\caption{(Color online) Comparisons between (a) quasiparticle dispersion $E_{\mathbf{k}}$, (b) spectral weight $Z_{\mathbf{k}}$, from the YRZ Green's function(lines) and the values obtained from the fits of the ARPES data(points) \cite{YA09}.  There is an obvious particle-hole asymmetry in the dispersions above and below zero energy. Starting from the maxima in the dispersion, the spectral weight is strongly suppressed.
 }
 \label{fig:ph-asymmetry}
\end{figure}

In the YRZ ansatz, the energy gap lies above $\mu$ causing particle-hole asymmetry in the quasiparticle spectra in the pseudogap state. The spectra are symmetric only along the nodal directions where $\Delta_{R}(\mathbf{k}) = 0$. The asymmetry increases on the hole pockets away from the nodal directions. To make a comparison with the quasiparticle dispersion and weight with the experiments, we fit the experimentally observed quasiparticle spectra with
\begin{eqnarray}
A(\mathbf{k}, \omega) = Im [Z_{\mathbf{k}}/\omega - E_{\mathbf{k}} + i\gamma_{\mathbf{k}}] + A^B(\mathbf{k}, \omega),
\end{eqnarray}
with $\gamma_{\mathbf{k}}$ the inverse life time and $A^B(\mathbf{k},\omega)$ a smooth background spectrum.
 $E_{\mathbf{k}}$  ($Z_{\mathbf{k}}$) are the dispersion (weight)  of the lower quasiparticle branches. The fits to the ARPES spectra are shown in  Fig.\ref{fig:arpes-fit}. For cut 1, MDC (momentum distribution curve) along the nodal direction, the fitting with an almost constant background spectrum works well for all the energies. For cuts 2 and 3 EDC (energy distribution curves), as $\mathbf{k}$ moves away from the Fermi wavevector, a good fit to the spectra requires a  background component which increases at lower energies. This increase is attributed to contamination from adjacent $\mathbf{k}$ values in the spectra due to the enhanced broadening at low energies. Both quasiparticle dispersion and weight were shown to agree well with the YRZ Greens' function prediction.
In Fig.\ref{fig:ph-asymmetry}, we show the quasiparticle energy as a function of
$\mathbf{k}$. The quasiparticle dispersion, $E_{\mathbf{k}}$ increases linearly with $\mathbf{k}$ along the nodal cut 1, and particle-hole asymmetry is evident in cuts 2 (3) with a maxima in $E_{\mathbf{k}}$ lying above (below) $\mu$. We remark that the maxima do not lie on the boundary of the reduced AF Brillouin zone as one would expect in a state with a superlattice which doubles the unit cell \cite{YAPDJ08}.

\begin{figure}[tbp]
\includegraphics[width=12.0cm,height=6.0cm]{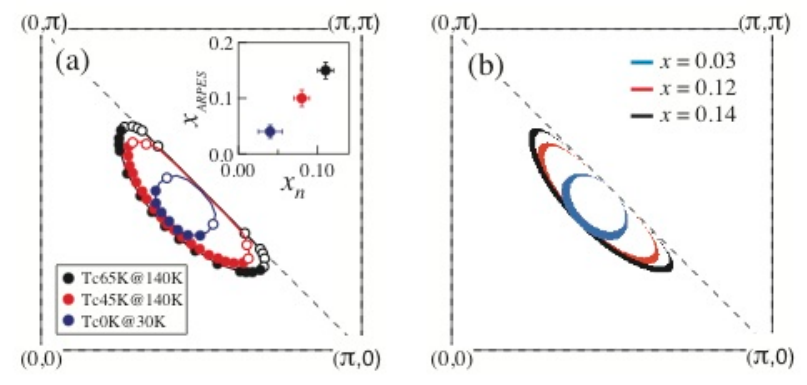}
\caption{
(Color online) (a) The pockets determined by ARPES on Bi2212 by H. B. Yang et al. \cite{YAPDJ10} for three different doping levels obtained by extrapolating the dispersion beyond the maxima which are observed to lie inside the AF reduced BZ.  Inset: The area of the pocket obtained in this way scales with the nominal doping level $x_n$. (b) For comparison, the Fermi pockets derived from the YRZ ansatz with different doping $x$.
 }
 \label{fig:hong-bo-pocket}
\end{figure}

Recently H. B. Yang et al. \cite{YAPDJ10} have refined their ARPES spectra of thermally excited states above the Fermi level obtaining the Fermi surface topologies of underdoped BSCCO. They showed that in the pseudogap phase of underdoped
samples a simple extrapolation of the quasiparticle dispersion beyond the maxima leads to Fermi surfaces which are composed of fully enclosed hole pockets as shown in Fig.\ref{fig:hong-bo-pocket}. The spectral weight of these pockets is vanishingly small near the umklapp surface. The measured area of the pockets is consistent with the hole density. Their results provide very strong support to the YRZ ansatz. The YRZ ansatz is formulated at $T=0K$ but a recent examination of the effects of finite $T$ on the spectral function found that it is quite robust against thermal fluctuations \cite{Khodas10}.

The area of the hole pockets predicted in the YRZ propagator is also consistent with recent STM data of  Kohsaka \textit{et al} \cite{KOH08}.
They reported an abrupt end to the coherent quasiparticle dispersion in the
STM experiments when $\mathbf{k}$ reaches the umklapp lines, and the area enclosed between the Fermi arcs and the umklapp lines is found to be close to $x$, the same as in the YRZ theory.

\begin{figure}[tbp]
\includegraphics[width=12.0cm,height=5.0cm]{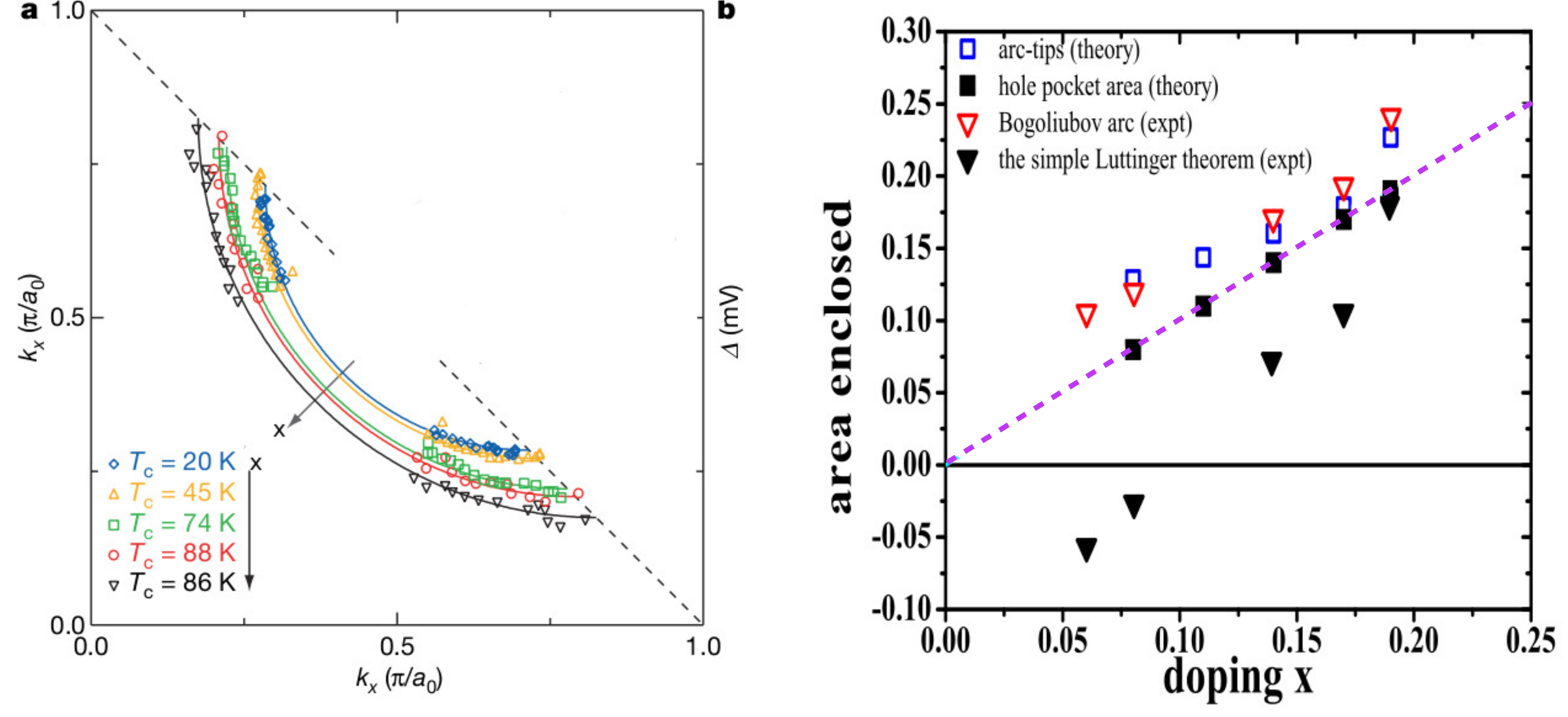}
\caption{
(Color online) (a) Loci of the Bogoliubov band turning points as the voltage increases extracted from STM quasiparticle interference. The loci end at the AF reduced BZ boundary enclose an area close to the doping level, from Kohsaka et al. \cite{KOH08}. (b) The hole pocket area predicted in YRZ Green's function and the area obtained from the STM experiments as defined in (a), along with areas obtained by other methods. The down triangles denote the hole densities calculated from the Luttinger theorem extrapolating the arcs to give a full Fermi surface.  There is good agreement between the YRZ ansatz and the areas in (a).
 }
 \label{fig:pocket-STM-area}
\end{figure}

Meng et al. \cite{Zhou} reported laser ARPES study on an underdoped
La-Bi2201 sample and concluded that their experiments revealed
Fermi pockets of hole carriers in the normal state. Their experiment qualitatively agrees with the Fermi pocket prediction of YRZ ansatz.  Note, there are different interpretations to the observed Fermi pockets due to possible contamination by superlattice order on the sample surface \cite{Shen-comment}.

\subsection{Comparison with AIPES}

\begin{figure}[tbp]
\includegraphics[width=13.0cm,height=10.0cm]{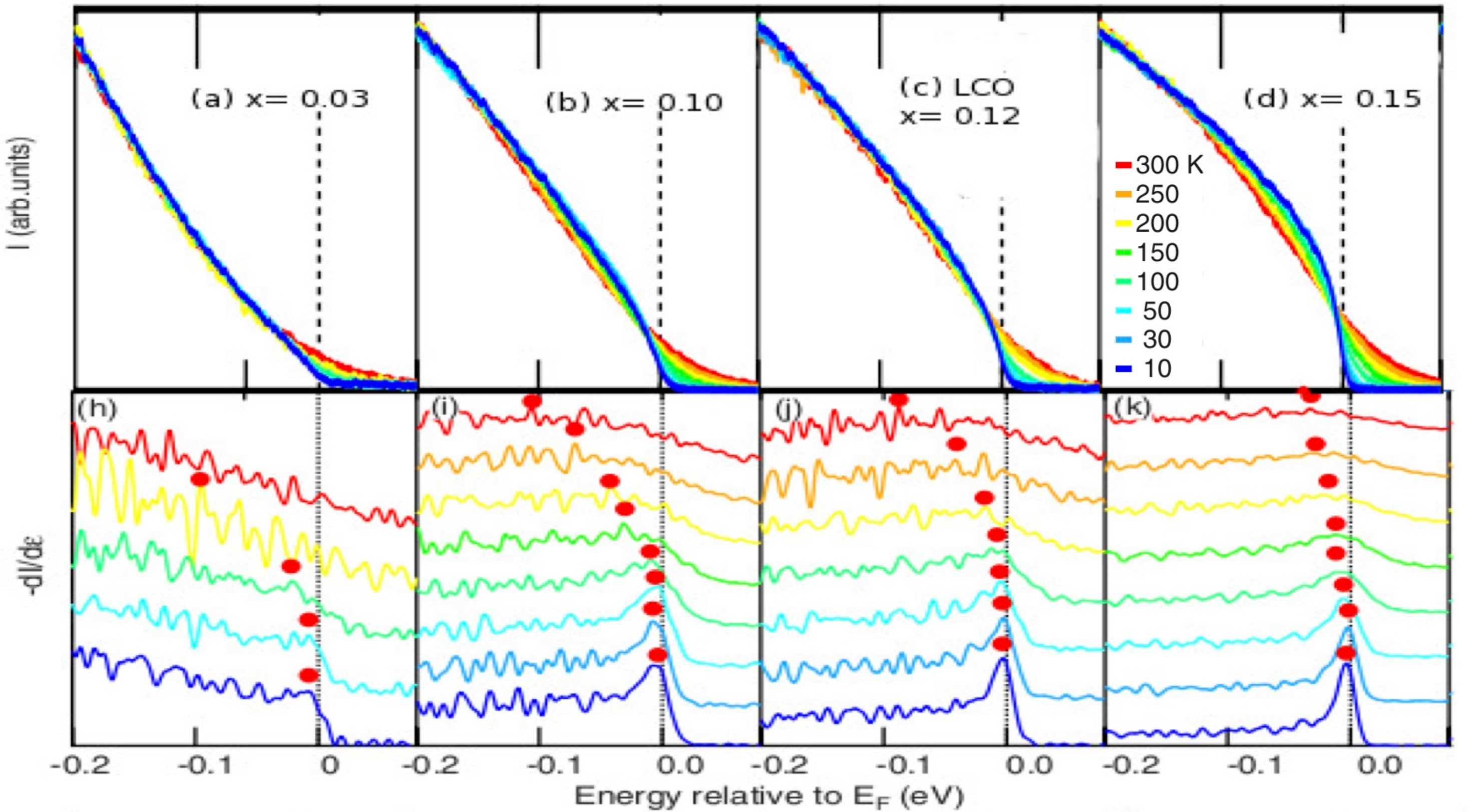}
\caption{ (Color online) {(a-d) Doping $x$ and temperature
dependences (ranging from $300-10K$) of AIPES spectra.
(h-k)
 The first derivative curves of the AIPES spectra with respective to energy of the spectra (2nd row) with the maxima
indicated by the red dots \cite{HA09}. Each spectrum is shifted vertically so that one can see the peaks clearly.}
}
 \label{fig:AIPES1}
\end{figure}

\begin{figure}[tbp]
\includegraphics[width=13.0cm,height=10.0cm]{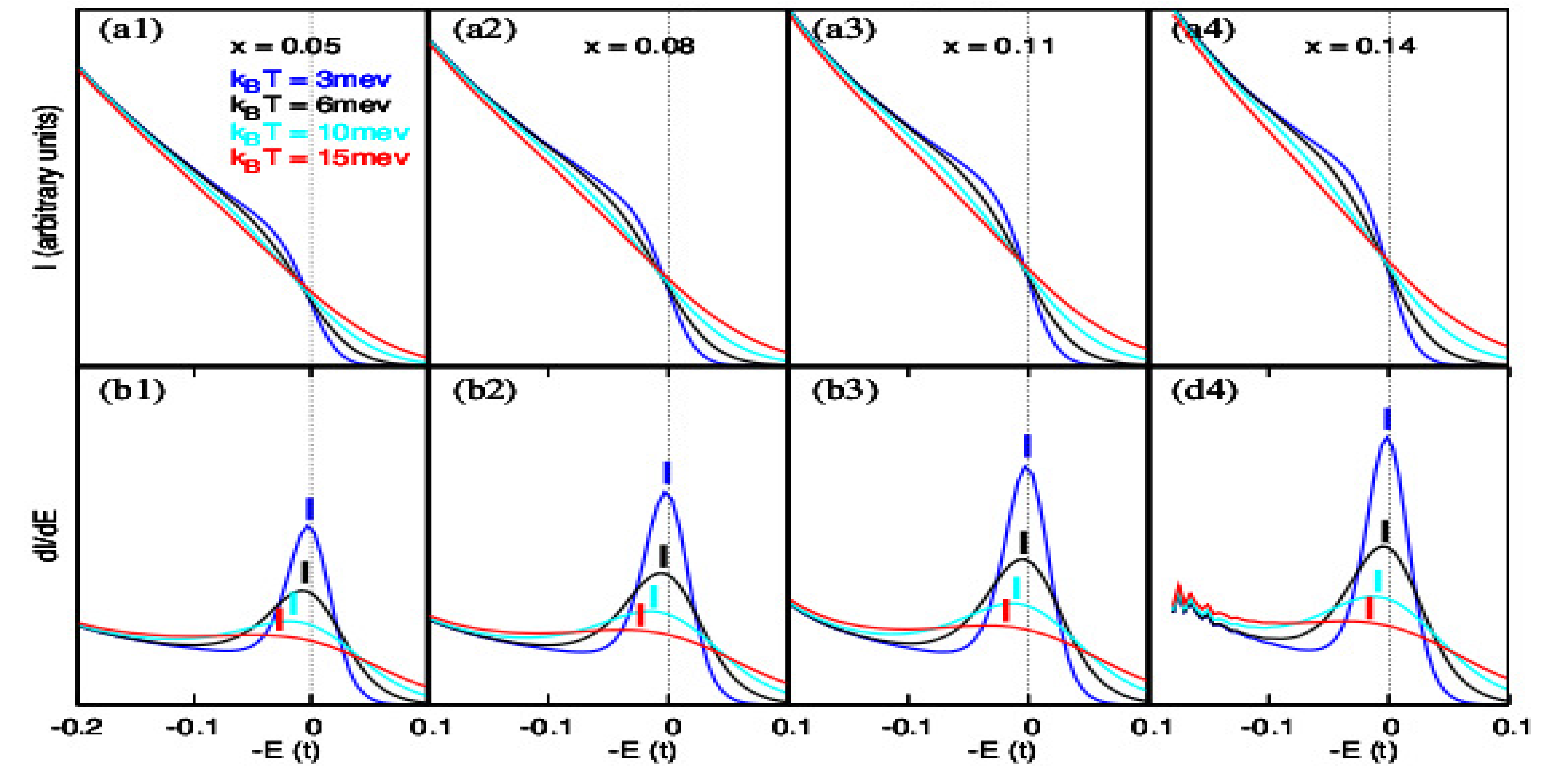}
\caption{ (Color online) 
 Calculated spectra (1st row) and its first derivative (2th row) within the YRZ
model \cite{YA09}. The maxima of the derivatives are indicated in the
2nd row and unusually move to lower energies with increasing $T$. Note,
 both experiment and theory show an approximate linear DOS extrapolating to zero at $x=0$.
 (a1-a4) are the angle integrated spectra $I(E, T, x)$    near    the chemical    potential   (at zero temperature) for various doping (0.05 - 0.14). (b1-b4) are the first   order   derivative  of  $I(E, T, x)$,   with    the peak position $E_{p}$ shifting towards negative electron energy (i.e. positive hole energy), opposite to the shift in the chemical potential. Blue, black, cyan and red curves are for the temperatures $k_{B}T$ = 3, 6, 10, 15 meV.
 }
 \label{fig:AIPES2}
\end{figure}

In this subsection we will test the YRZ against recent angle integrated photoemission (AIPES) experiments on hole doped
La$_2$CuO$_4$ by Hashimoto and coworkers \cite{HA09}. These showed several anomalous features at underdoping in the normal state of the
pseudogap phase. The data show an approximate linear dependence of the total density of states (DOS) on energy, $E <0$, at small hole doping $x$, with only a small quadratic correction which grows with increasing $x$ as shown in Fig. \ref{fig:AIPES1}.  In 2-dimensions and for quadratic dispersion, the hole DOS would be a constant near the valence band maximum, independent of the carrier density. The linear dependence of the DOS on energy, observed in the AIPES experiments,
is reminiscent of a Dirac fermion dispersion. Secondly, the maximum in the first derivative of the angle integrated spectra $dI/dE$ moves to positive not negative, hole energies. The shift rises quadratically with temperature. Writing this shift as $ T^2/T_{coh}$ one finds that the value of the coherence temperature $T_{coh}(x)$, increases with $x$.

The quasiparticle dispersion in the YRZ Green's function is different from the usual 2D case, but is similar to a Dirac fermion at small doping. The vanishing of $\Delta_{R}(\mathbf{k})$ along the nodal directions leads to 4 Dirac points where the upper and lower quasiparticle band touch. In Fig. \ref{fig:AIPES2}, the calculated $I(E, T, x)$ obtained by multiplying the  DOS at energy E by the Fermi function is shown. An approximate linear dependence on $E$ is clear. The energy of the maximum $E_p$, in the derivative $|dI/dE|$ moves to positive hole energies, in contrast to the negative shift of the chemical potential. The $T$-dependence is found to be roughly quadratic, and the resulting characteristic coherence temperature $T_{coh}$ increases with $x$. Therefore, the key anomalous properties in the AIPES are well reproduced by the YRZ ansatz.

\subsection{YRZ Ansatz in the Superconducting State}
The phenomenological form $G^{RVB}(\mathbf{k}, \omega)$ for a hole doped
RVB spin liquid may be generalized to a \textit{d}-wave SC state.
Phenomenologically, the gap at the antinodal region does not change substantially across the SC transition.  This leads us to consider a theory where the SC gap opens up within a small shell of energy around the Fermi level, predominantly near the hole Fermi pockets in the $\mathbf{k}$-space. Since the higher energy band lies well above the Fermi level, we will consider a SC gap function only on the states in the lower $\alpha(-)$ band with a $d_{x^2-y^2}$-wave form in accordance with experiment \cite{Tsuei-RMP-00},
\begin{eqnarray}
\Delta_{sc}^{-}({\mathbf{k}})= \Delta_{sc}^{0} (\cos k_{x} - \cos
k_{y}),
\end{eqnarray}
and $\Delta_{sc}^{+}({\mathbf{k}}) =0$. The Green's functions for the SC state in  the underdoped and the overdoped or optimal doped cases read, in Nambu's spinor representation, 
\begin{eqnarray}
&&G^{sc}_{UD}{(\mathbf{k}, \omega)}= \sum_{\alpha = \pm}
g_t(x)  Z^{\alpha}_{\mathbf{k}} \frac{\omega +
E^{\alpha}({\mathbf{k}}) \tau_{z} -
\Delta_{sc}^{\alpha}({\mathbf{k}}) \tau_{x}} {\omega^{2} -
(E^{\alpha}({\mathbf{k}}))^{2} -
(\Delta_{sc}^{\alpha}({\mathbf{k}}))^{2}},
\nonumber \\
&&G^{sc}_{OD}{(\mathbf{k},\omega)} = g_t(x) \frac{\omega +
\xi({\mathbf{k}}) \tau_{z} - \Delta_{sc}({\mathbf{k}})
\tau_{x}} {\omega^{2} - (\xi({\mathbf{k}}))^{2} - (\Delta_{sc}({\mathbf{k}}))^{2}},
\label{eq:sc}
\end{eqnarray}
where $\tau_{x/y/z}$ are the Pauli matrices, and $g_t(x)Z^{\alpha}({\mathbf{k}})$ and $E^{\alpha}({\mathbf{k}})$ refer to the spectral weight and the dispersion of the quasiparticle bands in the UD region shown in  Eq.\ref{eq:PG-band}. $G^{sc}_{OD}(\mathbf{k}, \omega)$ has a BCS form. The higher energy band lies far above the chemical potential, and the Green's function $G^{sc}_{UD}$ for that band is essentially the same as that in the normal state. 

We remark that in the original paper, the YRZ Green's function was extended to include superconductivity by adding a pairing self-energy \cite{YRZ06}. A similar extension was put forward by LeBlanc et al. \cite{CAR09}. Here we adopt the formalism from a later paper \cite{YA09} which extends the YRZ Green's function to the SC state by introducing a direct pairing interaction between quasiparticles. The formalism presented here preserves the asymmetry of the quasiparticle and quasi-hole spectra, consistent with ARPES experiments~\cite{YAPDJ08}.

  At this point, we should comment on the difference between the YRZ ansatz and the preformed pair model proposed by Norman et al. \cite{Norman-PRB-98}
 and elaborated further in recent work by Banerjee et al. \cite{Banerjee10}. The greatly reduced phase stiffness of the superconducting state in the underdoped region led Emery and Kivelson \cite{EK95} some years ago to propose that resulting strong phase fluctuations would greatly reduce the superconducting $T_{c}$. Norman et al. \cite{Norman-PRB-98} proposed a pairing form which opens  \textit{d}-wave gap on a full Fermi surface in the superconducting state. In the normal state phase fluctuations strongly reduce the gap near the nodal directions but the larger gap near antinodal directions survives. In addition, they included finite lifetime corrections for both single quasiparticles and Cooper pairs. Banerjee et al. \cite{Banerjee10} give a more detailed treatment,  which agrees with the recent calculations on the effect of thermal fluctuations on the single particle propagator in a \textit{d}-wave superconductor by Khodas and Tsvelik \cite{Khod10}. The biggest difference to the YRZ ansatz is that particle-hole symmetry is maintained at all angles around the Fermi surface, which disagrees with the marked asymmetry found by H. B. Yang and coworkers on the Fermi arcs as discussed above.

A closely related issue is the angular dependence of the measured gap in the SC phase of underdoped cuprates. Chatterjee and collaborators \cite{Campuzano} reported that the SC gap obtained from ARPES experiment on Bi-2212 accurately fits the simple \textit{d}-wave form of $\Delta(\mathbf{k}) \sim \cos(k_{x}) -\cos(k_{y})$ for all the angles including the antinodal region.  On the other hand, Yazdani et al. \cite{Yazdani} concluded that the SC gap should have a kink at the end of the Fermi arc from their STM experiments. Yazdani et al. measured the edge in the DOS whereas Chatterjee et al. measured a small peak at the antinodes in the SC phase. Kondo et al. \cite{Kaminski} used ARPES to study the spectral weights associated with the pseudogap and SC spectra  features. They found that the weight of the SC coherent peak increases away from the node following the trend of the SC gap, but starts to decrease in the antinodal region. They concluded that the pseudogap competes with the superconductivity by depleting the spectral weight available for pairing. Their observation supports a two gap scenario with different origins for the antinodal pseudogap $\Delta_{R}$, and the SC gap $\Delta_{sc}$, on the Fermi arcs. A careful theoretical treatment of the spectral line shape in a long range ordered SC state in the presence of substantial disorder is lacking at present.

\subsection{ Comparison to STM Spectra}

  Kohsaka \textit{et al}.\cite{KOH08} determined the dispersion of the coherent Bogoliubov quasiparticles at very low temperatures in the superconducting state. They used STM data taken in a wide field of view in BSCCO samples, analyzing the spatial interference patterns induced by the disorder. The samples ranged from near optimal doping ($x$ = 0.19) to strongly underdoped ($x$ = 0.06). In this way they determined a set of wavevectors connecting the turning points in
the constant energy contours of the Bogoliubov quasiparticle dispersion. These contours
can be compared to the corresponding  YRZ contours.

\begin{figure}[tbp]
\includegraphics[width=8.0cm,height=8.0cm]{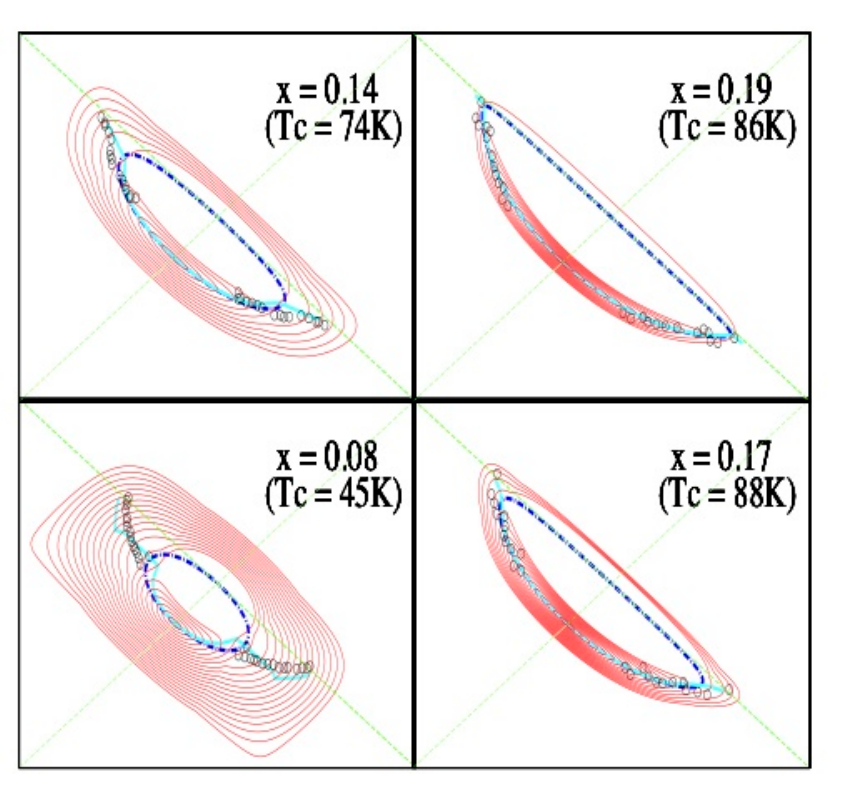}
\caption{
(Color online) The underlying Fermi surface determined from the constant energy curves of Bogoliubov quasiparticles. The locus of the tips of the iso-energy contours predicted by the YRZ ansatz, is shown as a gray curve. It lies very close to the corresponding curve derived from the STM quasiparticle interference data \cite{YA09, KOH08} shown as open circles.
}
\label{fig:iso-energy-stm}
\end{figure}

In Fig.\ref{fig:iso-energy-stm}, we display the evolution of the contours of constant quasiparticle energy near the chemical potential. Note the strong variation in quasiparticle weight in our Green's function means that only the inner contours closer to the zone center are most relevant. The turning points of these contours are determined by the minima in the velocity $dE(\mathbf{k})/dk_{\perp}$ along these contours. The comparison to the STM data is shown in the figure for four representative hole densities. The overall agreement is quite satisfactory. We remark that the calculated contours defined from the turning points at finite voltage are slightly larger than the underlying hole Fermi pockets. This small discrepancy is evident in Fig. 3(a) of the paper by Kohsaka \textit{et al} \cite{KOH08}.  In Fig.\ref{fig:pocket-STM-area}, we plot the areas enclosed by various proposals as function of hole doping.

\begin{figure}[tbp]
\includegraphics[width=8.0cm,height=8.0cm]{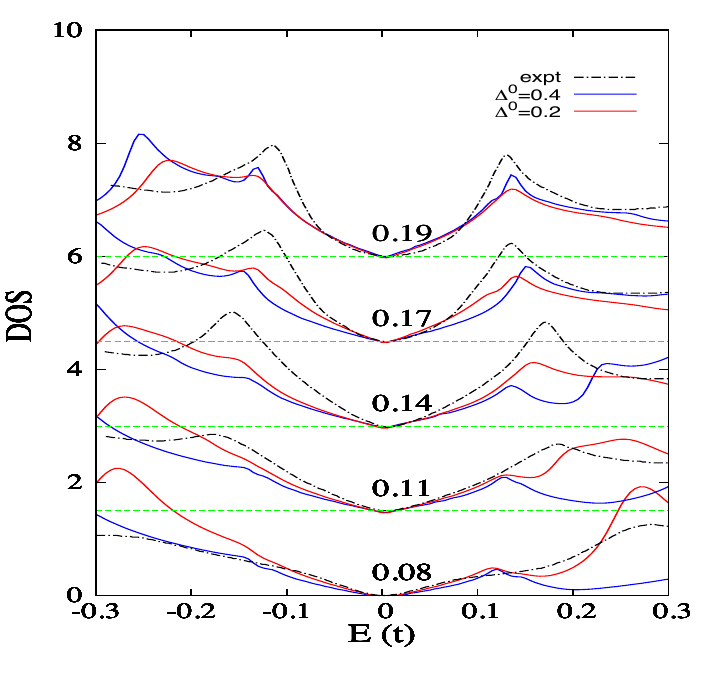}
\caption{
(Color online) 
A comparison between calculated DOS from the YRZ ansatz and STM experiments in the superconducting phase at different densities $x= 0.08-0.19$. At low energies the symmetric profile comes from the SC gap. The YRZ Green's function predicts a stronger particle-hole asymmetry than observed by STM at the underdoped side. The inelastic scattering processes not included in the theoretical DOS, may lead to a
 smoother DOS in better agreement with the STM experiments. A more detailed discussion is given in the text.
 }
\label{fig:DOS-EPL}
\end{figure}

The DOS obtained from the YRZ SC Green's function for an underdoped sample is
given by $N(\omega) = \sum_{\alpha = \pm, i=1, 2,
\mathbf{k}}Z^{\alpha}_{i,\mathbf{k}}\delta{(E^{\alpha}_i(\mathbf{k})
-\omega)}$ ($i=1,2$ refers to the BCS quasiparticle band originated
from the same quasiparticle branch of the pseudogap Green's function $Z^{\alpha}_{i}$ and $E^{\alpha}_{i}$ are the spectral weight and dispersion of the BCS quasiparticle bands).
This may be compared with the STM data reported by Kohsaka et
al. \cite{KOH08}.  In Fig.\ref{fig:DOS-EPL}, we plot the comparison with the experiments for $x=0.08 - 0.19$. At higher hole densities, $0.1 < x < 0.19$, the spectra are dominated by the maximum SC gap. As a result the quasiparticle bands near the Fermi energy are all split by $\Delta_{sc}(\mathbf{k})$ with symmetric low energy DOS. At lower hole density the magnitude of the RVB gap $\Delta_{0}(x)$ rises and exceeds the maximum of $\Delta_{sc}(\mathbf{k})$, leading to two structures in the DOS. One at lower energy related to $\Delta_{sc}(\mathbf{k})$ at the tips of arc, and second at higher energies associated with the RVB gap. If we set the RVB gap coefficient $\Delta_{0}$, to the value used in the ARPES fits, the DOS shown in blue in Fig.\ref{fig:DOS-EPL} displays a much stronger particle-hole asymmetry than the experiments. However, if we reduce the value of $\Delta_{0}$ to 0.2, the agreement is much improved, suggesting that the two gaps are not independent of each other. Note that the resulting energy gap near the antinodal points does not change so much. We do not claim a quantitative fit to the STM DOS but the main features are reproduced at least qualitatively, by the YRZ ansatz. We shall return to this topic when we discuss the comparison of Andreev and Giaever tunneling spectra in Sect. 5.5.

\section{General Properties of the Pseudogap Phase - Comparison Between Experiment and the Predictions of the YRZ Ansatz}
The anomalous features of the pseudogap phase show up in a wide variety of properties of the superconducting and normal phases at underdoping. In a recent series of papers by Carbotte and Nicol and their collaborators, and the study of Andreev reflection by us, it is found that the YRZ ansatz gives a consistent description of these anomalies as a function of both doping $x$ and temperature $T$ with a reasonable parameter choice.

\subsection{Infrared Optical Properties}
    The optical properties of the cuprate superconductors in the overdoped region agree qualitatively with a BCS form for a \textit{d}-wave superconductor, but as the hole density is reduced into the underdoped pseudogap region a number of anomalous non-BCS features appear. Illes, Nicol and Carbotte \cite{IL09} have analysed these anomalies in the context of the YRZ ansatz. The key anomaly is the appearance of an absorption band in the planar conductivity in the normal pseudogap state, which persists into the superconducting phase. This has the effect of reducing the percentage of the carriers that condense in the superconducting phase compared to a BCS model.

\begin{figure}[tf]
\centerline
{
\includegraphics[width = 12.5cm, height =8cm, angle= 0]
{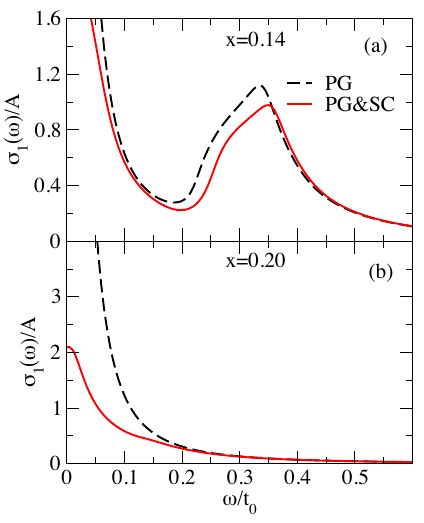}
}
 \caption[]
 {(Color online)
  Calculated optical conductivity curves with (solid red curve) and without (dashed black curve) a superconducting gap \cite{IL09}.
  The hump at finite frequency in the upper panel results from the opening of the pseudogap in the underdoped region and acts to reduce the condensate fraction in the superconducting state. Note because of the \textit{d}-wave symmetry a Drude peak at low frequencies remains in the superconducting phase.
  (Plots reproduced from Ref.\cite{IL09})
  }
\label{fig:optical-1}
\end{figure}

\begin{figure}[tf]
\centerline
{
\includegraphics[width = 16cm, height =8cm, angle= 0]
{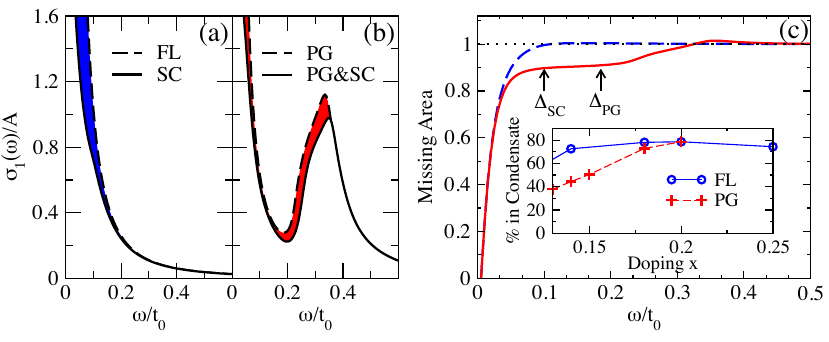}
}
 \caption[]
{(Color online)
The evolution of the missing area shaded in plots of the conductivity vs. frequency transferred to the superconducting condensate calculated in the absence (FL, shaded blue in panel (a)) and the presence (PG, shaded red in panel(b)) of the pseudogap at $x= 0.14$ \cite{IL09}. Panel (c) shows the integrated missing area, normalized to its high frequency value, as a function of frequency in both cases. In the absence of the pseudogap, all the missing area is below the superconducting gap, $\Delta_{sc}$. but not in the presence of the pseudogap. The reduction in the total weight in the condensate due to the pseudogap with hole density is shown in the insert in panel (c) .
 (Plots reproduced from Ref.\cite{IL09})
  }
\label{fig:optical-2}
\end{figure}

   Illes \textit{et al.} \cite{IL09} evaluate the conductivity $\sigma(T, \omega)$ using a Kubo formula as
\begin{eqnarray}
\sigma(T,\nu) &=& \frac{e^{2}}{2v}\sum_{\mathbf{k}} v_{\mathbf{k}}^{2} \int^{\infty}_{-\infty} \frac{d\omega}{2\pi} [f(\omega) - f(\omega + \nu)] \nonumber \\
&&[A(\mathbf{k},\omega) A(\mathbf{k}, \omega +\nu) +B(\mathbf{k},\omega) B(\mathbf{k}, \omega +\nu)  ] \\
A(\mathbf{k},\omega) &=& -2 \mbox{Im} G(\mathbf{k},\omega+i0^{+}) \mbox{;  } B(\mathbf{k},\omega) = -2 \mbox{Im} F^{\dag}(\mathbf{k},\omega)
\label{eq:conductivity}
\end{eqnarray}
where  the spectral functions are related to the superconducting generalization of the YRZ propagators (with a slightly different form from Eq.\ref{eq:sc})
\begin{eqnarray}
G(\mathbf{k},\omega) &=& g_{t} /(\omega - \xi(\mathbf{k}) - \Sigma(\mathbf{k},\omega)) \nonumber \\
\Sigma(\mathbf{k},\omega) &=& \Sigma^{RVB}(\mathbf{k},\omega) + \frac{|\Delta_{sc}(\mathbf{k})|^{2}} {\omega +  \xi(\mathbf{k}) + \Sigma^{RVB}(\mathbf{k},-\omega)} \nonumber \\
F^{\dag}(\mathbf{k},\omega) &=& \frac{-[\Delta_{sc}(\mathbf{k})]^{\dag}G(\mathbf{k},\omega)}{\omega + \xi(\mathbf{k}) + \Sigma^{RVB}(\mathbf{k},-\omega)}
\label{eq:YRZ-SCG}
\end{eqnarray}
The presence of a finite pseudogap in the YRZ propagator leads to an infrared absorption band, which increases in strength as the hole density is reduced in the pseudogap phase. This is illustrated in Fig.\ref{fig:optical-1} which shows the conductivity for two hole densities, one underdoped with a pseudogap and a second beyond the critical concentration with a full Fermi surface. It is clear that the presence of the pseudogap reduces the weight lost to the condensate. This is illustrated further in Fig.\ref{fig:optical-2} which compares the results for the $x=0.14$ concentration with and without a pseudogap. The insert in Fig.\ref{fig:optical-2} (c) shows how the percentage of the finite frequency weight lost to the condensate is progressively reduced as the hole density is decreased.

\begin{figure}[tf]
\centerline
{
\includegraphics[width = 14.5cm, height =8.0cm, angle= 0]
{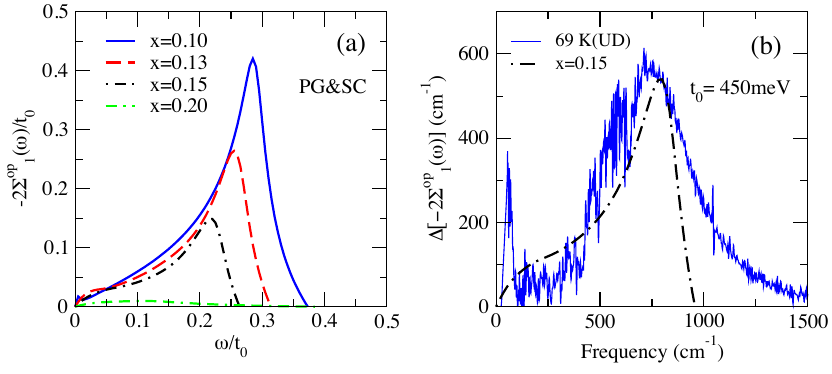}
}
 \caption[]
 {(Color online)
 The optical self-energy, -2$\Sigma^{op}_{1}(\omega)$ at underdoping calculated in the presence of the pseudogap ($ x< 0.2)$ and superconductivity is shown in panel(a). Panel (b) compares the optical self-energy with the experimental values deduced by Hwang et al \cite{Hwang-PRB-08} in an underdoped BSCCO(2212) sample with a $T_{c}$ = 67K. [For details see LeBlanc et al \cite{IL09}) ]
}
\label{fig:optical-sigma}
\end{figure}

The anomalous absorption band can also be expressed in terms of an optical self energy $\Sigma^{op}(T,\omega)$  defined in a generalized Drude form
\begin{eqnarray}
\sigma(T,\omega) = (i\omega_{p}^{2}/4\pi) /[\omega - 2\Sigma^{op}(T,\omega)]
\label{eq:optical-sigma}
\end{eqnarray}
 In Fig.\ref{fig:optical-sigma}(b) the calculated and experimental forms for $\Sigma^{op}(T, \omega)$ for underdoped BSCCO are shown in good qualitative agreement with each other.
Illes \textit{et al} \cite{IL09} conclude that the YRZ form gives a good description of the experimental anomalies in the infrared planar conductivity in the pseudogap phase of underdoped BSCCO.
 Note the presence of the infrared peak in both the normal and superconducting state argues in favour of the interpretation of the pseudogap near antinodal as a consequence of precursor insulating behaviour rather than as arising from finite temperature phase fluctuations in the normal phase which suppress $T_{c}$. The latter should be absent in the low temperature ordered superconducting state.

\subsection{ Specific Heat  }
\begin{figure}[tf]
\centerline
{
\includegraphics[width = 12.5cm, height =8.5cm, angle= 0]
{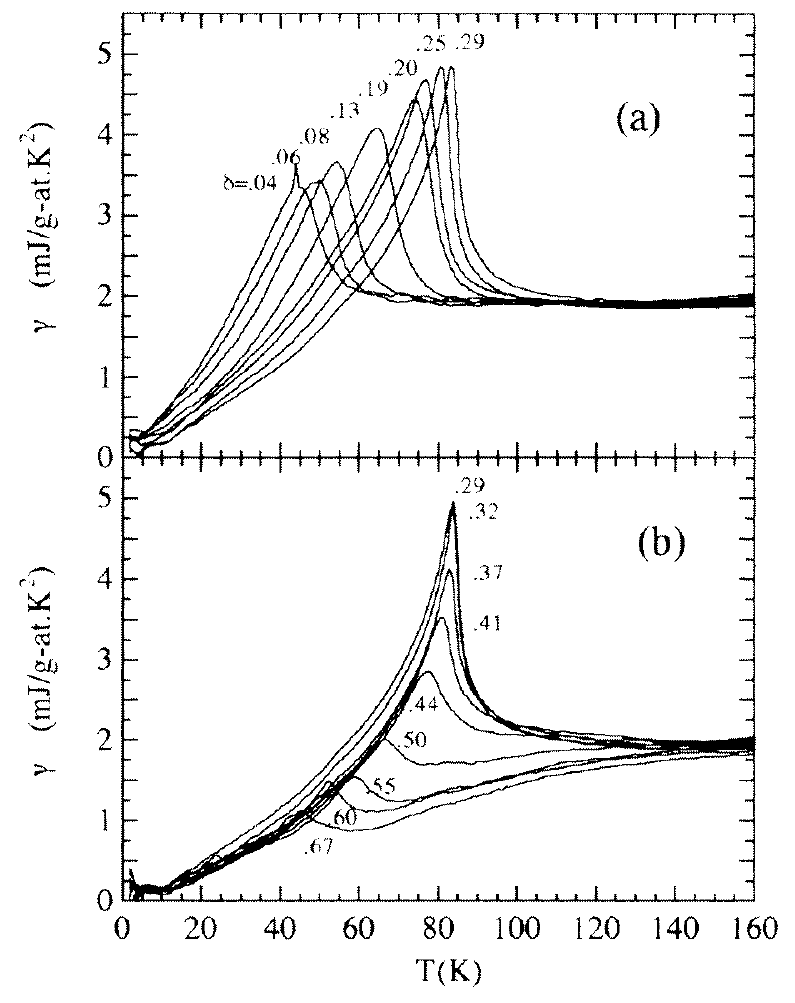}
}
 \caption[]
 {
 (Color online) Specific heat coefficient $\gamma(T,x) $ for (a) overdoped and (b) underdoped Y$_{0.8}$Ca$_{0.2}$Ba$_{2}$Cu$_{3}$O$_{7-\delta}$. In the overdoped material a gap, signaled by a depression in $\gamma$, opens up below $T_{c}$. In the underdoped samples a gap starts to form in the normal state below $140 K$ \cite{Lo97}.
 }
\label{fig:specific-heat-loram}
\end{figure}

In a series of papers going back to the early days of research on the cuprates, Loram and coworkers \cite{Lo94,Lo01, Lo97} extracted the electronic component of the specific heat, $C_{v}$, in the superconducting and normal states in the underdoped region of the phase diagram. They found that $C_{v}$ was highly anomalous and deviated strongly from a simple BCS form. These anomalies are apparent in Fig.\ref{fig:specific-heat-loram} when we compare the ratio $\gamma = C_{v} /T$ in the underdoped (lower panel) to the overdoped (upper panel) samples which where it behaves standardly. The anomalies are a marked reduction in the size of the specific heat jump $\Delta \gamma$, at $T_{c}$ and a depression in the normal state above $T_{c}$, when the hole density was reduced below a critical concentration, $x_{c}$. These grow more pronounced with further decreases in $x$. The linear $T$ term in $\gamma$ at low temperature however does not vary much with $x$. The reduced effect of the onset of superconductivity on the electronic specific heat translates into a reduced value of the condensation energy of the superconducting state $\Delta U$ which can be calculated by integrating the difference in the entropy between the superconducting and extrapolated normal states from $T_{c}$ down to 0K. As is apparent in Fig.\ref{fig:specific-heat-loram}, $\Delta \gamma$ drops dramatically in the pseudogap phase, which is reflected in a similar drop in $\Delta U$. Loram and coworkers \cite{Lo94,Lo01,Lo97} interpreted their results as evidence that in underdoped cuprates the pseudogap is an intrinsic feature of the normal state density of states that competes with the superconducting condensate for low energy spectral weight and is not related to precursor superconducting fluctuations. This interpretation is of course qualitatively consistent with the partial truncation of the Fermi surface observed in the ARPES experiments in the pseudogap region discussed above that is a key feature of the YRZ ansatz for the single particle propagator.

\begin{figure}[tf]
\centerline
{
\includegraphics[width = 14.5cm, height =8.5cm, angle= 0]
{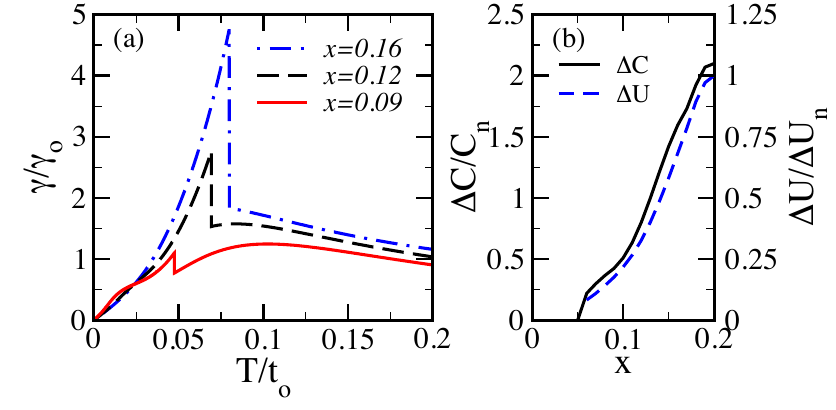}
}
 \caption[]
 {(Color online) (a) The calculated values of the normalized ratio  $\gamma /\gamma_{0}$ vs $T/t_{0}$. (b) The normalized jump, $\Delta C/C_{n} = \Delta C/ \gamma(T_{c})T_{c}$, and the normalized internal energy, $\Delta U/ \Delta U_{n}$ calculated in the superconducting state as functions of doping, $x$. The pseudogap strongly suppresses low energy DOS in underdoped samples leading to the dramatic drop of the specific jump across $T_{c}$. Note that the Fermi arc model also predicts similar results (plots reproduced from Ref.\cite{LeB10})}.
\label{fig:specific-heat}
\end{figure}

  Recently LeBlanc, Nicol and Carbotte \cite{LeB10} calculated the specific heat as a test of the YRZ ansatz. They used the extension of the YRZ ansatz to the superconducting state discussed earlier. This leads to a set of four coherent quasiparticle bands in the superconducting state, which reduces to two bands in the normal. The entropy $S(x,T)$  was calculated using the independent fermion form with each quasiparticle band weighted by its contribution to the single particle propagator 
 \begin{eqnarray}
 S = -2k_{B} \sum_{\alpha = \pm} \sum_{\mathbf{k}} Z^{\alpha}_{\mathbf{k}} \Big\{ f(E^{\alpha}_{sc}) \ln f(E^{\alpha}_{sc}) + f(-E^{\alpha}_{sc}) \ln f(-E^{\alpha}_{sc})  \Big\}
 \label{eq:entropy}
 \end{eqnarray}
where $E_{sc}^{\alpha} = \sqrt{E^{\alpha}(\mathbf{k})^{2} + \Delta_{sc}(\mathbf{k})^{2}}$. 
Typical results for  $\gamma(T)$ are shown in Fig.\ref{fig:specific-heat}.  As the hole density passes through the critical concentration, $x_{c} = 0.16$, the size of the jump $\Delta \gamma$ in $\gamma(T)$ at $T_{c}$ drops dramatically as the effect of the superconducting gap is essentially restricted to the Fermi pockets in the pseudogap phase. The anomalous reduction in $\gamma(T)$ above $T_{c}$ is also evident in their calculations and arises from the truncated Fermi surface in the normal pseudogap phase. Recently Borne and collaborators \cite{BoCv10} extended these calculations to examine the density dependence of the  low temperature specific heat in the superconducting phase. Experimentally this is found to be independent of the density in the underdoped region. At low $T$ excitations are confined to the near nodal region and the entropy depends on the quasiparticle velocity perpendicular to the nodal direction. Bourne et al\cite{BoCv10} reproduce the density independence through a cancellation between the quasiparticle weight, which enters 
as a prefactor in their formula for the entropy, Eq.\ref{eq:entropy}, and the  inverse magnitude of the superconducting gap $\Delta_{sc}^{0}$, which they assume scales with $T_{c}$.  However this cancellation may be misleading since the prefactor does not appear in the Landau formula for the entropy of quasiparticles and corrections due to disorder may enter.

Although this behavior agrees well with experiment, it should not be interpreted as a unique test of the YRZ form. LeBlanc \textit{et al} \cite{LeB10} found the behavior of $\gamma(T)$ near $T_{c}$ difficult to distinguish from  a simpler Fermi arc model in which the pseudogap is set to zero  inside a finite arc around the nodal directions, jumping to a finite value at a critical angle. Note, Banerjee et al \cite{BanerjeeGL10} found similar behavior for the specific heat near $T_{c}$ in a model based on preformed pairs in the normal phase which also gives a finite Fermi arc at $T>T_{c}$.
The limit of a vanishing arc length, i.e. a nodal liquid, leads to bigger discrepancies. Also we note that recent high magnetic field data show a finite DOS in the normal state at low $T$ \cite{Zheng10} in disagreement with a vanishing arc length under these conditions. 

The conclusion is that the YRZ ansatz is qualitatively consistent with experiment but the detailed form of the truncation of the Fermi surface in the pseudogap regime is not uniquely determined from the specific heat.

\subsection{Penetration Depth,  $\lambda(T,x) $}.
The transition into the underdoped region of the phase diagram has a strong effect on the penetration depth  $\lambda(T,x) $. In BCS theory the penetration depth at $T=0K$ is determined by the electron density in the full normal state Fermi surface. The partial truncation of the Fermi surface with a reduced carrier density leads to an enhancement of $\lambda$. Early experiments by Uemura and collaborators \cite{UE89} established that $1/\lambda^{2}(0,x)$ decreased with decreasing values of the doping $x$ consistent with a hole doped Mott insulator rather than a full metal in the underdoped cuprates.

\begin{figure}[tf]
\centerline
{
\includegraphics[width = 10.5cm, height =5.5cm, angle= 0]
{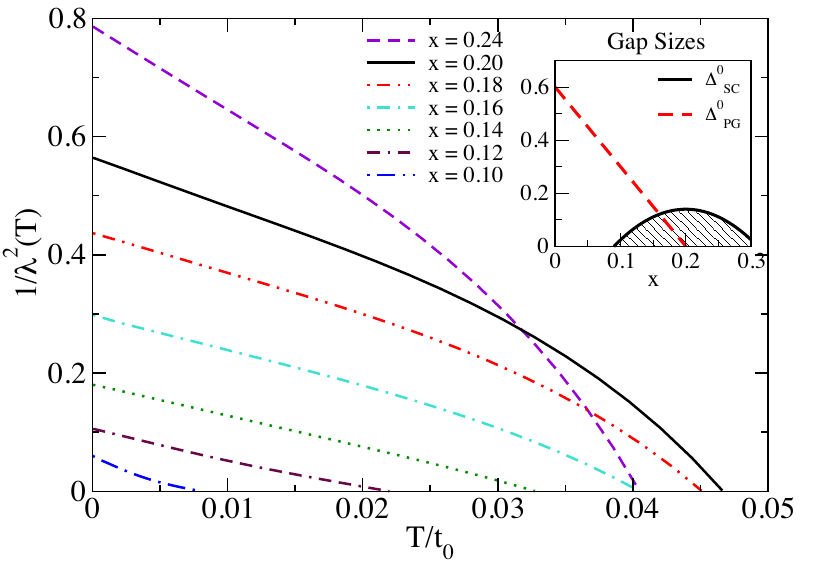}
}
 \caption[]
 {(Color online) The temperature dependence of the inverse square of the penetration depth for various dopings calculated by Carbotte et al \cite{CAR10} based on the YRZ ansatz. The superconducting and RVB energy gaps are set at the values shown in the insert.
 }
\label{fig:penetration}
\end{figure}

   In a recent paper Carbotte and coworkers \cite{CAR10} used the YRZ ansatz for the coherent part of the single particle propagator generalized to the superconducting state as input to examine the $T$ and $x$ dependence of $\lambda (T,x)$. They started from the formula
\begin{eqnarray}
\frac{1}{\lambda^{2}(T)} &=&
\lim_{\Omega \to 0} \frac{16\pi e^{2}} {c^{2}} \sum_{\mathbf{k}} v^{2}_{k_{x}}
\times \int d\omega^{\prime}
d\omega^{\prime \prime} \nonumber \\
&& \lim_{\mathbf{q} \to 0} \Big( \frac{f(\omega^{\prime\prime}) - f(\omega^{\prime})}  {\omega^{\prime \prime} -\omega^{\prime}}\Big) B(\mathbf{k+q},\omega^{\prime}) B(\mathbf{k}, \omega^{\prime \prime})
\label{eq:penetration}
\end{eqnarray}
 where $B (\mathbf{k},\omega )$ is the anomalous spectral density of $F(\mathbf{k},\omega)$ in the superconducting phase in Eq.\ref{eq:YRZ-SCG}.  The partial truncation of the Fermi surface to the Fermi pockets reduces $1/\lambda^{2}(0,x)$ progressively as the hole density is reduced. This is illustrated in Fig.\ref{fig:penetration} which illustrates the evolution of $1/ \lambda^{2} (T,x)$. The inset shows the forms of the RVB and superconducting gaps they used as input. Several features of these calculations are noteworthy.
 \begin{figure}[tf]
\centerline
{
\includegraphics[width = 10.5cm, height =5.5cm, angle= 0]
{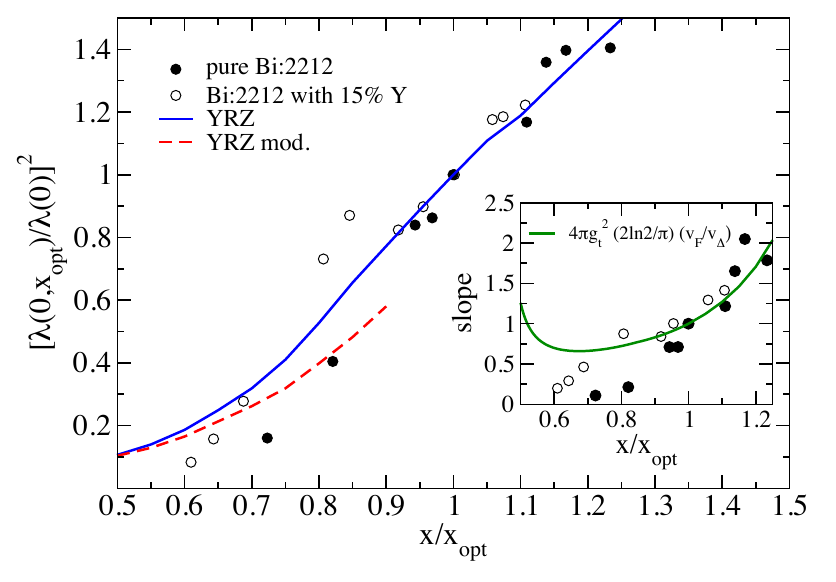}
}
 \caption[]
 {
(Color online) A comparison of the superfluid density at $T=0$ with the experimental values of Anukool et al \cite{Anukool-PRB-09} on BSCCO. The inset compares values of the initial slope, $d(1/\lambda^{2}(T,x) ) / dT$ at $T=0$, with the theoretical form deduced from the YRZ ansatz. Note the slope is larger in the overdoped region at $x >  x_{opt}$. }
\label{fig:lambda0}
\end{figure}

  First, in addition to the reduction of the $T=0K$ value, the shape of the curve evolves from concave downward in the upper two curves at the optimal and overdoping, to an almost linear form at underdoping. Carbotte and coworkers \cite{CAR10} note that a similar evolution has been reported recently in several experiments. Second, the initial slope at low T, $d(1/ \lambda^{2} (T,x) ) / dT$, is relatively constant as the hole doping, $x$, is reduced at $x_{c}$. This is easily understood as the initial slopes are dominated by thermally excited quasiparticles which are concentrated near the nodal direction and this region of the Fermi pockets is rather insensitive to $x$.
In Fig.\ref{fig:lambda0} the comparison with experiment is shown. In the main panel the value of $1/\lambda^{2}(0,x)$ is compared to the BSCCO data of Anukool et al. \cite{Anukool-PRB-09} with the doping scaled to $x_{c}$ the optimal concentration. The curve labelled ``YRZ mod'', sets $\Delta^{0}_{sc}=0$ outside the Fermi pockets. The insert compares the initial slope to the experiment. Note in Fig.\ref{fig:penetration}, the optimal concentration  $x= x_{opt}$ is set at $x_{c} = 0.2$. The doping of each sample was determined from the value of $T_{c}$ by fitting a parabolic form for $T_{c} (x)$ with a maximum at, $x= x_{opt}$. The overall agreement is satisfactory. However we should point out that there is a question regarding the neglect of vertex corrections in Eq.\ref{eq:penetration}.  In Eliasberg theory the Migdal theorem justifies this step. Generally the quasiparticle weight factors are cancelled by the vertex corrections, which would affect the $x$-dependence of $1/ \lambda^{2} (T,x) $. Also the fine details of the YRZ ansatz are not tested. For example, the lower curve in Fig.\ref{fig:lambda0} labelled YRZ mod, was calculated by Carbotte et al \cite{CAR10} using a modification of the YRZ form which omitted the region of $\mathbf{k}$-space outside the FS pockets and it does not differ so much from the full ansatz. But they find the fits to a nodal liquid are less good. So we can conclude that the YRZ ansatz is generally consistent with the $T$ and $x$ dependence of the penetration depth in the pseudogap phase.

\subsection{Raman Spectra in the Pseudogap Phase}
   The anomalous features of the pseudogap phase show up also in the Raman spectra especially in the comparison between the spectra in the experimental polarization configurations that measure the symmetry projections B$_{\mbox{1g}}$ and B$_{\mbox{2g}}$ of the D$_{\mbox{4h}}$ group relevant for the tetragonal cuprates. For more details see the recent review by Devereaux and Hackl \cite{DH07}. The comparison between these two spectra is particularly informative since they weight electronic transitions concentrated near the antinodal and nodal regions respectively. Several years ago Valenzuela and Bascones \cite{VB07} found that the YRZ ansatz described the main features in this comparison. Recently LeBlanc, Carbotte and Nicol \cite{LeB10} extended the earlier calculations and compared their results to the Raman spectra for the single layer tetragonal cuprates, HgBa$_{2}$CuO$_{4+\delta}$ (Hg-1201) \cite{Guyard-PRL-08}. In the superconducting state they used the extension of the YRZ form discussed above. The Raman spectra are determined by the standard formula
\begin{eqnarray}
\chi^{\prime\prime}_{\eta}(\Omega) &=& \frac{\pi}{4} \sum_{\mathbf{k},\alpha = \pm} (\gamma_{\mathbf{k}}^{\eta})^{2} g_{t} \Big(
Z^{\alpha}({\mathbf{k}}) [f(E_{sc}^{\alpha}) - f(E_{sc}^{\alpha} + \Omega)]
\times \nonumber \\
&& [(u^{\alpha})^{2} A(\mathbf{k}, E_{sc}^{\alpha}+\Omega) + \frac{\Delta_{sc}(\mathbf{k})}{2E_{sc}^{\alpha}}B(\mathbf{k}, E_{sc}^{\alpha}+\Omega)] \nonumber \\
&&+
Z^{\alpha}
({\mathbf{k}})[f(-E_{sc}^{\alpha}) - f(-E_{sc}^{\alpha} + \Omega)]
\times \nonumber \\
&& [(v^{\alpha})^{2} A(\mathbf{k}, -E_{sc}^{\alpha}+\Omega) - \frac{\Delta_{sc}(\mathbf{k})}{2E_{sc}^{\alpha}}B(\mathbf{k}, -E_{sc}^{\alpha}+\Omega)]
 \Big)
 \label{eq:Raman}
\end{eqnarray}
where $(u^{\alpha})^{2}= 1/2[1 + E^{\alpha}(\mathbf{k})/E^{\alpha}_{sc}]$, and $ (v^{\alpha})^{2}= 1/2[1 - E^{\alpha}(\mathbf{k})/E^{\alpha}_{sc}]$.
The vertex strengths in the two polarizations are obtained from the energy dispersion of the underlying band structure leading to
\begin{eqnarray}
\gamma_{\mathbf{k}}^{B_{1g}} &=& 2g_{t}t(\cos{k_{x}} - \cos{k_{y}}) + 8 g_{t}t^{\prime\prime} (\cos{2k_{x}} - \cos{2k_{y}})  \\
\gamma_{\mathbf{k}}^{B_{2g}} &=& -4g_{t}t^{\prime} \sin{k_{x}} \sin{k_{y}}
\label{eq:Raman-vertex}
\end{eqnarray}
The coefficients $g_{t}, t, t^{\prime}$ and $t^{\prime \prime}$ are defined earlier in Sect. 4.2.
 Several options for the form of the superconducting and RVB gaps which are input parameters in the YRZ ansatz were examined by LeBlanc \textit{et al.} \cite{LeB10}. The simplest form are full \textit{d}-wave gaps in the Brillouin zone with doping dependent strengths
\begin{eqnarray}
\Delta^{0}_{sc}(x, T=0) &=& 0.14 t_{0} [1- 82.6(x-0.16)^{2}] \\
\Delta_{0}(x) &=& 3 t_{0}(0.2-x)
\label{eq:gap}
\end{eqnarray}

 \begin{figure}[tf]
\centerline
{
\includegraphics[width = 12.5cm, height =12.5cm, angle= 0]
{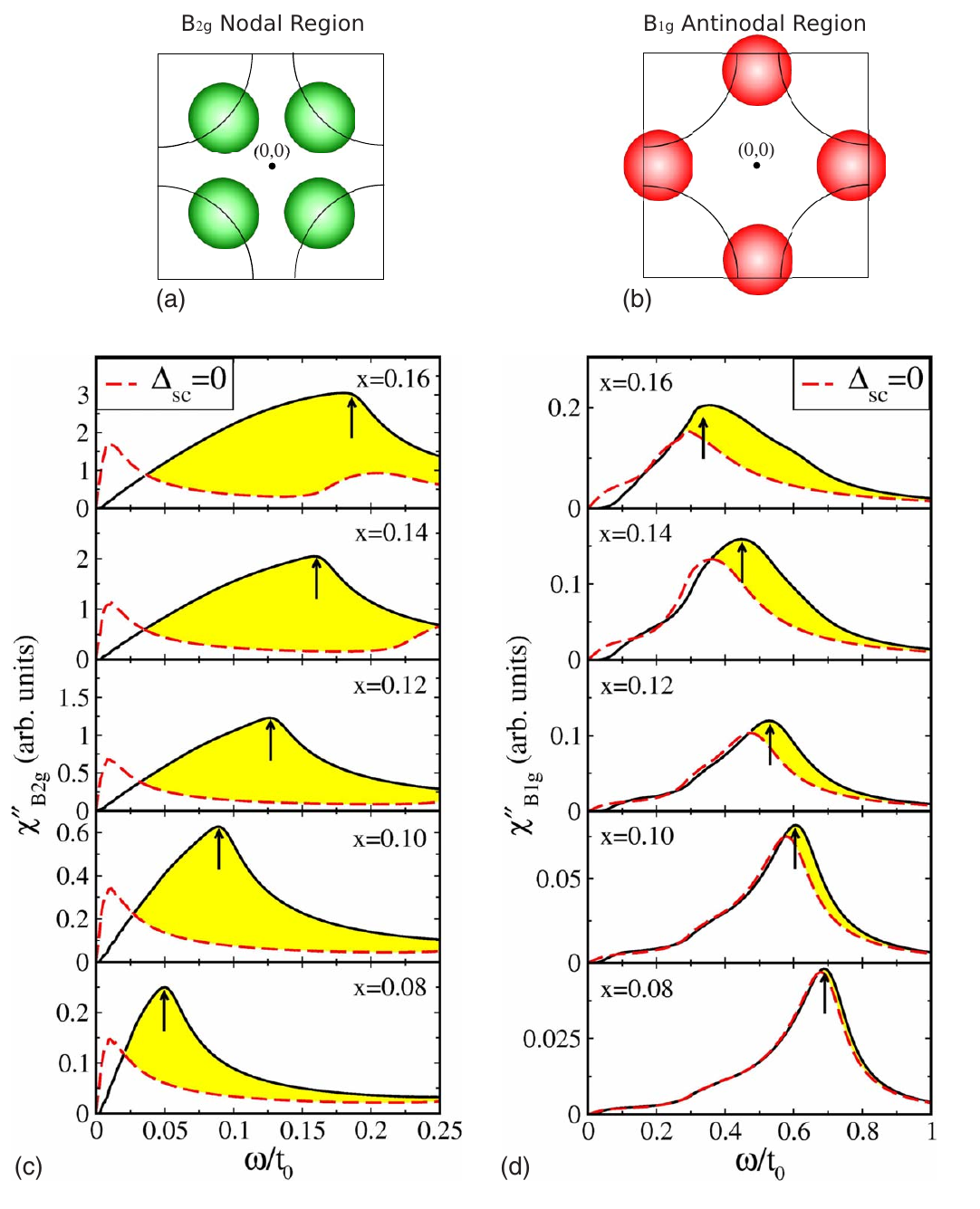}
}
 \caption[]
 {(Color online) (a) and (b) show the strongly weighted regions of $\mathbf{k}$-space  for polarizations B$_{2g}$ and B$_{1g}$.  (c) Raman response in the YRZ model for the B$_{2g}$ (left frames) and B$_{1g}$ (right frames) polarizations \cite{LeB10} with and without a SC gap.
 For B$_{2g}$ the response at low energy is dominated by the SC gap in the SC state, while for B$_{1g}$ in both SC and PG states, the response is dominated by pseudogap. Note the scale change between panels (c) and (d).
  }
\label{fig:Raman}
\end{figure}

 \begin{figure}[tf]
\centerline
{
\includegraphics[width = 10.5cm, height =6.5cm, angle= 0]
{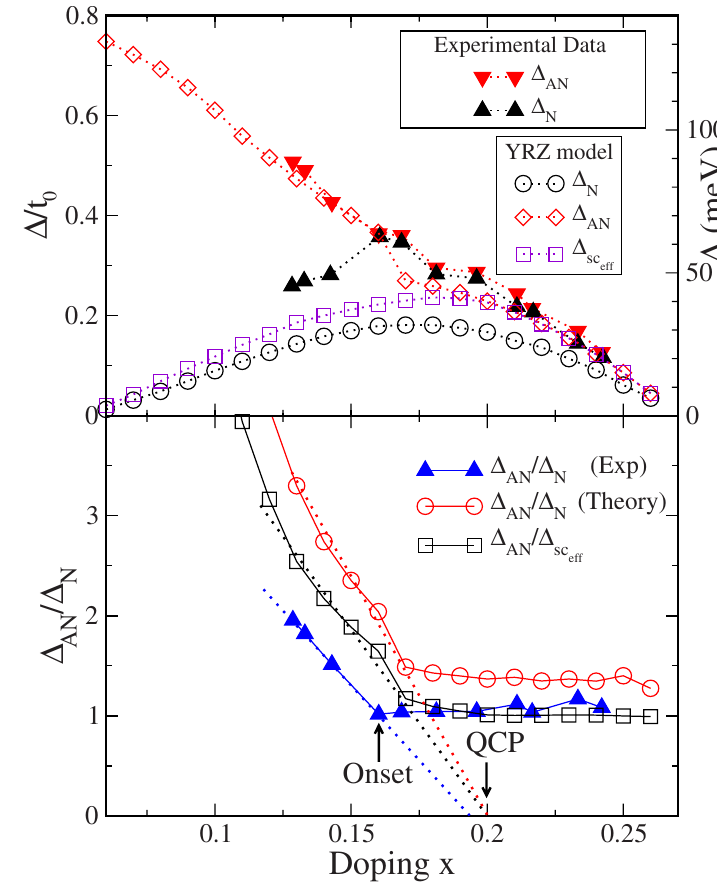}
}
 \caption[]
 {
 (Color online) The hole doping dependence of nodal ($\Delta_{N}$) and antinodal ($\Delta_{AN}$) peaks in the Raman spectra in Fig.\ref{fig:Raman} \cite{LeB10}.
 The upper panel shows the data of Guyard et al \cite{Guyard-PRL-08} as solid points and the theoretical results of LeBlanc et al. \cite{LeB10}) as open points. Ratio values are in the lower panel.
   ($\Delta_{sc_{eff}}$ is the SC gap at the tip of hole pockets in the YRZ model.) }
\label{fig:Raman-2}
\end{figure}

 The results of LeBlanc et al. \cite{LeB10} for both symmetries are summarized in Fig.\ref{fig:Raman}. The B$_{2g}$ spectrum is dominated by a peak whose energy is determined by the superconducting gap at the ends of the Fermi pockets of the YRZ propagator.  This contrasts with the B$_{1g}$ spectrum in the right panel where the peak is determined by the antinodal gap (note the energy scale is much larger in this diagram). In Fig.\ref{fig:Raman-2} the doping dependence of the peaks in the nodal B$_{1g}$ ($\Delta_{AN} $) and antinodal B$_{2g}$ ($\Delta_{N}$) symmetries in theory and in the experiments by Guyard et al.\cite{Guyard-PRL-08} is compared. The ratio of the two peaks is constant in the overdoping region but changes rapidly in the underdoped region. This behavior is reproduced well by the YRZ ansatz. The crossover between the two regions shows some deviations, which can be traced to the small enhancement of the nodal peak, $\Delta_{N}$ near optimal doping in the upper panel. Also shown is the effective superconducting gap calculated at the end of the pockets where the value on the pocket is largest.

 The main qualitative features of the Raman spectra observed in the single layer tetragonal cuprates, HgBa$_{2}$CuO$_{4+\delta}$ are well reproduced by the YRZ ansatz. However it would be premature to conclude that all details of the ansatz have been verified, e.g. the precise evolution of the Fermi surface through the QCP separating the overdoped and underdoped regions of the phase diagram.

\subsection{Andreev Tunneling Spectra in the Pseudogap Phase}
   A decade ago Deutscher \cite{De99} pointed out the contrast between the Andreev tunneling in the overdoped and underdoped regions of the phase diagram. The Andreev tunneling involves injecting a Cooper pair into a superconductor at low voltages and occurs in transparent junctions with a normal metal. In a standard BCS \textit{s}-wave superconductor the tunneling current is enhanced at voltages below the gap \cite{BO82}. In the opposite limit of a high tunneling barrier the Andreev process is suppressed and the tunneling spectrum is dominated by single particle processes which give the Giaever spectra proportional to the single particle DOS of the superconductor. The cutoff of the low voltage Andreev signal is equal to the gap in the Giaever spectrum. Deutscher noted that in the case of tunneling into cuprates with the barrier perpendicular to the antinodal direction, the Andreev cutoff and the Giaever gap coincide in overdoped samples. (This choice of tunneling orientation avoids the complication of Andreev bound states which arise when regions of $\mathbf{k}$-space with different signs of the pairing order parameter are sampled; see Kashiwaya and Tanaka \cite{Ka00}).  But in underdoped samples Deutscher pointed out that the Andreev cutoff was at a lower voltage than the peak in the Giaever spectrum and the discrepancy increased with decreasing doping. In a subsequent review \cite{De05}, he concluded that this discrepancy tilted the balance against the preformed pair scenario as an explanation for the pseudogap in the normal phase of underdoped cuprates.

\begin{figure}[t]
\centerline
{
\includegraphics[width = 6cm, height =10.0cm, angle= 270]
{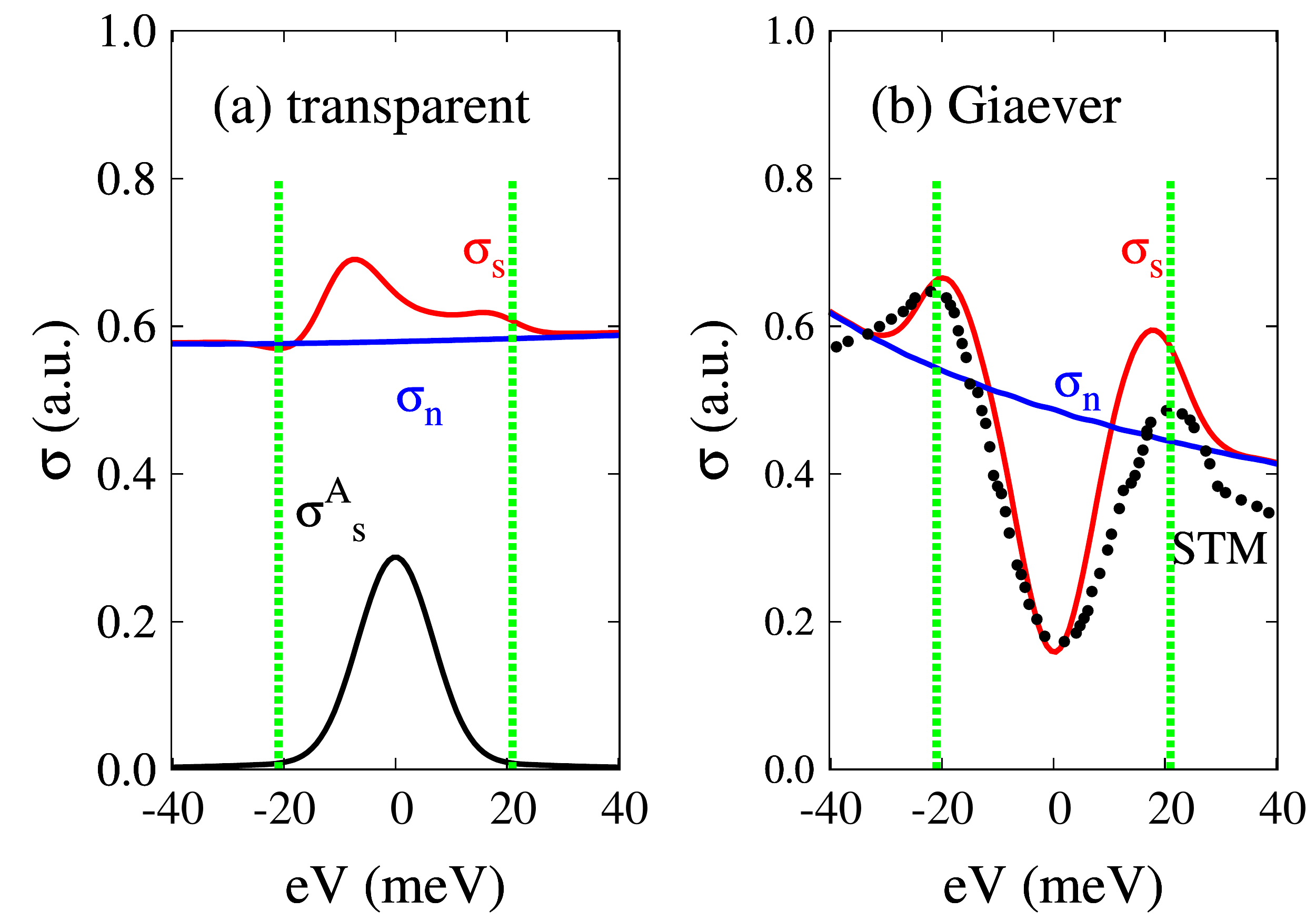}
}
 \caption[]
{(Color online) 
Conductance of an OD cuprate ($x = 0.25$) in the transparent
(a) and in the high barrier tunneling limits (b) \cite{KYY10}. Both limits have the same characteristic gap magnitude.
$\sigma_{s}$ ($\sigma_{n}$) for NS (NN) junctions respectively.
$\sigma_s^{A}$ is the contribution to $\sigma_{s}$ from Andreev reflection. Experimental STM data are from Ref.\cite{Yazdani-science-08}.
}
 \label{fig:OD}
\end{figure}
\begin{figure}[t]
\centerline {
\includegraphics[width = 6cm, height =10.0cm, angle= 270]
{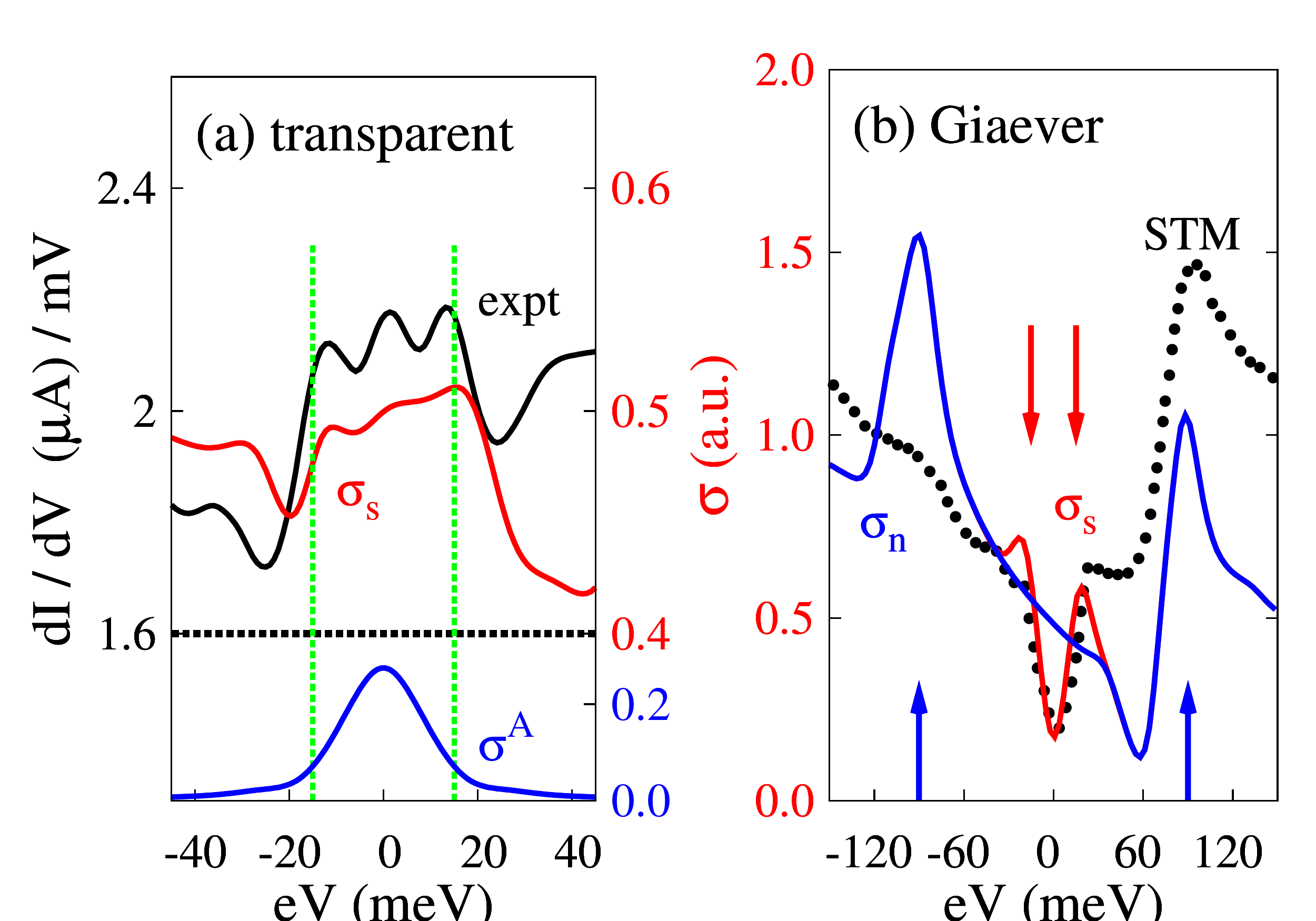} }
 \caption[]
{
(Color online) Conductance of an UD cuprate ($x = 0.1$) in the transparent
(a) and in the high barrier tunneling limits (b) \cite{KYY10}.
Two different gaps are unmasked in the two limits. Note that for single particle tunneling, the peak at negative energy has partial contribution from the van Hove singularity. Experimental STM data are from Ref.\cite{KOH08}.
}  \label{fig:UD}
\end{figure}

  Recently Yang \textit{et al} \cite{KYY10} examined this issue within the YRZ ansatz. For overdoped samples they treated the superconducting state as \textit{d}-wave BCS state over the full Fermi surface. In the underdoped region they used the generalization of the YRZ to the superconducting state described earlier in Sect. 4.5. It is most convenient to treat this problem using the Keldysh formulation of Andreev reflection developed by Sun \textit{et al} \cite{SU99} and Wang \textit{et al} \cite{Wa09}.  Consider a tunneling process along y-direction, then the conductance at zero temperature can be expressed as
\begin{eqnarray}
 \sigma_{s}(eV) = \int_{-\pi}^{\pi} \frac{d k_{x}} {2\pi} e^{2}\sum_{s =
\pm}\big[ T_{N}^{ss}(k_{x}, seV)
 + 2 T_{A}(k_{x}, seV) \big].
 \label{eq:conductance}
\end{eqnarray}
where  $T^{ss}_{N}$  and $T_{A}$  are the diagonal components in Nambu notation for single particle tunneling matrix and the Andreev tunneling coefficient respectively. $s=\pm$ corresponding to the electron (hole) channel. The contribution to $\sigma_{s}$ from $T^{ss}_{N}$ traces the behavior of density of states. For the overdoped region typical results are shown in Fig.\ref{fig:OD}. The energy scales in the transparent limit and in the high barrier limit agree. In panel (b) the DOS is compared to STM results on overdoped BSCCO samples with the magnitude of the superconducting gap on the full Fermi surface adjusted to agree with experiment. In the underdoped case shown in Fig.\ref{fig:UD} there is a marked discrepancy between the energy scale of the cutoff of the Andreev signal and the peak in the DOS. The former is determined by the maximum value of the pairing gap on the Fermi pockets while the latter is controlled by the antinodal RVB gap in the YRZ propagator. Note the single particle gap here has contributions from both the RVB and pairing gaps but the former dominates and acts to suppress Andreev processes.
   Such a suppression was discussed by Deutscher in his review \cite{De05} for the case of tunnelling into a semiconductor with an attractive interaction strong enough to create a weak pairing amplitude but not enough to close the charge gap. This discrepancy poses difficulties for the preformed pair scenario which assumes the antinodal pseudogap is a pairing gap which persists for $T>T_{c}$ in a normal state with strong phase fluctuations. However at low $T$ in the presence of superconducting order it is not clear why the Andreev spectra at over- and underdoping should be so different. On the other hand it is straightforwardly explained in scenarios such as YRZ which describe the antinodal pseudogap as a precursor to the Mott insulating state at stoichiometry. Finally we note that the theoretical and experimental DOS in the high barrier limit at underdoping (see in Fig.\ref{fig:UD}) show similar features but also significant differences.

\section{Relation to Microscopic Theories of Superconductivity and Magnetism }
   Up till now we have introduced superconductivity into the phenomenological YRZ ansatz by simply adding a \textit{d}-wave pairing term but without discussing its origin. The YRZ ansatz was motivated by the earlier FRG calculations on a moderately coupled 2-dimensional Hubbard model and by the analysis by Konik et al.  of an array of 2-leg weakly coupled Hubbard ladders near half-filling with a specially chosen inter-ladder hopping \cite{KRT06}. In this section we will discuss the implications of the model for a microscopic description of the pairing and magnetic properties.

\subsection{\textit{d}-wave Pairing Induced by AF Fluctuations in the Presence of  the Pseudogap}
Recently Schachinger and Carbotte \cite{SC10} examined the effect of the pseudogap on the superconducting state when a pairing interaction mediated by AF spin fluctuations with the form proposed by Millis, Monien and Pines \cite{MMP90} is introduced. They did a careful analysis of the resulting gap equation to determine the interplay between the two gaps --- the RVB gap in the YRZ propagator and the SC gap. They chose the coupling constants to give a parabolic form for $T_{c}(x)$, typical of many cuprate superconductors with a maximum in $T_{c}(x)$ at $x_{c}$ --- the QCP density.   Several interesting conclusions were obtained. The momentum dependence of the SC gap showed substantial deviations from the simple \textit{d}-wave form at hole densities near optimal. At UD the behavior is sensitive to the choice of the magnetic coherence length $\xi_{M}$, which determines the width of the AF spin fluctuation peak. A short $\xi_{M}$ (=2.5a) gives a simple \textit{d}-wave form for the SC gap but a longer value of $\xi_{M}$ (=10a) (i.e. sharply peaked AF spin fluctuations) gives rise to a marked change with a non-monotonic SC gap as $\mathbf{k}$ approaches the antinodal direction. In general though, the presence of a strong RVB gap near antinodal does not act to suppress the SC gap much. Another feature of their results is the increased strength of the overall coupling constant in the UD region that is necessary to reproduce the parabolic form for $T_{c}(x)$ as a result of the smaller size of the Fermi pockets. An enhanced value of the ratio $\Delta_{SC} /T_{c}$ follows at UD.

\subsection{The Nozieres-Pistolesi Model for a Semiconductor --- Superconductor Transition}
\begin{figure}[t]
\centerline {
\includegraphics[width = 8cm, height =6.0cm, angle= 0]
{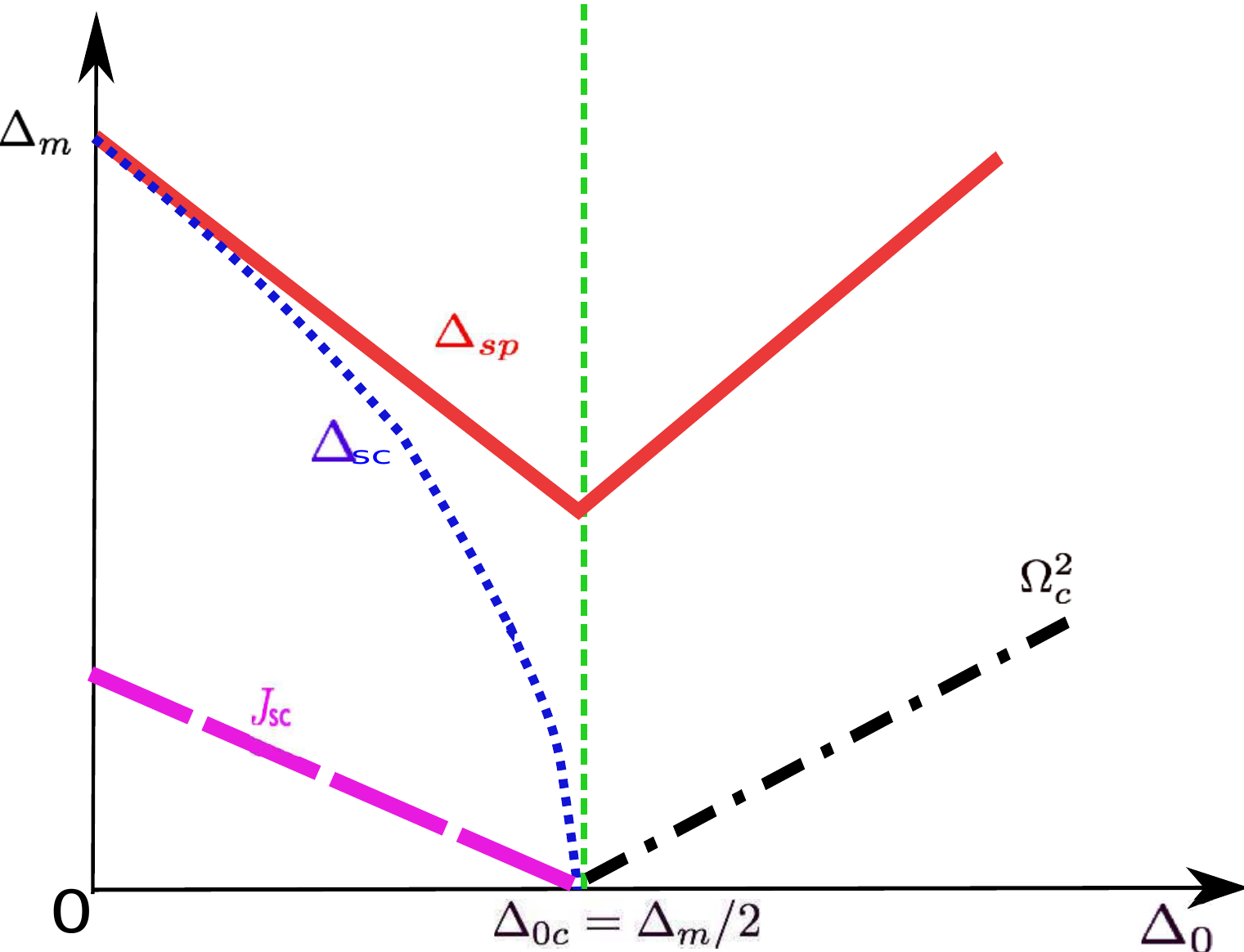} }
 \caption[]
{
(Color online) The Nozieres-Pistolesi model: the single particle gap 
$\Delta_{sp}(= \sqrt{\Delta_{0}^{2} + \Delta^{2}_{sc}}$), 
the superconducting parameter $\Delta_{sc}$, the phase stiffness $J_{sc}$, and the energy of the Cooperon excitation $\Omega_{c}$ at $T=0$. (Parts of the plot reproduced from Ref.\cite{NP99})
} 
 \label{fig:SM-SC}
\end{figure}

  To begin the discussion of the consequences of opening up an energy gap in the antinodal region as the hole density moves through the QCP into the pseudogap phase, it is worthwhile to recall a simple model for the pseudogap phase put forward some years ago by Nozieres and Pistolesi \cite{NP99}. They considered a 2-dimensional semiconductor with a local attraction between the electrons and examined the direct transition between a semiconductor and a superconductor as the size of the semiconducting gap is reduced. In a mean field description the zero temperature BCS gap equation for this model reduces to the form
\begin{eqnarray}
1& =& \rho V \log \big[ \frac{2\omega_{m}} {\Delta_{0} + \sqrt{\Delta_{0}^{2} + \Delta^{2}_{sc}} } \Big]
\label{eq:gap-NP}
\end{eqnarray}
where V is the attraction, $\rho$ is the constant density of states per spin,  $\Delta_{0}$ the semiconducting gap and $\omega_{m}$ is the energy cutoff. The pairing gap $\Delta_{sc}$ is
\begin{eqnarray}
\Delta_{sc} &=& \sqrt{\Delta_{m}(\Delta_{m} - 2\Delta_{0})}
\label{eq:pair-gap-NP}
\end{eqnarray}
where $\Delta_{m}$ is the gap in a metal ($\Delta_{0}=0$) with the same parameters.  Note in this model as illustrated in Fig.\ref{fig:SM-SC} the single particle gap is always finite. The pairing gap is finite only for $\Delta_{0} < \Delta_{m}/2$ where the ground state is superconducting \cite{NP99}.

   As the QCP is approached from the superconducting side Nozieres and Pistolesi found that the phase stiffness $J_{sc}$, vanishes continuously as, $J_{sc} = (\Delta_{m}- 2\Delta_{0}) /{4\pi}$.
On the semiconducting side the pairing vertex takes the form 
\begin{eqnarray}
\Gamma(\mathbf{q},\omega) &=& \frac{V}{1-V\gamma^{0}(\mathbf{q},\omega)} \nonumber \\
\gamma^{0}(\mathbf{q},\omega) &=& \frac{m}{4\pi} \log \left[ \frac{\omega_{m}^{2} } {\Delta_{0}^{2}-(\omega/2)^{2}} \right]
+ \frac{\mathbf{q}^{2}}{16\pi} \left[ \frac{\Delta_{0}}{(\omega/2)^{2} - \Delta_{0}^{2}}  \right]
\end{eqnarray}
by assuming that the semiconductor consists of a conduction and valence band with the dispersion $\epsilon(\mathbf(k)) = \pm (\Delta_{0} + k^{2}/2m)$.
The pairing vertex $\Gamma(\mathbf{q},\omega)$ has poles on the real axis when $\Delta_{0} > \Delta_{m}/2$ at an energy ($\omega = \Omega_{c} \ll \Delta_{0}$)
\begin{eqnarray}
\Omega_{c}^{2} = 4 (\Delta_{0}^{2} -\Delta_{0c}^{2}) + \frac{\Delta_{0c}^{2}}{ \Delta_{0} m} \mathbf{q}^{2}
\end{eqnarray}
These poles describe a bound pair or Cooperon excitation lying below the energy gap in the semiconducting phase. As the QCP is approached, the transition from the superconductor occurs through vanishing phase stiffness. The semiconductor on the other side of the QCP has a well-defined Cooperon excitation which goes soft as the QCP is approached.

\subsection{Do Cooperons Play a Role in the Cuprates?}
   The cuprates are clearly more complex than the Nozieres-Pistolesi model with a transition from a \textit{d}-wave superconductor to an AF Mott insulator not at a fixed el/atom  ratio but driven by doping. The semiconducting phase in the Nozieres-Pistolesi model is characterized by a soft Cooperon excitation. Recently Konik, Rice and Tsvelik \cite{KR10} proposed an important role for a Cooperon excitation in the pseudogap phase of the cuprates.
    As we noted above, the early FRG studies on a 2-dimension Hubbard model found that the crossover from overdoping to underdoping was characterized by the growth of a divergence in the AF channel as the hole density decreases in addition to a diverging \textit{d}-wave Cooper channel present at overdoping. This combination of two diverging channels with similar critical energy scales in the FRG flow raised the question as to the nature of the strong coupling phase that is stabilized at lower scales. If just a single channel diverges, the flow can be interpreted as a precursor to long range order in that channel.  Honerkamp \textit{et al} \cite{CH01} noted the similarity of this diverging flow in the AF and \textit{d}-wave pairing channels to the flow in the half-filled 2-leg Hubbard ladder where a similar multichannel flow pattern is seen. In this case the ground state is known from bosonization and it is an insulator with finite energy gaps in the single particle, charge and spin sectors. The response is enhanced in the AF and ``\textit{d}-wave'' pairing channels but the correlations in the ground state in these channels remain strictly short range. This led Honerkamp and coworkers \cite{CH01,CH02} to conjecture that at the QCP marking the transition to underdoping, a gapped region with similar characteristics could appear near the antinodal points in the square lattice.

 To test this conjecture exact diagonalizations of a reduced Hamiltonian retaining only the leading interactions in the FRG flow, were carried out by Laeuchli and collaborators \cite{Laeuchli}. To make the problem tractable the scattering processes, both normal and umklapp, were discretized to those connecting a finite grid of $\mathbf{k}$-points chosen in the antinodal regions. The low energy excitation spectrum was found to exhibit energy gaps in the single particle, charge and spin sectors similar to the 2-leg ladder.  Further, the pattern of the enhanced response functions in the ground state was found to be the same as that of the half-filled 2-leg Hubbard ladder.  In short these calculations offer support to the conjecture of Honerkamp and colleagues \cite{CH01,CH02} that a 2-dim. analog to the ladder forms initially in the antinodal region as a precursor to the Mott insulting state at  half-filling.  Laeuchli and collaborators \cite{Laeuchli} further suggested that in this case although the antinodal gap is insulating, it will be connected to the remnant Fermi surface in the nodal region through pair scattering processes in the Cooper \textit{d}-wave channel. Such processes can stabilize superconductivity primarily on the remnant Fermi pockets because of the enhanced ``\textit{d}-wave'' pairing response in the latter regions.

  Recently Konik, Rice and Tsvelik \cite{KR10} went further. They examined an array of coupled 4-leg Hubbard ladders near half-filling. As discussed earlier in wider ladders the bands pair up on a hierarchy of energy scales with the pairing of the outermost and innermost bands having the largest gap etc \cite{LeH09,LE00,CA07}. Now this band pair exhibits enhanced responses similar to the half-filled 2-leg Hubbard ladder. These are accompanied by the appearance of two collective excitations in the energy gap, a Cooperon pair and a triplet magnon excitation. Konik \textit{et al} \cite{KR10} proposed that in the 4-leg ladder array inter-ladder hopping at low doping can lead to Fermi pockets from the two central bands with the smaller energy gap, similar to the case of the array of 2-leg Hubbard ladders these authors considered earlier \cite{KRT06}. For a suitable parameter choice, the band pair with the larger energy gap remains half-filled  and insulating for a finite density range but with a Cooperon collective excitation.  Under these conditions, a pairing attraction is generated on the Fermi pockets with the central band pair through coupling in the Cooper channel to the Cooperon of the insulating band pair. Although this is clearly an artificial model, Konik \textit{et al} \cite{KR10}  proposed it as a tractable model which contains the essential features of the cuprates, namely \textit{d}-wave pairing on remnant Fermi pockets generated by coupling to the Cooperon excitation living on the remaining gapped bands.

   However in the antinodal region the YRZ ansatz assumes that the single particle gap, which is due to \textit{d}-wave pairing in the overdoped region, transforms into an insulating gap at the QCP. In the underdoped region the analogy was drawn to the 2-leg Hubbard ladder at half-filling, which shows a charge gap with a Cooperon lying below the single particle gap. The simplest scenario then to connect the two doping regions, is to assume that the QCP is analogous to that in the Nozieres-Pistolesi QCP associated with the emergence of a finite energy Cooperon. In the present case this could occur through the increasing strength of the umklapp scattering between the antinodal regions, rather than an increase in an underlying band gap. In this scenario the pseudogap phase is characterized by a reduced Fermi pocket centred on the nodal directions coexisting with a finite energy Cooperon excitation. Note, such a Cooperon can decay into 2-particle excitations of the Fermi pockets leading to  a Cooperon resonance that appears as the pseudogap suppresses the low energy DOS. The close similarity to the 4-leg Hubbard ladder array analyzed by Konik \textit{et al.} is clear.

\begin{figure}[t]
\centerline {
\includegraphics[width = 8cm, height =6.0cm, angle= 0]
{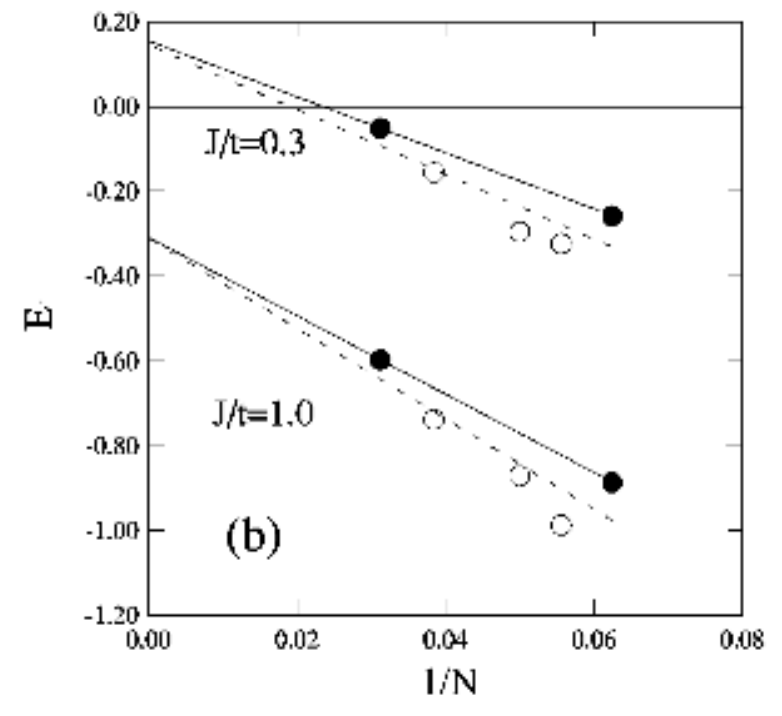} }
 \caption[]
{(Color online) The binding energy of the two particle bound state vs inverse cluster size for two representative J/t values, J/t=0.3 (upper) and J/t=1.0 (lower). Solid points denote square symmetric clusters, open other clusters \cite{Chernyshev-PRB-98}.}  \label{fig:ED_TJ}
\end{figure}

  So far the models and their interpretation have been restricted to weak to moderate coupling. But the Mott insulating state at zero doping shows that the onsite Coulomb repulsion is strong in underdoped cuprates. Recently, Konik \textit{et al} \cite{KR10} proposed that a Cooperon exists also at strong onsite repulsion. They based their conclusion on the results of  exact diagonalization studies of small clusters. The only limitation in these calculations is the finite cluster size which currently is limited to small clusters containing up to 32 sites and 1, 2 and 4 holes. Leung and his collaborators \cite{Chernyshev-PRB-98, Leung-PRB-02, Leung-PRB-06} have reported a series of calculations for these clusters using the strong coupling $t-J$ model, which describes the Hubbard model near half-filling and its extensions to include longer range hopping and interactions. The main conclusions of these cluster calculations are as follows. The allowed set of $\mathbf{k}$-points in a 32-site cluster with periodic boundary conditions contain both the 4 nodal ($\pi/2, \pi/2$) and 2 antinodal points ($\pi$, 0) and (0, $\pi$).  A single hole enters at a nodal point. For two holes there are two different states that are possible ground states, depending on the parameter values. For the plain $t-J$ model with only nearest neighbor hopping, a two-hole bound pair state with $d_{x^{2}-y^{2}}$ symmetry is the ground state on the 32-site cluster for $J/t > 0.28$, (see Fig.\ref{fig:ED_TJ}).
The binding energy is quite small at J/t = 0.3 but grows with increasing J/t. An extrapolation from finite size clusters to the infinite lattice however suggests that the pair state is no longer the ground state at J/t =0.3, but an excited state with an energy of approximately $0.17t$ \cite{Chernyshev-PRB-98}. The inclusion of longer range interactions and hopping in the $t-J$ model increases the energy of the pair state further and confirms the conclusion that for parameter values relevant to cuprates the ground state of the cluster has two unbound holes in the nodal states \cite{Leung-PRB-02}. Extending the calculations to the 32-site clusters with 4 holes, which corresponds to a doping of 1/8, shows all 4 holes entering into nodal states with no signs of pairing correlations \cite{Leung-PRB-06}. In view of the prominent bound pair excited state for two holes, a low energy excited state with two of the holes in a bound state may also be expected here. However at present there is no information on this question to the best of our knowledge. Leung and collaborators \cite{Chernyshev-PRB-98, Leung-PRB-02, Leung-PRB-06} concluded from these calculations that at low densities holes entered the nodal regions, consistent with nodal Fermi pockets. The low energy \textit{d}-wave pair excited state is interpreted as a finite energy Cooperon in the strong coupling  $t-J$ model and its extensions. Note, an earlier study of two holes on smaller clusters by Poilblanc and collaborators \cite{Poilblanc-PRB-94} concluded in favor of the interpretation of the two hole bound state as a quasiparticle with charge 2e and spin 0, which would be an actual carrier of charge under an applied electric field. In other words they concluded that a Cooperon is present in the strong coupling $t-J$ model at low doping. A more detailed analysis of the origin of the pairing in this Cooperon state was published recently by Maier et al \cite{Maier-PRL-08}. 

  In the calculations on 32-site clusters the doped holes enter into the nodal ($\pi/2, \pi/2$) points but the coupling to the \textit{d}-wave Cooperon cancels completely exactly at the nodal directions. If we consider what will happen if the cluster size is expanded keeping a small, but finite, hole density, then finite Fermi pockets centered on, but extending away from, the nodal directions will appear. These  will contain $\mathbf{k}$-states which do couple to the \textit{d}-wave Cooperon resonance. The coupling to the Cooperon will be weak since the Fermi surface of the pockets extends only a small distance away from the nodal directions, but it will lead to a finite, but small, \textit{d}-wave pairing order parameter. Recent numerical calculations using a  novel fidelity approach also found evidence for such a small but finite pairing order parameter at low doping \cite{Rigol-PRB-09}.

  Note also the hole density in the case of two holes in a 32-site cluster is very low so that the superconducting order we are  postulating should coexist with long range antiferromagnetic order. There is considerable evidence both numerical, in variational Monte Carlo calculations, and experimental, in favor of such coexistence, as discussed in the recent review by Ogata and Fukuyama \cite{OgFu08}. We conclude that there is strong evidence that the pairing mechanism through coupling to finite energy Cooperon resonances applies also at strong onsite repulsion.

  Phenomenological models based on coupled fermions and bosons similar to that derived here, have been proposed much earlier \cite{Friedberg-PRB-89, Ranninger-PRL-95} to describe the high temperature superconductors. The closest similarity are to the models proposed by Geshkenbein, Ioffe and Larkin \cite{Geshkenbein-PRB-97} and by Chubukov and Tsvelik \cite{Chubukov-PRL-07, Chubukov-PRB-07}. Both these phenomenological models examined Fermi arcs centered on the nodal directions, coupled in the \textit{d}-wave channel to Cooperons associated with the antinodal regions but were not elaborated to describe the full complexity of the cuprate phase diagram.

\subsection{Magnetic Excitation Spectrum }
  We turn now to the interpretation of the magnetic excitation spectrum that is observed in the pseudogap phase. A series of neutron scattering experiments on several cuprates have found an unusual spectrum which in keeping with the enhanced AF response is centred at the wavevector ($\pi,\pi$). It has been named as a hourglass spectrum consisting of  two components: a triplet mode centred at ($\pi, \pi$) which disperses upward symmetrically around ($\pi,\pi$), and legs which stretch down to zero energy at a ring of wavevectors displaced slightly away from the wavevector ($\pi, \pi$) to an incommensurate value \cite{Tranquada-04, Hayden-nature-04, Stock-PRB-04, Fauque-PRB-07}. There has been much discussion in the literature on the interpretation of this unusual spectrum. However, we will not attempt a full review here of this literature, which is covered in the reviews by Timusk and Statt \cite{TS99} and by Ogata and Fukuyama \cite{OgFu08}. Instead, we limit the discussion to the question of whether this hourglass spectrum observed at low temperatures, is compatible with our expectations based on the analogy we have been using between the pseudogap phase and the half-filled 2-leg Hubbard ladder. A recent review by Barzykin and Pines discusses the magnetic properties at higher temperature \cite{Barzykin-AdvPhys-09}.
  \begin{figure}[t]
\centerline {
\includegraphics[width = 8cm, height =6.0cm, angle= 0]
{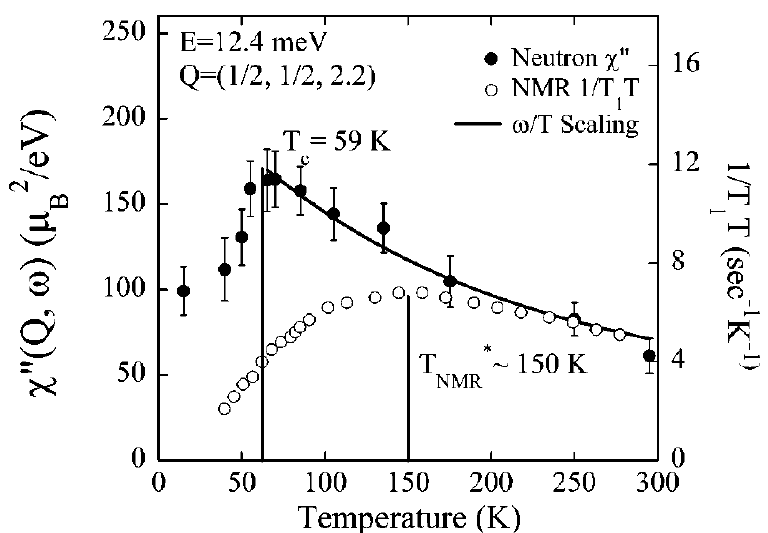} }
 \caption[]
{(Color online) The spectral weight at an energy of 12.4 meV,below the resonance energy, is suppressed at temperatures below superconducting $T_{c}$ \cite{Stock-PRB-04}.}  \label{fig:Stock1}
\end{figure}
\begin{figure}[t]
\centerline {
\includegraphics[width = 8cm, height =6.0cm, angle= 0]
{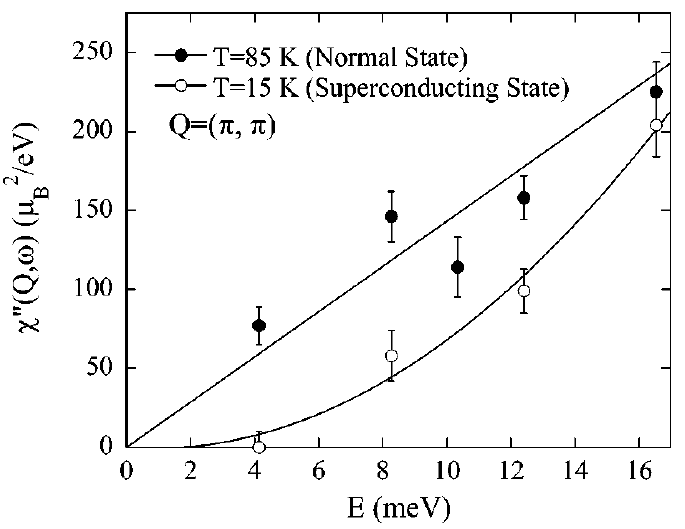} }
 \caption[]
{(Color online) Comparison of the low energy spectral weight in the normal and superconducting states, showing the strong suppression below linear behavior in the latter \cite{Stock-PRB-04}.}  \label{fig:Stock2}
\end{figure}

\begin{figure}[t]
\centerline {
\includegraphics[width = 12cm, height =8.0cm, angle= 0]
{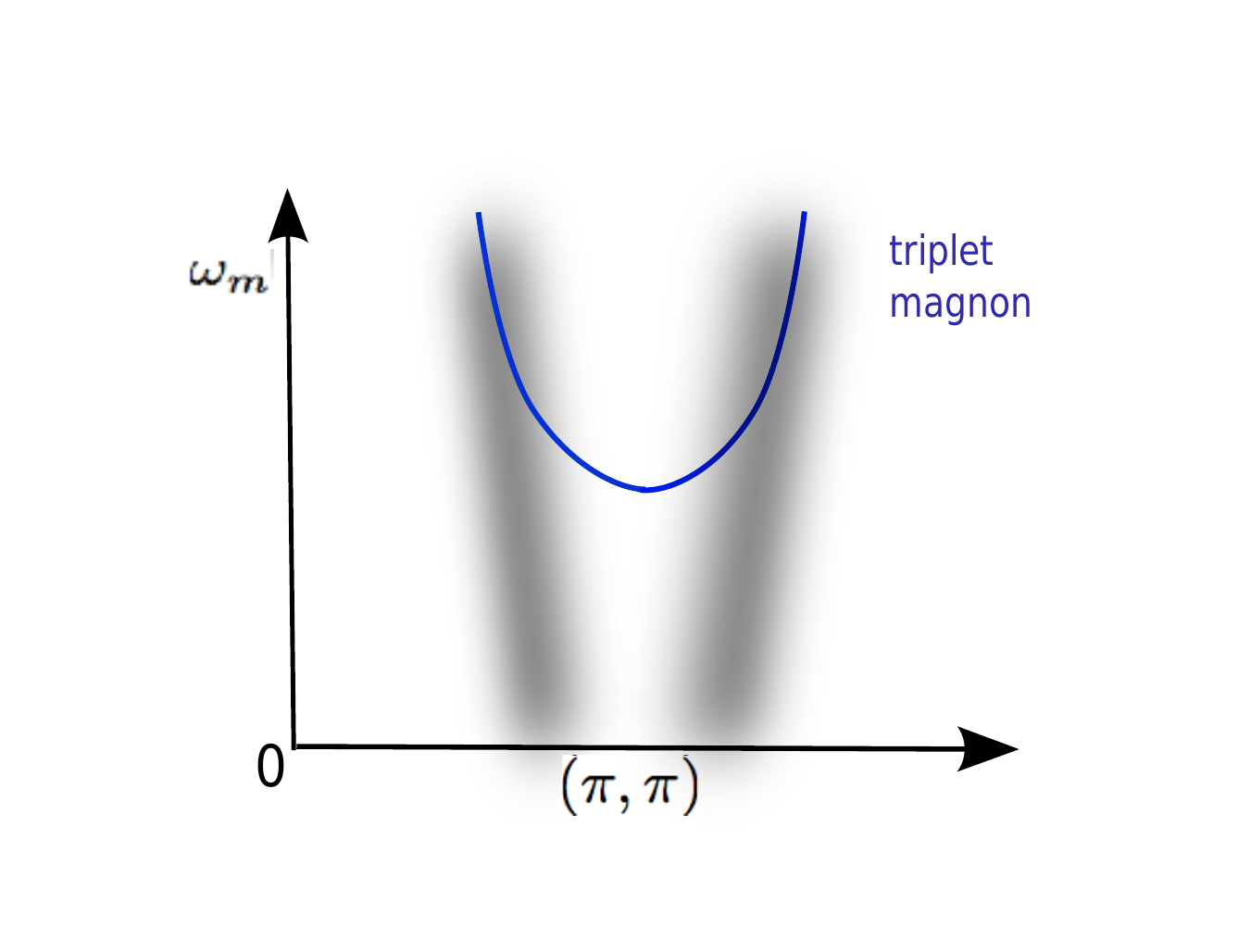} }
 \caption[]
{(Color online) Schematic demonstration of the hourglass spectrum. The upper branch is postulated to come  from a triplet magnon (blue curve), and the lower legs (shaded gray) from quasiparticle excitations between nodal arcs whose intensity is enhanced by the high magnetic polarizability near  ($\pi, \pi$). }  \label{fig:neutron-scattering}
\end{figure}

  This analogy also covers  enhanced AF response functions centered  on ($\pi, \pi$) \cite{Laeuchli}. This enhanced response in the half-filled 2-leg Hubbard ladder is a consequence of a finite energy triplet magnon and so it is reasonable to assume that a similar excitation should exist in the pseudogap phase. In the overdoped and optimally doped region the triplet magnon at  ($\pi, \pi$)  appears only in the superconducting state at temperatures $T<T_{c}$. A theoretical description using a RPA works well, as discussed in the review by Ogata and Fukuyama \cite{OgFu08}. This description can also be extended into the underdoped region as a collective mode appearing inside the reduced particle-hole DOS associated with the  pseudogap. The transition to long range AF order occurs as the energy of this triplet magnon collapses when the hole density is reduced. The RPA calculations also include the low energy excitation branch which arises from particle-hole transitions between opposite Fermi arcs. This interpretation of the low energy part of the spectrum as particle-hole excitations is confirmed by the neutron scattering experiments performed on YBa$_{2}$Cu$_{3}$O$_{6.5}$-OrthoII samples. This cuprate is among the best ordered underdoped cuprates and has the advantage that large single crystals are available. The experiments of Stock et al \cite{Stock-PRB-04} show a clear suppression of the low energy spectrum as $T$ drops below $T_{c}$ consistent with a suppression of the low energy  particle-hole excitations as the superconducting gap opens up on the nodal Fermi arcs-- see Figs.\ref{fig:Stock1} and \ref{fig:Stock2}. A sketch of the combined spectrum is shown in Fig.\ref{fig:neutron-scattering}.
  
   While the general features of the interpretation of the magnetic excitation spectra in the pseudogap phase fits within the theoretical framework, there are a number of features which remain to be clarified. In particular, the origin of
 broken translational and rotational symmetries in some cuprates at low temperatures covered in Vojta's recent review \cite{Vo09}, is an important issue. A careful study showing  anisotropic behavior in the low energy spectrum and a nematic transition has been reported recently by Hinkov et al \cite{Hinkov-science-08} in underdoped YBa$_{2}$Cu$_{3}$O$_{6+x}$ samples.
Bascones and Rice \cite{Bascones-PRB-06} a few years ago pointed out that the low energy branch is composed of  quasiparticle excitations in the presence of a highly polarizable background coming from the triplet magnon centred at  ($\pi, \pi$). They examined the incommensurability reported in YBa$_{2}$Cu$_{3}$O$_{6.5}$ OrthoII samples but with limited success. This question of the stability and incommensurability of the low lying particle-hole excitation branch in the presence of a highly polarizable background requires theoretical study.

  Another area with open questions is the role of impurities and disorder. A comprehensive review of this topic was recently published by Alloul and coworkers \cite{Al09} and we will limit the discussion here to some points relevant to this review. The variational Monte Carlo results reviewed by Ogata and Fukuyama \cite{OgFu08} on Gutzwiller projected wavefunctions with coexisting \textit{d}-wave pairing and long range AF order show a wide range of coexistence of both orders up to hole concentrations of $x= 0.1$. Also Kitaoka and collaborators \cite{Mukuda-PhysicaC-09} have interpreted their NMR experiments on multilayer Hg-cuprates as evidence for a similar wide coexistence region in the phase diagrams. While the inner layers in these multilayer cuprates are very clean, surely the cleanest of all strongly underdoped cuprates, it is not obvious whether it is the perfection of these layers or other effects, e.g. the combination of interlayer superexchange and a reduced hole density in the inner layers, that is playing the decisive role in stabilizing the AF long range order. The comparison to the cases of low doping in single layer La$_{2}$CuO$_{4}$ and bilayer YBa$_{2}$Cu$_{3}$O$_{6 }$ suggests that a region of incommensurate AF order occurs in a single layer. Sushkov \cite{Su09} has shown how such incommensurate AF order arises as carriers are introduced into a 2-dimensional AF ordered ground state, explaining the difference in the incommensurate order in a single layer with substantial disorder (La$_{2}$CuO$_{4}$) and a bilayer with interlayer superexchange and weak disorder (YBa$_{2}$Cu$_{3}$O$_{6 }$).

  One interesting form of disorder which has a strong effect is Zn substitution for Cu in the CuO$_{2}$ planes. In the undoped limit Zn substitution simply removes sites in the Heisenberg model \cite{Va02}. There is only a weak effect on long range AF order which is suppressed only when the Zn concentration reaches the percolation value. The situation changes completely when holes are introduced. In the underdoped region both NMR and neutron scattering experiments show additional very low energy magnetic fluctuations consistent with introduction of local moments although Zn$^{2+}$ is a closed shell ion \cite{Mahajan-PRL-04, Kakurai-PRB-93,Su-PRL10}. This behavior is similar to the effects of Zn doping in SrCu$_{2}$O$_{3}$ --- a 2-leg ladder cuprate \cite{Azuma-PRB-97}. In this case each Zn impurity adds a $S = 1/2$ moment and these Zn-induced moments show AF order already at very low Zn concentrations. This is easily understood intuitively for the 2-leg ladder as the breaking of a singlet bond on a rung leading to a free Cu$^{2+}$ moment. The observation of similar behavior in underdoped cuprates supports the close analogy between the pseudogap phase and the 2-leg ladders. A series of `a priori' numerical calculations reproduced the local moment character of Zn impurities in underdoped cuprates and interpreted their results in a similar manner \cite{Ri96,Ma97,Ma02,Pe04}.

\section{Relation to Other  Approaches}
    In this section we comment briefly on the relation of the YRZ  ansatz to other approaches to describe the anomalous properties of the pseudogap phase. A clear distinction is the starting point. Most  theories start from the undoped Mott insulating state and consider the consequences of introducing holes into a state where a strong onsite Coulomb repulsion has localized an electron on each site. Such a strong onsite repulsion is difficult to handle directly. The introduction of holes into the undoped Mott insulator is an essential perturbation which changes the low energy degrees of freedom from purely spin degrees of freedom to include fermionic degrees as well.
    The most popular approach to this problem has been to introduce a gauge field into the Hamiltonian to enforce the constraint of no double occupancy. Several reviews on these approaches have appeared recently where more details can be found \cite{LNW06,PAL08}. A factorization of fermions into spinons and holons is introduced with a local gauge condition on the spinon and holon operators which acts to enforce the onsite constraint. Typically in the subsequent analysis, it is too difficult to satisfy the gauge condition locally, and the condition is satisfied only on the average. This factorization leads to an expansion the Hilbert space with extra degrees of freedom. The analysis that gives results closest to the YRZ approach is the local SU(2) gauge theory introduced by Wen and Lee \cite{Wen-Lee}, later extended by Ng \cite{TK-Ng}. A key element is the postulate of an attraction between spinons and holons leading to a fermionic bound state of a holon and spinon which then carries both spin and charge. This leads to a single electron propagator that, as discussed earlier, has a form with close similarities to the YRZ form. 

Recently Patrick Lee \cite{PL11} pointed out to us that actually the hopping term, which in a spinon-holon representation is a product of spinon and holon hopping operators, acts as a strong attraction between spinons and holons. Further, a simple derivation of the YRZ form for the propagator of a spinon-holon bound state, follows if one takes the spinon propagator that follows from RVB at half-filling, e.g. as in Zhang et al \cite{RMF}. The existence of such a spinon-holon bound state is at variance with the assumption of spin-charge separation with separate spinons and holons interacting through transverse fluctuations of the gauge field. The existence of a bound state has the advantage that it cures the excess entropy problem associated with separate spinon and holon and transverse gauge field degrees of freedom. Hlubina et al. \cite{Hl92} some years ago compared the entropy under the latter approximation to the entropy of the  t-J model including the onsite constraint, calculated in a high temperature expansion. A substantial excess over the high temperature expansion entropy results, even at moderate temperatures. They found that including the longitudinal gauge field gave rise to strong spinon-holon interactions which reduce the discrepancy. The presence of  spinon-holon binding is a simple way to resolve the excess entropy difficulty. 

   A derivation of the YRZ form starting from the strong coupling t-J model complements the approach from weak coupling and the breakdown of Landau Fermi liquid theory presented here. Further the YRZ ansatz in this case offers a continuous connection between the weakly doped Mott insulator at strong underdoping and the Landau Fermi liquid at strong overdoping. Continuous crossovers between strong and weak coupling are a feature of one dimensional systems e.g. Hubbard chains and ladders. As we have stressed earlier the YRZ ansatz was motivated by analogies to Hubbard ladders. In this sense it may be viewed as a generalization of ladder physics to two dimensions. An important feature of the Hubbard 2-leg ladder is that strong qualitative changes in the low energy properties are brought about by short range, not long range correlations. This points out the need for theoretical methods to handle such qualitative changes in physical properties due to short range correlations.  We believe this is a key challenge for the 
future.

An alternative approach to the gauge theories for the strong coupling $t-J$ model has been pursued in recent years by Anderson and his collaborators and is reviewed in a very recent paper \cite{PWA10}. In this approach the crossover between the overdoped and underdoped regions is attributed to a crossover from a ground state dominated by the kinetic energy of the holes, i.e. the t-term, at overdoping, to one dominated by interactions i.e. the J-term, at underdoping. In the overdoping region the ground state can be written as a Gutzwiller projected Fermi sea with a full Fermi surface. The effect of the constraint on double occupancy imposed by the Gutzwiller projector is to enhance the low energy scattering vertex leading to strong inelastic damping of quasiparticles. The FRG calculations discussed in Sect. 2 show similar behavior appearing at weak to moderate coupling and it is certainly reasonable to expect the such effects are enhanced as the coupling strength increases. The interaction dominated underdoped region is more difficult to analyze.  Anderson \cite{PWA10} argues that the nodal-antinodal dichotomy arises from a difference in the effective Gutzwiller projector which acts as a strong constraint in the antinodal region but only as a weak constraint near the nodes.  The detailed form of the angular dependence is left open but it leads to a truncation of the Fermi surface qualitatively similar to the YRZ ansatz. The holes are concentrated in the nodal region while in the antinodal region the interactions dominate, leading to a self energy induced gap, similar to that in the RMFT of Zhang et al \cite{RMF} for an undoped RVB state.  Hopefully these various approaches built on phenomenological inputs, which give properties in line with the wide range of experiments on the pseudogap phase available today, will guide the way to a fully microscopic solution to this very challenging many body problem.

Another strong coupling approach is based on the FL$^{*}$ --- a concept introduced in the analysis of the Kondo lattice model for heavy fermions. In that model the FL$^{*}$ state is a state with a small Fermi surface of conduction electrons moving in a background of  localized electrons which under appropriate conditions, do not order magnetically due to strong fluctuations. Sachdev and collaborators \cite{QS10,MS10} have recently applied a similar approach to model the lightly doped Mott insulator in underdoped cuprates as a Fermi sea of holes moving in a background with local AF order. Long range AF order is suppressed by strong fluctuations in the direction of the AF moment but not in its magnitude. In the normal state the hole quasiparticles are coupled to fractionalized excitations of the fluctuating AF background to form a fractionalized FL$^{*}$ state. Fermi pockets form which are centered on the nodal directions while the spectrum near antinodal remains gapped. This leads to a nodal-antinodal dichotomy --- the central feature of the pseudogap phase. Note, the FL$^{*}$ pockets are not identical to the YRZ pockets e.g. the anomalous self-energy remains finite at the back end of the pockets in the nodal directions and as a result the back end of the pockets does not touch the umklapp surface there, as they do in the YRZ ansatz. This difference, however may not be easy to detect experimentally. A bigger difference would seem to be in the magnetic properties. The FL$^{*}$ state, at least in its simplest version, has a spin gap with fractional spin excitations above it, which would seem to differ from the hourglass spectrum discussed in Sect. 6.4.  In addition there are gapless collective nonmagnetic excitations. At present, such excitations, which should contribute to the specific heat, have not been reported in experiments.

  Numerical calculations on the planar Hubbard model using cellular and cluster DMFT (Dynamic Mean Field Theory) schemes which give Mott metal insulator transitions, have been making progress on improving the $\mathbf{k}$- space resolution \cite{Civelli-PRL-08,Ferrero-PRB-09,Werner-PRB-09,Imada-PRB-10,Gull10}. A recent series of calculations have found a nodal-antinodal dichotomy with single particle energy gaps in the antinodal but not in the nodal $\mathbf{k}$--space regions, at small hole dopings into a Mott insulating state. This dichotomy has been described as a $\mathbf{k}$--space selective Mott transition in analogy to the orbitally selective Mott transition proposed for certain transition metal compounds \cite{Ferrero-PRB-09}. At present the detailed $\mathbf{k}$--space behavior is not available in most calculations so that one can only say that the numerical results are consistent with the central characteristics of the YRZ ansatz. The calculations by Sakai et al \cite{Imada-PRB-10} obtain a specific  form for the Green's function with Fermi nodal pockets and Luttinger zeroes but these do not lie on the umklapp surface.

\section{Conclusions}
   In this article we have concentrated on our recent theoretical ansatz to describe the so-called pseudogap phase in underdoped cuprates.  The theory is motivated by examining the functional renormalization group flows of the scattering vertex in the Fermi liquid in the optimally to overdoped density range. At these densities both ARPES and quantum oscillation experiments, show that the full Fermi surface is maintained. But the low energy scattering vertex develops strong structure in $\mathbf{k}$--space. The strongest scatterings as discussed above, are those connecting the antinodal regions in $\mathbf{k}$--space. This $\mathbf{k}$--space structure shows up in angular dependent magnetoresistance experiments which measure the anisotropic scattering rates for quasiparticles in the normal state. The leading instability in this density region is towards \textit{d}-wave pairing. However the growing strength of umklapp scattering processes as the hole density is reduced leads eventually to a breakdown of the full Fermi surface. The ansatz for the single particle propagator was proposed to describe this breakdown. The key ingredient is a self-energy whose form is chosen by analogy with the form in half-filled 2-leg Hubbard ladders where similar umklapp processes cause a charge gap, even at weak coupling. The close similarity between the FRG flows in the simplified 2-patch model where 2-dimensional Fermi surface is reduced to the two antinodal points, and the Hubbard ladder is the main motivation behind this choice. In addition it was shown that in a 2-leg ladder array, the anomalous self-energy remains and leads to Fermi pockets. In the cuprates the onsite repulsion is strong, not weak as in the cases discussed above. The ansatz incorporates the strong coupling effects through Gutzwiller renormalization factors introduced by Zhang et al. \cite{RMF} earlier, to describe the strong coupling effects within a mean field approximation treatment of the strong coupling $t-J$ model. Note this renormalized mean field theory yielded a similar self energy to that used in the ansatz.

  The effectiveness of any phenomenological ansatz is measured by its success, or failure, to describe a wide range of experiments. In the present case there is a large body of experiments which determine the many anomalous properties of the pseudogap phase. The challenge to theory is to develop a consistent description covering this wide range. The account given above shows that the ansatz has had considerable success in describing these anomalies, both in a wide range of direct spectroscopic experiments and also in thermodynamic and other properties. 

   Our ansatz  has of course its limitations. It aims to describe the pseudogap state as an unstable fixed point, different from but analogous to  the Fermi liquid unstable fixed that describes the normal phase of standard metals. The key role of such unstable fixed points in condensed matter physics was emphasized some years ago by P.W. Anderson \cite{PWA02}. The fixed points are called unstable because at the lowest temperature scales, broken symmetry instabilities generally appear. We have not attempted to address possible instabilities here, apart from some discussion on the \textit{d}-wave pairing instability towards superconductivity. Experiments have shown that other instabilities appearing at temperatures below the onset temperature  $T^{*}$,  in high magnetic fields but also in zero fields in some cases as well. In addition there are recent reports of broken time reversal symmetry setting in at or near $T^{*}$ whose origin may well require going beyond the single band description. These are some of the open issues that remain in the challenge that the spectacular properties of the high $T_{c}$ cuprates pose to condensed matter theory. 

 We are grateful to our many collaborators on the physics of cuprates over the past quarter century. In particular the work discussed has benefited from our discussions with Elena Bascones, Jules Carbotte, Weiqiang Chen, Yan Chen, Carsten Honerkamp, Robert Konik, Andreas Laeuchli, Urs Ledermann, Patrick Lee, Karyn Le Hur, Elisabeth Nicol, Tai Kai Ng, Matthias Ossadnik, Manfred Salmhofer, Manfred Sigrist, Alexei Tsvelik and Qiang-Hua Wang. We would like to thank the many experimentalists for explanations and insights into their results, especially Peter Johnson, Jon Rameau, John Tranquada and Hong-Bo Yang at Brookhaven National Labs. Support from the MANEP program of the Swiss National Fund (KYY and TMR), the US Department of Energy under Contract No. DEAC02-98CH10886 (TMR), Boston College and DOE-DE-SC0002554 (KYY) and RGC grant 705608 and 707010 of HKSAR (KYY and FCZ) is gratefully acknowledged.

\end{document}